\documentclass[twocolumn,twoside]{article}

\usepackage{a4}
\usepackage{amssymb}
\usepackage{amsmath}
\usepackage[numbers,sort&compress]{natbib}
\usepackage{graphicx}
\usepackage{blindtext}

\begin{document}

\title{{\bf Radio pulsars: already fifty years!}\\
{\normalsize\sc 100 years of Uspekhi Fizicheskich Nauk Journal}}

\date{{\normalsize\textit{P N Lebedev Physical Institute, Russian Academy of Sciences,\\
Leninskii prosp. 53, 119991, Moscow, Russian Federation,\\
Moscow Institute of Physics and Technonogy (State University), \\
Institutsky per. 9, 141700, Dolgoprudny, Moscow Region, Russian Federation}}\\[1ex]
{\small \textit{Usp.\ Fiz.\ Nauk} \textbf{188}, 
377-408 (2018) 
[in Russian]\\
English translation: \textit{Physics -- Uspekhi}, \textbf{61}, 353-370 (2018)}
\\{\small Translated by K A Postnov; edited by A Semikhatov}
}

\author{V S Beskin}

\maketitle

\begin{abstract}             
Although fifty years have passed since the discovery of radio pulsars, there 
is still no satisfactory understanding of how these amazing objects operate. 
While there has been significant progress in understanding the basic properties 
of radio pulsars, there is as yet no consensus on key issues, such as the nature 
of coherent radio emission or the conversion mechanism of the electromagnetic 
energy of the pulsar wind into particle energy. In this review, we present the 
main theoretical results on the magnetosphere of neutron stars. We formulate a 
number of apparently simple questions, which nevertheless remain unanswered 
since the very beginning of the field and which must be resolved before any 
further progress can be made.
\end{abstract}

\setcounter{secnumdepth}{3}
\setcounter{tocdepth}{2}

\tableofcontents

\section{Pulsar chronicles}

\subsection{Prehistoric period (before 1967)}
\label{chronicle1}

The possible existence of neutron stars (with a mass of the order of the solar mass, 
$M \approx M_{\odot} = 1.99 \times 10^{33}$ g, having a radius $R$ of only 10--15 km) 
was known long before the discovery of radio pulsars. They were predicted by Baade 
and Zwicky [1] as early as the mid-1930{\footnote{There is a paper by Landau [2] 
discussing the possible existence of a star with a nuclear density, which was published 
a few months before the discovery of the neutron [3].}}. However, due to their very 
small diameter, it had been thought for a long time that it would be very difficult 
to detect single neutron stars. Just several months before the discovery of pulsars 
did Pacini [4] surmised that such stars should have very short rotational periods, 
\mbox{$P \sim 1$ s,} and superstrong magnetic fields, $B_{0} \sim 10^{12}$ G, and therefore 
should be powerful energy sources; however, it was not stated in [4, 5], of course, 
that neutron stars should be bright cosmic radio sources. For this reason, no dedicated 
searches for these objects were organized and pulsars were serendipitously detected 
in 1967 by Bell and Hewish during a different observational program [6]. We recall 
that the possibility of detecting thermal X-ray emission from accreting neutron stars 
in close binary systems was clearly formulated in [7-9], and X-ray pulsars were indeed 
discovered soon after the launch of the first X-ray telescope [10].

\subsection{Hellas (1968--1973)}
\label{chronicle2}

The period 1968-1973 is a remarkable era of simple and visual ideas, which nevertheless 
enabled intuitive understanding of the nature of physical processes in pulsars. Indeed, 
what can be apparently simpler than a magnetized ball rotating in a vacuum? Such a simple 
model turned out to be sufficient for describing the main properties of radio pulsars [11]. 
For example, it soon became quite clear that the rotation of a neutron star underlies the 
extremely stable pulse arrival time [12, 13] and the kinetic energy of rotation is the 
energy reservoir for the activity of radio pulsars.

Simultaneously, as mentioned above, the main idea was formulated according to which the 
energy release mechanism should be related to electrodynamic processes [4]. To date, 
the canonical expression for magnetic dipole energy losses,
\begin{equation}
W_{\rm tot}^{({\rm V})} = -I_{\rm r} \Omega \dot\Omega 
=  \frac{1}{6} \, \frac{B_0^2 \Omega^4 R^6}{c^3} \sin^2 \chi
\approx  10^{32} \frac{B_{12}^2}{P^{4}} \, \, {\rm erg}\, {\rm s^{-1}}
\label{Eqn01}
\end{equation}
(where $B_{0}$ is the polar magnetic field, $I_{\rm r} \sim M R^2$ is the moment of inertia, 
and $\chi$  is the magnetic dipole inclination angle with the spin axis), has been used to 
estimate radio pulsar energy losses. Here and below in similar formulas, we use the notation 
$B_{12} = B_{0}/(10^{12}$ G) and ${\dot P}_{-15} = {\dot P}/10^{-15}$, and express the period
$P = 2 \pi/\Omega$ in seconds. By inverting equation (1), we obtain a very simple expression 
for the magnetic field estimate
\begin{equation}
B_{12} \approx (P {\dot P}_{-15})^{1/2}.
\label{Eqn01''}
\end{equation}

We recall that the eureka moment was related exactly to expression (1), because after 
the discovery of a pulsar in the Crab Nebula ($P \approx 33$ ms, ${\dot P}_{-15} \approx 
420$){\footnote{Now $P \approx 34$ ms.}} two values already known by that time --- the 
total power of the Crab nebula $L_{\rm tot} \approx 5 \times 10^{38}$ erg s$^{-1}$ (this 
energy should be permanently injected into the nebula to provide its optical emission due 
to synchrotron losses) and the so-called dynamical age \mbox{$\tau_{\rm D} = P/2{\dot P} 
\approx 1000$ years} corresponding to the historical supernova 1054AD --- were naturally 
explained.

By the way, this simple model enabled the first step towards understanding that radio pulsars 
can be sources of cosmic rays; for millisecond pulsars (the fastest pulsar among the more than 
2600 known to date indeed has the period $P \approx 1.39$ ms) and sufficiently high magnetic 
fields $B_{0} \sim 10^{13}$ G, the potential difference between the pole and the equator of a 
rotating magnetized sphere, 
\begin{equation}
\Delta V \approx \left(\frac{\Omega R}{c}\right)B_{0}R, 
\end{equation}
reaches $10^{20}$ eV, i.e., corresponds to the maximum energy observed in cosmic rays. 
A similar estimate was also made by Ostriker and Gunn [14] for the energy of particles 
accelerated in an electromagnetic wave propagating from a rotating neutron star. If all 
radio pulsars were born with sufficiently shot periods (now it is fully clear that this 
is not the case [15]), the problem of the origin of cosmic rays could be solved [16,17].

However, in a few years, it had already become clear that a magnetized sphere rotating 
in a vacuum is too far away from reality. In 1971 Sturrock showed in [18] that in the 
super strong magnetic field of a pulsars, the process of single-photon conversion of hard 
gamma-ray photons into electron-positron pairs should play a crucial role
\begin{equation}
\gamma +(B) \rightarrow e^{+} + e^{-} + (B). 
\label{ggg}
\end{equation}
As a result, the neutron star magnetosphere is very quickly filled up with charged 
particles, which must inevitably result in the total restructuring of the braking 
mechanism of radio pulsars.

Indeed, the rotation of a magnetized neutron star in a vacuum inevitably leads 
to the appearance of a longitudinal (parallel to magnetic field) electric field 
$E_{\parallel} \sim (\Omega R/c)B$ outside the neutron star. Any primary charged 
particle entering this region would be accelerated up to ultrarelativistic energies,  
${\cal E}_{\rm e} \gg m_{\rm e} c^2$. Due to the extremely short time of synchrotron
losses, which in magnetic fields $B_{0} \sim 10^{12}$ G for electrons (and positrons) 
is only
\begin{equation}
\tau_{\rm s} \sim
\frac{1}{\omega_B}\left(\frac{c}{\omega_B r_{\rm e}}\right) \sim 10^{-15} \,
{\rm c},
\label{tau}
\end{equation}
(here and below $\omega_{B} = eB/m_{\rm e}c$ is the gyrofrequency and 
\mbox{$r_{\rm e} = e^2/m_{\rm e}c^2$} is the classical radius of an 
electron), the particles can move along magnetic field lines 
only. But because the magnetic field lines are curved, the 
primary particles must emit hard gamma-ray quanta with the 
characteristic energy (so-called `curvature radiation' [19])
\begin{equation}
{\cal E}_{\rm ph} = \hbar \omega_{\rm cur} \sim \frac{\hbar c}{R_{\rm c}} \, 
\left(\frac{{\cal E}_{\rm e}}{m_{\rm e}c^2}\right)^3,
\label{cur}
\end{equation}
where $R_{\rm c}$ is the curvature radius of a magnetic field line. 
Just the curvature radiation losses impose a bound on the particle
energy ${\cal E}_{\rm e}$. Indeed, writing the energy equation in
the form
\begin{equation}
\frac{{\rm d}{\cal E}_{\rm e}}{{\rm d}l} =
eE_{\parallel} - \frac{2}{3}\,\frac{e^2}{R_{\rm c}^2}
\left(\frac{\cal E_{\rm e}}{m_{\rm e}c^2}\right)^4,
\label{d10}
\end{equation}
we obtain the maximum energy of the primary particles
\begin{equation}
{\cal E}_{\rm max} \sim 
\left(\frac{R_{\rm c}^2 E_{\parallel}}{e}\right)^{1/4} m_{\rm e}c^2 
\sim (10^{7} - 10^{8}) \, {\rm MeV}.
\label{d10'}
\end{equation}

Next, by propagating almost rectilinearly in the curved magnetic field, 
the curvature radiation quanta start moving at increasingly large angles
$\theta_{\rm b}$ to the magnetic field, to ultimately produce secondary 
electron-positron pairs (the synchrotron photons generated by the secondary 
particles from their transitions to the ground Landau levels also play a 
role here). This is because for a photon with an energy ${\cal E}_{\rm ph}$
propagating at an angle $\theta_{\rm b}$ to the magnetic field ${\bf B}$, far 
from the threshold ${\cal E}_{\rm ph} = 2 m_{\rm e}c^2$ the probability of 
single-photon conversion (4) has the form [20]
\begin{equation}
w = \frac{3 \sqrt{3}}{16 \sqrt{2}} \,
\frac{e^3 B\sin\theta_{\rm b}}{\hbar m_{\rm e}c^3}
\exp\left(-\frac{8}{3}\frac{B_{\rm cr}}{B\sin\theta_{\rm b}}
\frac{m_{\rm e}c^2}{{\cal E}_{\rm ph}}\right),
\label{d8}
\end{equation}
where
\begin{equation}
B_{\rm cr} = \frac{m_{\rm e}^2c^3}{e\hbar} \approx
4.4 \times 10^{13} \, {\rm G}
\label{bh}
\end{equation}
is the critical magnetic field at which the energy gap between two adjacent 
Landau levels is of the order of the rest-mass energy of the electron:
$\hbar \omega_{B} = m_{\rm e}c^2$. Hence, for not too long mean free paths 
of photons, $l_{\gamma} < R$, we obtain the simple
estimate 
\begin{equation}
l_{\gamma} = \frac{8}{3\Lambda} R_{\rm c} \frac{B_{\rm cr}}{B}\,
\frac{m_{\rm e}c^2}{{\cal E}_{\rm ph}},
\label{d9}
\end{equation}
where the logarithmic factor for the characteristic parameters near the 
neutron star surface is
\begin{equation}
\Lambda \approx \ln\left[
\frac{e^2}{\hbar c}\,\frac{\omega_B R_{\rm c}}{c}
\left(\frac{B_{\rm cr}}{B}\right)^2
\left(\frac{m_{\rm e}c^2}{{\cal E}_{\rm ph}}\right)^2\right] \sim 20.
\end{equation}

We recall that unlike an electric field, the magnetic field cannot 
create particles. However, it can play the role of a catalyst, 
enabling the energy and momentum conservation for the process 
being considered. This is why the creation of a pair is prohibited 
for a photon propagating along the magnetic field lines: the 
probability of pair creation (9) is zero in this case.

As a result, the process of secondary particle creation precipitously 
increases due to the acceleration of secondary particles and their 
emission of curvature photons, etc. This process stops only when the 
secondary electron-positron plasma screens the longitudinal electric 
field $E_{\parallel}$. Incidentally, this means that radio pulsars 
should also be sources of positrons, albeit much less energetic than
follows from estimate (3).

Thus, it became clear that pulsar magnetospheres should be filled with 
electron-positron plasma. Therefore, since the 1970s, a more realistic 
magnetosphere model has been considered, in which the longitudinal 
electric field is fully screened ($E_{\parallel} = 0$). Indeed, the 
appearance of a longitudinal electric field in some region would 
immediately lead to abrupt plasma acceleration and hence to an explosive 
creation of secondary particles.

On the other hand, it is well known that in a fully screened longitudinal 
electric field, the plasma starts rigidly rotating together with the star, 
as in Earth's and Jupiter's magnetospheres [21]. It is this plasma property 
that proved to be crucial to understanding the pulsar activity.

First of all, the rigid rotation becomes impossible at sufficiently large 
distances from the rotation axis, \mbox{$r_{\perp} > R_{\rm L}$,} where 
the rotation velocity exceeds the speed of light (Fig. 1). Here,
\begin{equation}
R_{\rm L} = \frac{c}{\Omega} 
\label{RL}
\end{equation}
is the so-called light cylinder radius. For ordinary radio pulsars with periods 
of $0.1$--$1$ s, we have \mbox{$R_{\rm L} \sim$ $10^{9}$--$10^{10}$ cm,} which 
means that the light cylinder lies at a distance of thousand times the neutron 
star radius. Next, it is easy to estimate the size of the polar cap, 
$R_{\rm cap} \approx R_{0}$, where
\begin{equation}
R_{0} = \left(\frac{\Omega R}{c}\right)^{1/2}R, 
\label{R0}
\end{equation}
which is the region around the magnetic poles of a neutron star 
from which the magnetic field lines go beyond the light cylinder. 
For ordinary radio pulsars, the polar cap size is just several 
hundred meters. Hence, the polar cap area can be conveniently 
written as $s = f_{\ast} \pi R_{0}^2$, where $f_{\ast} \approx 1$.

The importance of polar caps stems from the fact that charged 
particles moving along magnetic field lines can escape the 
neutron star magnetosphere. As noted above, such a motion arises 
not only due to small Larmor radii compared to other characteristic 
scales but also because of an extremely short synchrotron cooling 
time. As a result, two sets of magnetic field lines appear. The open 
field lines coming out of the polar caps cross the light cylinder 
and go from the magnetosphere to infinity, whereas other field lines 
close inside the light cylinder. The plasma within the closed region
turns out to be trapped, while the particles moving along the open 
field lines can leave the magnetosphere.

That simple magnetospheric model allowed Radhakrishnan and Cooke [22] 
(and later Oster and Sieber [23]) to formulate the so-called `hollow
cone model', which remarkably explained all basic morphological 
properties of pulsar radio emission. Indeed, secondary plasma generation 
should be suppressed near the magnetic poles, where, thanks to almost 
rectilinear magnetic field lines, the curvature radiation intensity is 
also significantly reduced and, in addition, the curvature gamma quanta
emitted by relativistic particles propagate at small angles to the 
magnetic field, which also reduces the secondary pair creation 
probability.

\begin{figure}
\begin{center}
\includegraphics[width=220pt]{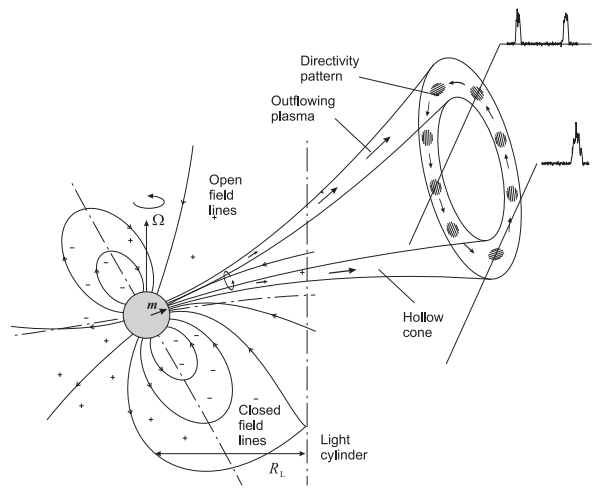}
\caption{Basic elements of the pulsar magnetosphere and the geometric
`hollow cone' model. The potential drop near the magnetic poles results in
additional rotation of the outflowing plasma around the magnetic axis.
Here and below, pluses and minuses indicate the sign of the charge density.
}
\label{fig01}
\end{center}
\end{figure}

As a result, as shown in Fig. 1, it is possible to assume that in the 
central part of the open field lines, the density of the outflowing 
plasma is strongly reduced. Now, assuming quite reasonably that the 
radio emission is directly related to the outflowing relativistic 
plasma, a significant decrease in the radio emission can be expected 
in the central part of the radio beam. As a result, for the lateral 
crossing of the directivity pattern, we should expect to observe a 
single mean pulse profile, while for the central crossing, a two-hump 
profile should be observed. Without going into the details, we note
that just such a situation occurs in reality [24].

In these first years of radio pulsar studies, three main parameters 
determining the key electrodynamic processes were defined. The first 
was the electric charge density that is needed to screen the longitudinal 
electric field near the neutron star surface,
\begin{equation}
\rho_{\rm GJ} = -\frac{{\bf \Omega \cdot B}}{2 \pi c}.
 \label{Eqn02a} 
\end{equation}
This quantity, introduced by Goldreich and Julian in 1969 [25], determines 
the characteristic particle number density $n_{\rm GJ} = |\rho_{\rm GJ}|/|e|$ 
(of the order of $10^{12}$ cm$^{-3}$ near the neutron star surface) and also 
the characteristic current density, $j_{\rm GJ} = c \rho_{\rm GJ}$, which is 
much more important. As we see in what follows, it is the longitudinal electric 
current circulating in the magnetosphere that will play the key role. The second 
parameter is the particle multiplication factor $\lambda$, 
\begin{equation}
\lambda = \frac{n_{\rm e}}{n_{\rm GJ}},
 \label{Eqn02b} 
\end{equation}
which shows how much the secondary particle number density exceeds the critical 
number density $n_{\rm GJ}$. Finally, the third parameter is the so-called 
magnetization parameter $\sigma_{\rm M}$ 
\begin{equation}
\sigma_{\rm M} = \frac{1}{4 \lambda} \, \frac{e B_{0}\Omega^{2} R^{3}}{m_{\rm e}c^4},
\label{Eqn02c} 
\end{equation}
also introduced in 1969 by Michel [26]. It is equal to a maximum possible 
Lorentz factor of particles, $\gamma_{\rm max}$, which is achieved if all 
the energy $W_{\rm tot}$ (1) is converted into the hydrodynamic particle flow 
${\dot N} m_{\rm e}c^2 \Gamma$. Here 
\begin{equation}
{\dot N} = \lambda \pi R_{0}^2 n_{\rm GJ}c 
\end{equation}
is the electron-positron pair injection rate, and $\Gamma$ here and below 
denotes the hydrodynamic Lorentz factor of the outflowing plasma. In formula 
(17), we added the subscript `M' because $\sigma$ is now commonly used for 
another quantity, $\sigma = \sigma_{\rm M}/\Gamma = W_{\rm em}/W_{\rm part}$, 
the ratio of the electromagnetic energy flux to the particle energy flux. Using 
formula (1), we can rewrite definition (17) in the very simple form [27]
\begin{equation}
\sigma_{\rm M} = \frac{1}{\lambda}\left(\frac{W_{\rm tot}}{W_{\rm A}}\right)^{1/2}.
\label{newsigma}
\end{equation}
where $W_{\rm A} = m_{\rm e}^{2}c^{5}/e^{2} \approx 10^{17}$ 
erg s$^{-1}$ is the minimum power of the `central engine' 
enabling particle acceleration to relativistic energies 
($\sigma_{\rm M} \sim 1$ for $\lambda = 1$ and $W_{\rm tot} = W_{\rm A}$).

Thus, during the first several years after the discovery of radio pulsars, 
answers to most of the key questions were obtained (the periodicity of 
pulses is related to rotation, the energy source is the kinetic energy 
of rotation, the energy release is due to the electrodynamic mechanism). 
Naturally, it remained to be understood how all this works. It was
necessary to understand how the energy is carried away from a rotating 
neutron star to infinity; what the energy spectrum of the outflowing plasma 
is; and, of course, what the mechanism of the observed coherent radio 
emission is (and this mechanism indeed must be coherent, because the 
brightness temperature is usually $T_{\rm br} \sim 10^{28}$ K [15] and 
in individual giant pulses can be as high as  $10^{38}$  K [28]). To 
date, the answers to most of these questions are unknown.

\subsection{Rome (1973--1983)}
\label{chronicle3}

In this period, the first rigorous laws were formulated concerning all 
the main topics of the field, including secondary plasma generation, 
the pulsar magnetosphere structure, and the pulsar wind problem. First 
of all, two detailed models of electron-positron pair generation near 
the neutron star surface were proposed. This process is made possible 
due to the continuous plasma outflow along open field lines, which 
results in the formation of a region with a longitudinal electric 
field (a `gap') above the polar cap. Below, we refer to this region as 
the inner gap, to distinguish it from other regions in the magnetosphere 
where longitudinal electric fields can appear. The height of the gap 
is determined by the secondary particle generation mechanism. At that
time, most of the secondary particles were thought to have been created 
outside the acceleration region of the primary particles, where the 
longitudinal electric field is already small and the secondary plasma 
can freely leave the neutron star magnetosphere.

The first model was proposed in 1975 by Ruderman and Sutherland [29], 
as well as by Eidman's group [30]. This model assumed that the particle 
ejection from the star surface is insignificant, because it was thought 
at that time [31+35] that the work function of particles of the neutron 
star surface $A_{\rm w} \sim 1$--$5$ keV, which, for example, determines 
the cold emission current [36, 37]
\begin{equation}
 j(E) = \frac{e^3 \, B_{0}}{8 \pi^2 \hbar c \, (2 m_{\rm e}A_{\rm w})^{1/2}} E \, 
\exp \left[-\frac{(8 m_{\rm e}A_{\rm w}^3)^{1/2}}{3 \hbar e E}\right]
\label{cold}
\end{equation}
is sufficiently high{\footnote{The difference between the pre-exponential factor 
and the classical Fowler-Nordheim formula [38] is due to the quantum effect of the
magnetic field, which changes the density of electron states in the neutron star 
crust.}}. Accordingly, thermal emission was ignored as well [39] (we discuss it 
in more detail in Section 2.1.2). As a result, from an analysis of Eqns (6) and
(9)-(11), it is easy to obtain an estimate of the potential drop needed for the 
secondary plasma creation [29]:
\begin{eqnarray}
\psi_{\rm RS} & \approx & \frac{m_{\rm e}c^2}{e}
\left(\frac{\hbar}{m_{\rm e}c}\right)^{-3/7} R_{\rm c}^{4/7}
R_{\rm L}^{-1/7} \left(\frac{B_{0}}{B_{\rm cr}}\right)^{-1/7} 
\nonumber \\
 & \approx & 6 \times 10^{12} \, B_{12}^{-1/7} P^{-1/7}R_{7}^{4/7} \, {\rm V}.
\label{psiRS}
\end{eqnarray} 
Here $R_{7} = R_{\rm c}/(10^{7}$ cm). Accordingly, the gap height is
\begin{eqnarray}
H_{\rm RS} & \approx & \left(\frac{\hbar}{m_{\rm e}c}\right)^{2/7}R_{\rm c}^{2/7}
R_{\rm L}^{3/7} \left(\frac{B_{0}}{B_{\rm cr}}\right)^{-4/7}
\nonumber \\
 & \approx & 10^{4} \, B_{12}^{-4/7} P^{3/7}R_{7}^{2/7} \, {\rm cm}.
\label{H_RS}
\end{eqnarray} 
We recall that the condition $\psi_{\rm RS} = \psi_{\rm max}$ (or, equivalently,
$H_{\rm RS} = R_{0}$), where
\begin{equation}
\psi_{\rm max} = \frac{1}{2} \, \left(\frac{\Omega R}{c}\right)^{2}RB_{0}
\label{d45}
\end{equation}
is the maximum possible potential drop in the inner gap region, is the 
mathematical expression for the `death line' on the $P-{\dot P}$ diagram 
(Fig. 2), below which secondary plasma generation (and hence the activity 
of a neutron star as a radio pulsar) is impossible. However, later on, 
when more exact calculations in [40-43] showed that the work function of 
electrons is a factor of 10 lower than assumed before ($A_{\rm w} \sim 100$ eV), 
Arons's group proposed an alternative model in which the potential drop
was noticeably smaller,
\begin{eqnarray}
\frac{\psi_{\rm A}}{\psi_{\rm RS}} & \approx & \left(\frac{\Omega R}{c}\right)^{1/2}. 
\label{psiAA}
\end{eqnarray}
It is this model with free particle ejection that has been considered as the most 
appropriate one in the subsequent thirty years, despite the obvious problems that 
it has (for example, in its original version, particles were generated only in the 
half of the polar cap closest to the pole).

\begin{figure}
\begin{center}
\includegraphics[width=250pt]{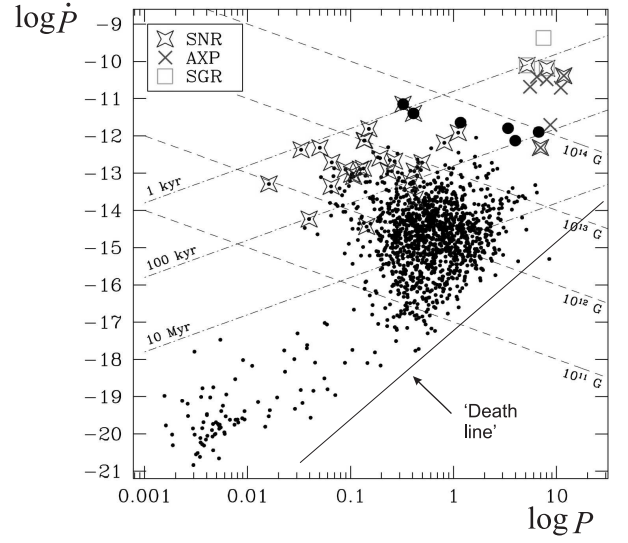}
\caption{The `death line' on the period $P$---period derivative ${\dot P}$
plot corresponding to the Ruderman-Sutherland model [29]. SNR---pulsars
in supernova remnants, AXP---anomalous X-ray pulsars, SGR---soft
gamma-ray repeaters.
}
\label{fig02}
\end{center}
\end{figure}

These models have been important because they approximately determined the 
multiplicity factor of secondary particles l and their energy spectrum. 
Starting from the first paper by Daugherty and Harding in 1982 [47], it 
became clear that the plasma multiplicity factor cannot exceed $10^{4}$--$10^{5}$. 
In particular, this meant that the particle number density near the neutron 
star surface, $\lambda n_{\rm GJ} \sim 10^{16}$ cm$^{-3}$, is too low for the
511 keV annihilation line to be detected. Accordingly, the magnetization parameter 
$\sigma_{\rm M}$ cannot exceed $10^{3}$--$10^{4}$ for most pulsars and can be 
as high as $10^6$ only for young pulsars (Crab, Vela).
 
We note that the values of $\lambda$ and $\sigma_{\rm M}$ given above can be 
easily estimated from the following simple considerations. An analysis of Eqn (7) 
shows that particles in the inner gap indeed reach energies comparable to the 
maximum value ${\cal E}_{\rm max}$ in (8). This means that the energy transferred 
to the curvature radiation photons is also of the order of 
${\cal E}_{\rm max} = e \psi_{\rm RS}$. Then the plasma multiplicity $\lambda$ 
can be estimated as
\begin{equation}
\lambda \sim \frac{{\cal E}_{\rm max}}{{\cal E}_{\rm min}},
\end{equation}
where ${\cal E}_{\rm \min}$ is the minimum energy of a photon capable of
creating an electron+positron pair. This value can be easily derived from 
Eqn (11) by setting $l_{\gamma} = R$ because for large photon mean free 
paths $l_{\gamma} > R$, the neutron star magnetic field starts rapidly 
decreasing along the gamma-quantum trajectory. As a result, for ordinary 
pulsars, we obtain
\begin{equation}
{\cal E}_{\rm min} \approx 
\frac{B_{\rm cr}}{\Lambda B} \, \frac{R_{\rm c}}{R} m_{\rm e}c^2
\sim 10^2 \, {\rm MeV},
\label{anaemin}
\end{equation}
which yields
\begin{equation}
\lambda \sim  10^{5} R_{\rm c, 7}^{-3/7} P^{-1/7} B_{12}^{6/7}.
\label{analambda}
\end{equation}
To estimate $\sigma_{\rm M}$, it suffices to use (19), whence we find
\begin{equation}
\sigma_{\rm M} 
\approx  \frac{{\cal E}_{\rm min}}{{\cal E}_{\rm max}}
\left(\frac{W_{\rm tot}}{W_{\rm A}}\right)^{1/2}
\sim 10^4.
\label{anasigma}
\end{equation}

We stress that the photon energy ${\cal E}_{\rm \min}$ does not generally 
coincide with the characteristic maximum energy in the energy spectrum of 
the secondary particles. This is related to already mentioned synchrotron 
losses due to which the secondary particles can rapidly lose most of their 
energy soon after creation. Passing to the reference frame in which the 
photon propagates perpendicular to the external magnetic field, it is possible 
to show that after the transition to low-lying Landau levels, the secondary 
particle energy is
\begin{equation}
{\cal E}_{\rm e} \approx  \frac{R_{\rm c}}{R} m_{\rm e}c^2,
\label{Ee}
\end{equation}
(which differs little from (26), however, for magnetic fields $B \sim 10^{12}$ G).

We see that both estimates (26) and (29) demonstrate that the secondary particle 
energy spectrum essentially depends on the curvature of magnetic field lines. For 
a dipole magnetic field with the characteristic curvature radius 
$R_{\rm c} \approx R \, (\Omega R/c)^{-1/2} \sim 10^{7}$--$10^{8}$ cm in the polar 
cap region , the 
characteristic particle Lorentz factor is $\gamma_{\rm min} \approx 100$. Exactly 
these values for the maximum in the energy spectrum of secondary particles were 
obtained in both the pioneering paper [47] and all subsequent papers [48-51] 
calculating the outflowing plasma energy spectrum. If, as is frequently assumed 
recently [52, 53], the near-surface dipole magnetic field is strongly distorted by
the multipole components, significantly reducing the curvature radius $R_{\rm c}$ 
(which is required for the `death line' in the Arons model to be consistent with 
observations), then the maximum in the secondary particle energy spectrum can
decrease to $\gamma_{\rm min} \approx 3$--$10$.

The above formulas can also be used to estimate the electron-positron plasma ejection 
rate ${\dot N}$ and the total number of particles $N = \int {\dot N}{\rm d}t$ ejected 
during the pulsar lifetime:
\begin{eqnarray}
{\dot N} & \approx & \lambda \, 
\frac{B_{0}\Omega^2 R^3}{ce} \sim 3 \times 10^{34} \, P^{-2} B_{12} \, {\rm s}^{-1}, 
\label{Eqn03a} \\
N & \approx & \lambda \, \frac{Mc^2}{e B_{0}R} \sim 10^{50} \, B_{12}^{-1}.
 \label{Eqn03b} 
\end{eqnarray}
Because the total number of neutron stars in the Galaxy does not exceed $10^9$ [15], 
we can conclude that radio pulsars cannot be the main source of cosmic ray positrons.

On the other hand, there are presently sufficiently reliable (although still indirect) 
observational data that radio pulsars are indeed the sources of electron-positron plasma. 
For example, an analysis of emission from the Crab Nebula related to the pulsar wind 
suggests the pair multiplicity from $\lambda \sim 10^{6}$ [54] to $\lambda \sim 10^{7}$ 
[55]; for the pulsar PSR B0833-45 (Vela), a similar estimate yields $\lambda \sim 10^{5}$ 
[56]. This agrees with the estimate $\lambda \sim 10^{6}$ obtained in the synchrotron 
absorption model of pulsar emission in the pulsar wind of a binary system containing the 
pulsar PSR J0737-3039 [57]. Finally, as is presently actively discussed, radio pulsars 
could be responsible for the positron excess in the energy range 1--100 GeV detected by 
the PAMELA experiment [58, 59].

\begin{figure}
\begin{center}
\includegraphics[width=150pt]{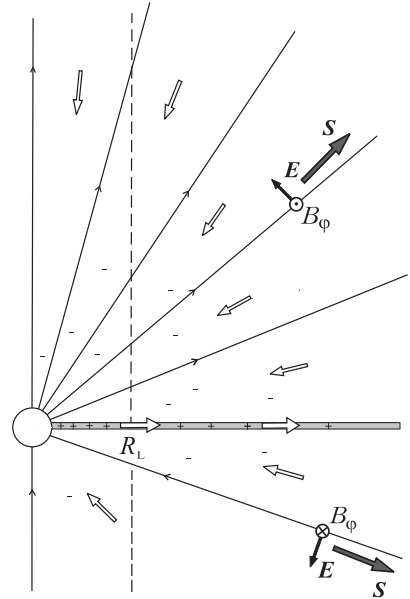}
\caption{Michel's monopole solution [66] in which the energy release is
due to the electromagnetic energy flux at a null frequency (dark arrows).
Light arrows indicate the direction of electric currents.
}
\label{fig03}
\end{center}
\end{figure}

Very important results have also been obtained in the theory of pulsar magnetospheres 
and pulsar wind. First of all, Mestel [60], Michel [61], Okamoto [62], and many others 
(see, e.g., [63, 64]) formulated an axisymmetric force-free `pulsar equation' for 
the magnetic flux  $\Psi(r_{\perp}, z)$ 
\begin{eqnarray}
-\left(1-\frac{\Omega_{\rm F}^2r_{\perp}^2}{c^2}\right)\nabla^2\Psi
+\frac{2}{r_{\perp}}\frac{\partial \Psi}{\partial r_{\perp}}
-\frac{16\pi^{2}}{c^2}I\frac{{\rm d}I}{{\rm d}\Psi}
\nonumber \\
+\frac{r_{\perp}^{2}}{c^2}\left(\nabla\Psi\right)^{2}
\Omega_{\rm F}\frac{{\rm d}\Omega_{\rm F}}{{\rm d}\Psi}=0,
\label{d39}
\end{eqnarray}
where the total electric current within the tube $I(\Psi)$ and the 
angular velocity $\Omega_{\rm F}(\Psi)$ determining the electric field 
$${\bf E} = - \frac{\Omega_{\rm F}}{2\pi c} \, \nabla \Psi$$  depend 
on the magnetic flux only. This nonlinear equation with a singularity 
at the light cylinder allowed determining the magnetic field structure because 
$${\bf B}_{\rm p} = \frac{[\nabla \Psi \times {\bf e}_{\rm \varphi}]}{2 \pi r_{\perp}}.$$
Therefore, it is not surprising that this equation established itself 
for many years as the main tool in theoretical studies of radio 
pulsars{\footnote{Equation (32) is the relativistic generalization of 
the Grad-Shafranov equation [65] for cold plasma.}}. 

In particular, one of the analytic solutions, which could be obtained only for a very 
special class of the functions $I(\Psi)$ and $\Omega_{\rm F}(\Psi)$ determining the 
poloidal current density $j_{\parallel} = j_{\rm GJ}$ and the electric field ${\bf E} $ 
showed that a monopole solution can be constructed for the Goldreich current 
$j_{\parallel} = j_{\rm GJ}$ with the  electric field being lower than the 
magnetic field up to infinity [66]:
\begin{eqnarray}
B_{r} & = & B_{\rm L} \frac{R_{\rm L}^2}{r^2}, \\
B_{\varphi} & = & E_{\theta} = 
- B_{\rm L} \, \frac{\Omega R_{\rm L}^2}{c r} \, \sin\theta.
\end{eqnarray}
Here, $B_{\rm L} = B_{0}(\Omega R/c)^3$ is the magnetic field on the light
cylinder, and we present the solution in the upper half-plane. The longitudinal 
current exactly corresponds to a purely radial motion of massless particles 
with the speed of light:
\begin{eqnarray}
j_{r} = \rho_{\rm GJ} c = 
- \frac{\Omega B_{\rm L}}{2 \pi} \, \frac{R_{\rm L}^2}{r^2}\, |\cos\theta|.
\end{eqnarray}
As shown in Fig. 3, in this solution the longitudinal electric currents 
(contour arrows) generate a toroidal magnetic field $B_{\varphi}$, which 
together with the induction electric field $E_{\theta}$ induced by rotation 
forms a radial electromagnetic energy flux (Poynting vector)
\begin{eqnarray}
S_{r} = 
\frac{B_{\rm L}^2\Omega^2 R_{\rm L}^4}{4 \pi c r^{2}} \sin^2\theta,
\label{Eqn04} 
\end{eqnarray}
that carries energy away from the neutron star. Thus, the possibility of 
a magneto-hydrodynamic wind leading to the neutron star spin-down was 
demonstrated. 

We emphasize that all energy losses for such an axisymmetric (and 
stationary) force-free solution are related to the Poynting  vector flux. 
However, in contrast to a magnetodipole wave, the energy flux occurs at 
the zero frequency. Of course, the presence of an equatorial current 
sheet separating the incoming and outgoing magnetic field fluxes appeared 
to be rather artificial initially. However, as we see in what follows, 
this solution played a fundamental role in later investigations.

On the other hand, our group (Beskin, Gurevich, and Istomin (BGI) [67]) 
obtained an analytic solution for an oblique rotator, but for a zero 
longitudinal electric current in the neutron star magnetosphere (Fig. 4a). 
In this case, energy losses were shown to be zero for any angle $\chi$ because,
for a zero longitudinal current, the condition $B_{\varphi}(R_{\rm L}, z) = 0$
must hold on the light cylinder for any inclination angle. This effect, 
later confirmed by Mestel's group [68] (Fig. 4b) occurs because the plasma 
filling the neutron star magnetosphere fully screens the magnetodipole 
radiation of the central star. Therefore, all energy losses must be related 
to the braking torque ${\bf K}$ caused by the action of the longitudinal 
currents in the neutron star magnetosphere.

\begin{figure}
\begin{center}
\includegraphics[width=\columnwidth]{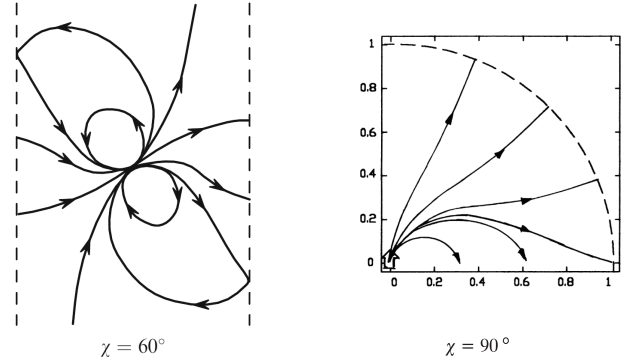}
\caption{Structure of magnetic field lines for an oblique rotator with a null
longitudinal current for the angles (a) $\chi = 60^{\circ}$  ($xz$ plane) [67] 
and (b) $\chi = 90^{\circ}$  ($xy$ plane) [68].
}
\label{fig04}
\end{center}
\end{figure}

It is convenient in what follows to decompose the braking torque 
${\bf K}$ into two components, parallel and perpendicular to the 
magnetic dipole ${\bf m}$. We also introduce the dimensionless current 
density $i = j_{\parallel}/j_{\rm GJ}$, and decompose it into the 
symmetric and antisymmetric components $i_{\rm s}$ and $i_{\rm a}$, 
depending on whether the corresponding component has the same or 
different sign in the north and south parts of the polar cap. It is easy to 
verify that $K_{\parallel} \propto i_{\rm s}$ and $K_{\perp} \propto i_{\rm a}$. 
Here and below, we normalize it to the `local' Goldreich current density 
$j_{\rm GJ} = |{\bf \Omega}\cdot{\bf B}|/2\pi$. In particular, the 
direct effect on the star of the Amp\`ere force 
\mbox{${\bf K} = \int [{\bf r} \times [{\bf J}_{\rm s} \times {\bf B}]/c] \, {\rm d} s$,} 
due to the surface currents ${\bf J}_{\rm s}$ closing the volume 
longitudinal currents in the magnetosphere can be written as [69]
\begin{eqnarray}
K_{\parallel}^{\rm cur}  & \approx & - c_{\parallel} \, \frac{B_{0}^{2}\Omega^{3}R^{6}}{c^{3}} 
i_{\rm s},
\label{16'} \\
K_{\perp}^{\rm cur}  & \approx & - c_{\perp} \, \frac{B_{0}^{2}\Omega^{3}R^{6}}{c^{3}}
\left(\frac{\Omega R}{c}\right)i_{\rm a}.
\label{17'}
\end{eqnarray}
Here, the coefficients $c_{\parallel} \approx 1$ and $c_{\perp} \approx 1$ (which 
we do not discuss in what follows) depend on the actual current distribution in 
the open line volume. For the `local' Goldreich current ($i_{\rm s} = i_{\rm a} = 1$), 
Eqn (38) yields
\begin{equation}
K_{\perp}^{\rm cur} \approx \left(\Omega R/c\right) K_{\parallel}^{\rm cur}.
\end{equation}

Returning to the evolution of the angular velocity $\Omega$  and
the inclination angle $\chi$, we can write the equations of motion
in the general form [69, 70]
\begin{eqnarray}
I_{\rm r} \, \dot{\Omega} & = &  
K_{\parallel}^{\rm A} + (K_{\perp}^{\rm A}-K_{\parallel}^{\rm A})\sin^2\chi, \\
I_{\rm r} \, \Omega \, {\dot\chi} & = &  
(K_{\perp}^{\rm A}-K_{\parallel}^{\rm A})\sin\chi\cos\chi,
\label{Eqn05}
\end{eqnarray}
where, again, $I_{\rm r} \sim M R^2$ is the neutron star moment of inertia,
and we set $K_{\parallel} = K_{\parallel}^{\rm A} \cos\chi$ and 
$K_{\perp} = K_{\perp}^{\rm A} \sin\chi$. Clearly, both equations involve
the factor $(K_{\perp}^{\rm A}-K_{\parallel}^{\rm A})$, and therefore the
angle $\chi$ evolves to $90^{\circ}$ (counter-alignment) if the total 
energy losses decrease at large inclination angles and to zero
(alignment) otherwise. For example, for the local Goldreich current 
($i_{\rm s}^{A} = i_{\rm a}^{A} = 1$), as assumed in the BGI model [69], the
angle $\chi$ increases with time:
\begin{eqnarray}
W_{\rm tot}^{({\rm BGI})} = 
i_{\rm s}^{\rm A}(\Omega, B_{0}, \chi) \, \frac{f_{\ast} ^{2}(\chi)}{4} \,
\frac{B_{0}^2\Omega^4 R^6}{c^3} \, \cos^2\chi, \\
{\dot \chi}^{({\rm BGI})} = 
i_{\rm s}^{\rm A}(\Omega, B_{0}, \chi) \, \frac{f_{\ast} ^{2}(\chi)}{4 I_{\rm r}} \,
 \frac{B_{0}^{2}\Omega^{2}R^{6}}{c^{2}} \sin\chi \, \cos\chi.
\label{Eqn07}
\end{eqnarray}
Here, the dimensionless factor $1.59 < f_{\ast}(\chi) < 1.96$ determines the polar 
cap area $s(\chi)$,
\begin{equation}
s(\chi) = f_{\ast}(\chi)  \pi R_{0}^2. 
\end{equation}
As regards the dimensionless currents $i_{\rm a}^{\rm A}(\Omega, B_{0}, \chi)$ 
and $i_{\rm s}^{\rm A}(\Omega, B_{0}, \chi)$, we consider them in more detail 
in Sections 2.2.1 and 3.1.1 below.

As we see, already in the first years, the main processes responsible for the 
evolution of radio pulsars were understood and the main laws describing pulsar 
magnetospheres were formulated. Moreover, several analytic solutions enabling 
the first cautious predictions were obtained. Apparently, all the remaining 
problems were to be fully clarified quite soon.

Unfortunately, that did not happen. Due to the absence of any significant progress 
in solving nonlinear equations describing pulsar magnetospheres (analytic solutions 
could be obtained only in some model cases, and numerical methods were not 
sufficiently developed at that time), the number of astrophysicists actively 
working in this field sharply decreased. Difficult times had come.

\subsection{Dark ages (1983--1999)}
\label{chronicle4}

This was a hard time indeed, especially for the theory of radio pulsar 
magnetospheres. At first glance, no significant results were obtained 
in the field in these 15 years. However, slowly, step by step, our 
understanding of processes in neutron star magnetospheres was becoming 
more and more clear. 

First of all, important results were obtained in the theory of strongly 
magnetized pulsar wind. We recall that the force-free approximation (i.e., 
the approximation postulating the vanishing of only the electromagnetic 
force) suggests nothing about the outflowing plasma energy, because such 
an approximation is equivalent to massless particles. Therefore, in this 
period, a full MHD theory of relativistic and nonrelativistic flows was 
actively being developed [71-76]. In particular, this theory showed that 
the acceleration of particles must be strongly suppressed in a quasi-spherical
magnetized wind. As first shown by Tomimatsu in 1994 [77], at long distances 
(more precisely, beyond the fast magnetosonic surface, $r \gg r_{\rm F}$, 
where $r_{\rm F} \sim \sigma_{\rm M}^{1/3} \, R_{\rm L}$), the particle 
energy cannot exceed $\sigma_{\rm M}^{1/3} \, m_{\rm e} c^2 \sim 1$--$10$ 
GeV, and hence the electromagnetic-to-particle energy flux ratio 
 $\sigma = W_{\rm em}/W_{\rm part}$ must be high: 
 $\sigma \sim \sigma_{\rm M}^{2/3} \gg 1$. 

Simultaneously, it was realized that the longitudinal electric current 
density $j_{\parallel}$, as well as the accretion rate in the Bondi 
solution, is not a free parameter but is fixed by the critical conditions 
on the fast magnetosonic surface. In the relativistic case, the longitudinal 
current density must be close to the Goldreich current density $j_{\rm GJ}$.
Another step forward was related to the recognition of the significant role 
played by general-relativity effects in the process of particle generation 
near the neutron star magnetic poles. Mathematically, they are due to an 
additional term appearing in the expression for the Goldreich-Julian charge
density 
$$\rho_{\rm GJ} \approx - \frac{(\Omega - \omega)B}{2\pi \, c},$$
related to the Lense-Thirring angular velocity $\omega$
\begin{equation}
\omega \approx \frac{2G I_{r}}{c^2 \, r^3} \, \Omega.
\end{equation}
According to general relativity, this is the angular velocity of the spacetime 
`drag' at a distance $r$ from any rotating body. Despite the smallness of this 
quantity, its spatial derivative could be sufficiently large. As shown in 1990 
in [78] (and later in [79, 80]), in the Arons model, secondary plasma generation
becomes possible inside the entire polar cap precisely due to general relativity 
effects.

\begin{figure}
\begin{center}
\includegraphics[width=200pt]{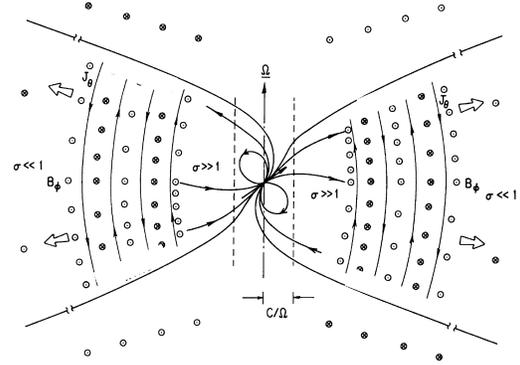}
\caption{Model of the `striped' pulsar wind with a wave-like current sheet
separating oppositely directed magnetic fluxes in the equatorial plane [84].
}
\label{fig05}
\end{center}
\end{figure}

Here, we also note papers [81-83], which showed that particle creation in the 
inner gap should be significantly affected by the inverse Compton scattering 
of thermal photons from the neutron star surface on the relativistic electrons 
and positrons that are accelerated inside the gap. Hard gamma quanta generated 
in this process should also lead to single-photon pair creation. Presently, 
this process is taken into account in most of the works devoted to particle
creation near the pulsar surface.

Next, important results were obtained in the pulsar wind theory. First of all, 
Coroniti [84] and Michel [85] drew attention to a wave-like (striped) current 
sheet that must arise in a strongly magnetized wind from an oblique rotator
by separating the incoming and outgoing magnetic fluxes (Fig. 5). Later, 
Kennel and Coroniti [86, 87], by analyzing pulsar wind interaction with the 
Crab Nebula, concluded that at long distances fromthe pulsar, comparable 
to the size of the nebula, the wind magnetization should be very weak: 
$\sigma \sim 10^{-2}$. This result was already in direct contradiction with 
the pulsar wind theory predictions discussed above. Since then, the
`$\sigma$-problem', i.e., the impossibility of sufficiently efficient particle 
acceleration in quasi-spherical flows, has become one of the major problems 
in the radio pulsar theory. In any case, in the MHD approximation, it remains 
unsolved so far.

\begin{figure}
\begin{center}
\includegraphics[width=200pt]{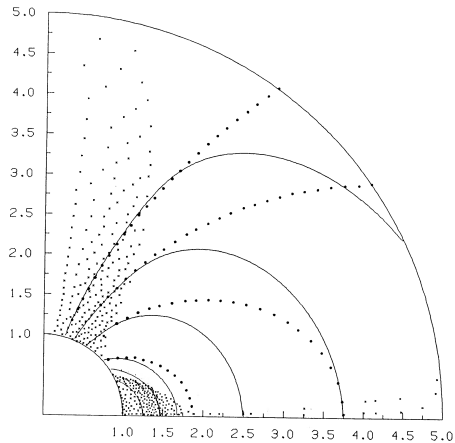}
\caption{First `crazy' idea [88]: the `disk-dome' model of an axisymmetric magnetosphere. Solid lines show the equipotential electric
surfaces, dashed lines are magnetic surfaces.
}
\label{fig06}
\end{center}
\end{figure}

As usual in difficult times, several `crazy' ideas were proposed to solve a 
heap of problems \mbox{(Figs 6 and 7).} First, Michel and Krause-Polsdorff [88] 
considered the so-called `disk-dome' structure of the axisymmetric pulsar 
magnetosphere, in which positive and negative charges are captured in
different parts of the neutron star magnetosphere separated by vacuum 
gaps{\footnote{Qualitatively, a similar structure was discussed earlier 
by Rylov [89] and Jackson [90].}}. At first glance, it was totally unclear 
how such a structure could be stable when taking secondary particle creation 
into account. However, later, as we see in what follows, this structure was 
indeed reproduced in numerical simulations.

\begin{figure}
\begin{center}
\includegraphics[width=200pt]{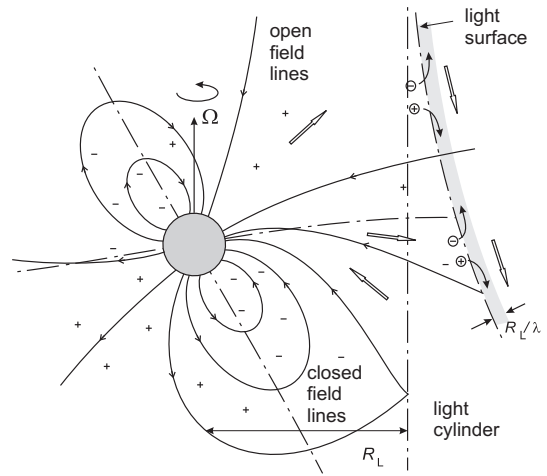}
\caption{Second `crazy' idea [69]: a model of the magnetosphere with the
light surface near which, in a narrow layer $\sim R_{\rm L}/\lambda$, 
effective particle acceleration occurs. Light arrows show the direction 
of electric currents; pluses and minuses are the signs of the electric 
charge density. 
}
\label{fig07}
\end{center}
\end{figure}

In addition, our group considered the case where the longitudinal current 
$j_{\parallel}$  is small enough to sustain the MHD flow up to infinity [69]. 
In this model, shown in Fig. 7, the pulsar magnetosphere must have a `natural
boundary' --- a light surface at which the electric field becomes equal to 
the magnetic field. This enables effective particle acceleration up to energies
${\cal E}_{\rm e} \approx \sigma_{\rm M} m_{\rm e}c^2$ (here, the $\sigma$-problem 
is solved as well!); as we see in what follows, this conclusion was also later
confirmed, albeit indirectly.

We stress that such a structure should inevitably arise in the Arons 
model, which postulates a `local' Goldreich longitudinal current 
$j_{\parallel} = c \rho_{\rm GJ} \approx \Omega B \cos\chi/2\pi$, 
i.e., a longitudinal current that is insufficient to sustain the 
pulsar wind. Indeed, as discussed in Section 1.3 (see Fig. 3), in 
the MHD wind, the toroidal magnetic field $B_{\varphi}$ should 
match the electric field on the light cylinder. But for an oblique 
rotator, the local Goldreich longitudinal current is too small to 
create the necessary toroidal magnetic field. As a result, in the 
BGI model, a weakly magnetized wind, i.e., a flow with low $\sigma$,
should be formed already close to the light cylinder.

\subsection{Renaissance (1999--2006)}
\label{chronicle5}

The Renaissance epoch started from two papers published in 1999 and is related 
to the recognition of the validity of simple models that had already been proposed 
to explain processes in pulsar magnetospheres. In the first paper, Contopoulos,
Kazanas, and Fendt [91] finally solved force-free `pulsar equation' (32) for an 
axisymmetric magnetosphere numerically (Fig. 8). This became possible due 
to an iterative procedure to avoid the light cylinder singularity (we already
stressed that the light cylinder is a singular surface for the pulsar equation).

As a result, as in the monopole Michel solution, the solution contained an 
equatorial current sheet at long distances $r > R_{\rm L}$. But in the inner 
regions, the solution, naturally, was matched to the dipole magnetic field 
of the neutron star. Several years later, this solution was reproduced
in many studies [92-99]{\footnote{Half of the authors of these papers were 
Russian-speaking astrophysicists.}}, which renewed interest in the theory
of pulsar magnetospheres. Of course, all these solutions corresponded to the 
axisymmetric case, which still did not allow obtaining the full information 
on real pulsar magnetospheres, 

\begin{figure}
\begin{center}
\includegraphics[width=200pt]{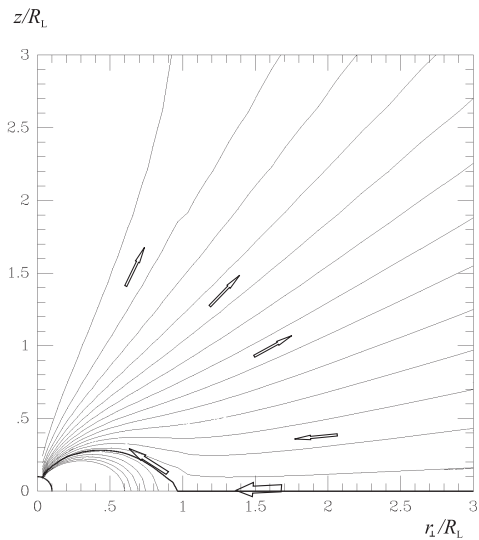}
\caption{Axisymmetric force-free solution found numerically in [91].
Light arrows show the direction of electric currents.
}
\label{fig08}
\end{center}
\end{figure}

In the same 1999, Bogovalov [100] found an analytic
solution for the so-called `inclined split monopole' \mbox{(Fig. 9):}
\begin{eqnarray}
B_{r} & = & B_{\rm L} \frac{R_{\rm L}^2}{r^2} \, {\rm Sign}(\Phi), \\
B_{\varphi} & = & E_{\theta} 
= - B_{\rm L} \, \frac{\Omega R_{\rm L}^{2}}{c r} 
\, \sin\theta  \, {\rm Sign}(\Phi).
\label{Eqn08}
\end{eqnarray}
In this solution, the current sheet near the neutron star separating radial 
magnetic fluxes is not orthogonal to the spin axis. The quantity
\begin{eqnarray}
\Phi = \cos\theta \cos\chi - \sin\theta \sin\chi 
\cos\left(\varphi - \Omega t + \Omega r/c \right),
\label{Eqn08new}
\end{eqnarray}
exactly defines the current sheet form. As a result, near the neutron star, the 
magnetic field has not a dipole but a monopole structure. However, this simple 
analytic solution was very important for the pulsar wind. In this solution, 
inside the cones \mbox{$\theta < \pi/2 - \chi$} and $\pi - \theta < \pi/2 - \chi$ 
around the rotation axis, electromagnetic fields remained time-independent and 
coincided with the Michel solution shown in Fig. 3. On the other hand, in the 
equatorial plane, all electromagnetic field components, as assumed before (see 
Fig. 5), changed sign jump-wise when the current sheet crossed a given point; at 
all other times, the field remained constant{\footnote{Already for this reason 
this solution did not include a magnetodipole wave.}}.

\begin{figure}
\begin{center}
\includegraphics[width=220pt]{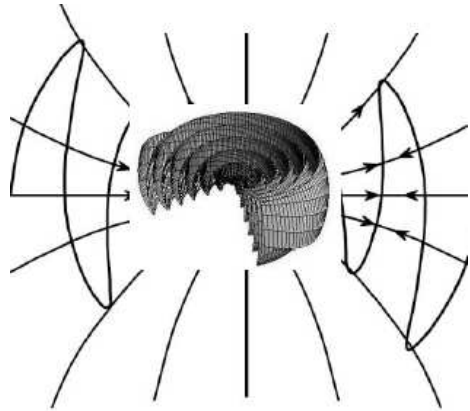}
\caption{Structure of Bogovalov's analytic solution for an `inclined split
monopole' [100] and the form of the current sheet reproduced on a 3D
printer.
}
\label{fig09}
\end{center}
\end{figure}

We note that the condition $\Phi = 0$ for the current sheet form has 
a purely kinematic nature. This is because in the free-force case, 
both particles and the current sheet move radially with the speed 
of light. As a result, in the limit  $r \rightarrow \infty$, the 
relation $E_{\theta}(\theta) = B_{\varphi}(\theta)$  must hold for 
an arbitrary dependence on the angle $\theta$ for any wind structure; 
such a stationary asymptotic solution of the `pulsar equation' was
first obtained by Ingraham as early as 1973 [101]. Therefore, it
is not surprising that a similar structure of the current sheet
has been constantly reproduced in subsequent numerical
calculations.

Thus, in this time period, real progress was made. First of all, 
the important role played by the current sheet in the wind dynamics 
was recognized. Therefore, studies of processes inside the current 
sheet became the focus of radio pulsar studies. In particular, 
Lyubarsky and Kirk [102] noted the importance of magnetic reconnection, 
which initiated many studies. Finally, almost all papers considering 
an axisymmetric magnetosphere confirmed the magnetosphere
structure obtained in [91], thus suggesting the existence of
some `universal' solution.

But the main point was apparently related to the change of the 
viewpoint on the nature of the longitudinal current. The value 
of $j_{\parallel}^{({\rm us)}}$ determined from the `universal 
solution' rather than defined by the particle creation process 
near magnetic poles was now taken as the true longitudinal 
electric current. In other words, in all subsequent numerical 
calculations, no constraints on the longitudinal electric 
current amplitude were imposed. If the particle density in some 
region became too low in the MHD approach, plasma was injected 
into that region artificially. It is not surprising therefore 
that all solutions obtained in this way satisfied both conditions
$\rho_{\rm e} = \rho_{\rm GJ}$ and $j_{\parallel} = j_{\parallel}^{({\rm us)}}$.

On the other hand, new questions arose. First of all, the spatial 
distribution of the longitudinal current $j_{\parallel}$ following 
from the 'universal solution' was significantly different from the local 
Goldreich-Julian current, which, as we especially mentioned above, 
was predicted by the Arons model. Nor did this distribution match 
the current from a rotating split monopole. In particular, as shown 
in Fig. 8, the `universal solution' obtained for an axisymmetric 
magnetosphere required the inverse current to flow not only along the
separatrix separating the closed and open magnetic field lines but 
also in a sizeable volume outside it. This directly contradicted all  
models of particle creation near magnetic poles that existed at that 
time, because, in this case, the sign of the potential drop and hence 
the direction of the outflowing current had to be the same over the 
entire polar cap area. 

Although this difficulty was not explicitly 
discussed at that time, the first attempts to overcome this problem 
were undertaken. In particular, Shibata [103] and later Beloborodov 
[104] showed that the secondary plasma creation should be significantly 
suppressed for a sufficiently small longitudinal current 
$j_{\parallel} < j_{\rm GJ}${\footnote{This statement, however, 
pertained only to the models with free particle ejection from the 
neutron star surface.}}. 

\subsection{Industrial revolution (2006--2014)}
\label{chronicle6}

The industrial revolution in the radio pulsar theory was related to the 
emerging possibility of carrying out 3D time-dependent calculations. 
Like any revolution, it resolved all accumulated problems in a purely 
technical way. 

This epoch started from the work by Spitkovsky [105], who was the first 
to obtain a numerical solution for the force-free magnetosphere of an 
oblique rotator. This was the first 3D solution with the pulsar wind 
outflowing to infinity. As shown in Fig. 10, despite the relative 
smallness of the computation domain, this solution confirmed the 
existence of the current sheet. 

\begin{figure}
\begin{center}
\includegraphics[width=240pt]{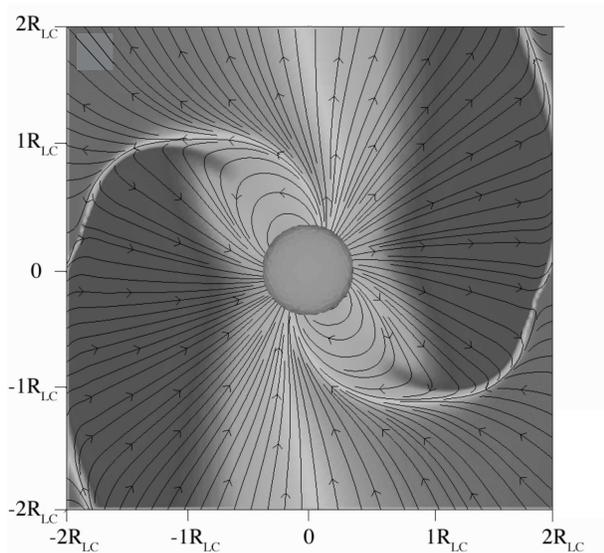}
\caption{ Structure of magnetic field lines in Spitkovsky's force-free
solution [105] for an oblique rotator ($\chi = 60^{\circ}$) in the $xz$ 
plane.
}
\label{fig10}
\end{center}
\end{figure}

As is well known, an industrial revolution is impossible in one individual 
country. After Spitkovsky's work, similar calculations were performed in 
many research centers, and not only in the force-free but also in the full 
MHD approximation [106-111]. As a result, the `universal solution' was also 
found for an oblique rotator, which, in particular, confirmed the formula 
for the total energy losses obtained by Spitkovsky{\footnote{More precisely, 
the interpolation of numerical results yields 
$W_{\rm tot} \propto (k_{1} + k_{2}\sin^2\chi)$,
where $k_{1} \approx 1.0 \pm 0.1$ and $k_{2} \approx 1.1 \pm 0.1$.}}: 
\begin{equation}
W_{\rm tot}^{({\rm MHD})} \approx  
\frac{1}{4} \, \frac{B_{0}^2\Omega^4 R^6}{c^3} \, (1 + \sin^2\chi). 
\label{WMHD}
\end{equation}
We see that the `universal solution' suggests an increase in
losses with the angle $\chi$. Therefore, according to general
relation (41), the inclination angle should decrease with time:
\begin{equation}
{\dot \chi}^{({\rm MHD})} \approx  
-\frac{1}{4I_{\rm r}} \frac{B_{0}^{2}\Omega^{2}R^{6}}{c^{2}} \sin\chi \, \cos\chi
\label{Eqn09}
\end{equation}
(this result was obtained somewhat later).

\begin{figure}
\begin{center}
\includegraphics[width=220pt]{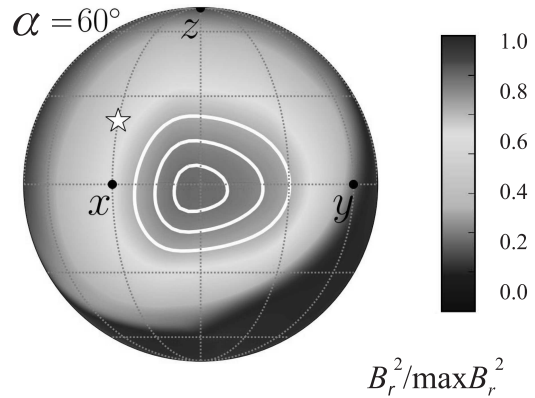}
\caption{MHD solution [112] demonstrating the magnetic field line
concentration ($r = 6 \, R_{\rm L}$) in the direction at an angle 
of approximately $30^{\circ}$ to the magnetic axis (the light star).
}
\label{fig11}
\end{center}
\end{figure}

On the other hand, the `universal solution' was shown to significantly 
differ from Michel-Bogovalov monopole solution (47). Notably, for 
sufficiently large angles \mbox{$\chi > 30^{\circ}$,} the radial magnetic 
field was not homogeneous but concentrated near the equatorial plane. 
The fields averaged over the angle $\varphi$ had the form
\begin{eqnarray}
<B_{r}>_{\varphi} & \approx & B_{\rm L} \frac{R_{\rm L}^2}{r^2} 
\, \sin\theta \, {\rm Sign}(\Phi),  
\label{Eqn10a} \\
<B_{\varphi}>_{\varphi} & = & <E_{\theta}>_{\varphi} \nonumber \\
&&\approx - B_{\rm L} \, \frac{\Omega R_{\rm L}^2}{cr} 
\, \sin^2\theta  \, {\rm Sign}(\Phi), \\
<S_{r}>_{\varphi} & \approx & 
 \frac{B_{\rm L}^2\Omega^2 R_{\rm L}^4}{4 \pi c r^2} \sin^4\theta,
\label{Eqn10}
\end{eqnarray}
whereas in the Michel-Bogovalov solution the radial magnetic field 
$B_{r} = B_{\rm L}(R_{\rm L}^2/r^2) \, {\rm Sign}(\Phi)$ in (47) does not 
depend on the angle $\theta$.

We stress that the fields in Eqns (51)-(53) are averaged over the angle 
$\varphi$. In fact, as shown in Fig. 11, there is a noticeable concentration 
of the force lines (and hence the electromagnetic energy flux) in the 
direction rotated longitudinally through approximately $30^{\circ}$ 
relative to the magnetic axis (the light star) [107, 112]. Thus, the 
magnetic field is not time-independent between the current sheet crossings 
of a given point. Moreover, for large angles $\chi$, the current sheet 
becomes more and more pronounced, and hence the asymptotic solution 
for an orthogonal rotator can be approximated with good accuracy as{\footnote{To 
reconcile the total losses with expression (49), we must set 
$B_{\rm L} = \sqrt{15/8}(\Omega R/c)^3 \, B_{0}$ here.}}
\begin{eqnarray}
B_{r} & \approx & B_{\rm L} \frac{R_{\rm L}^2}{r^2} \, 
\sin\theta \cos(\varphi -\Omega t + \Omega r/c - \varphi_{0}),  
\label{Eqn10new1} \\
B_{\varphi} & = & E_{\theta} 
\label{Eqn10new2} \\
& \approx & - B_{\rm L} \, \frac{\Omega R_{\rm L}^2}{cr} 
\, \sin^2\theta  \, \cos(\varphi  -\Omega t + \Omega r/c - \varphi_{0})
\nonumber 
\end{eqnarray}
(where $\varphi_0 \approx 30^{\circ}$). This is another significant 
difference from the Michel-Bogovalov solution, in which, as follows
from (47) suggests, the electromagnetic fields are time-independent 
outside the current sheet. In Section 2.3.1, we discuss the nature 
of such a pulsar wind in detail, emphasizing its variability.

As noted above, asymptotic solutions with an arbitrary dependence on the 
angle $\theta$ have been known since the 1970s. Surprising, however, was 
the fact that these simple solutions were reproduced in three-dimensional 
MHD simulations. In particular, the general relation between the energy 
flux averaged over the angle $\varphi$ and the radial magnetic field
\begin{equation}
S_{r}(\theta) \propto B_{r}^2(\theta) \, \sin^2\theta 
\label{S}
\end{equation}
holds with good accuracy (it can be easily obtained using the
definitions of the ${\bf E}$ and ${\bf B}$ fields).

In addition, an analysis of the obtained solutions shows that the 
dimensionless antisymmetric longitudinal current significantly 
exceeds the local Goldreich current: 
$i_{\rm a}^{\rm A} \sim (\Omega R/c)^{-1/2} \gg 1$ (see Section 
2.2.1 for more details). Due to the normalization introduced above, 
this means that the total current circulating in the magnetosphere 
of an oblique rotator is close to the total current flowing in an 
axisymmetric magnetosphere. This should be the case because, 
independently of the angle $\chi$, for the MHD wind outgoing to 
infinity to exist, the toroidal magnetic field on the light cylinder 
$B_{\varphi}(R_{\rm L})$  should be close to the poloidal field
$B_{\rm p}(R_{\rm L})$. This implies that the total current $I$ 
should be weakly dependent on $\chi$.

On the other hand, we stress that the value of this current
is insufficient to explain the energy losses in the `universal
solution' (49) for $\chi \approx 90^{\circ}$ only due to volume 
currents circulating in the magnetosphere. Indeed, according to
Eqn (38), this requires the antisymmetric current to be much
larger: $i_{\rm a}^{\rm A} \sim (\Omega R/c)$. Of course, we 
must not forget that in all numerical calculations, the neutron 
star size was not less than 10\% of the light cylinder radius 
(for ordinary pulsars, this value is hundredths of a percent). 
However, no significant dependence of the pulsar wind parameters 
on the stellar size was found. Anyway, the problem of the pulsar 
braking mechanism remained unsolved.

At the same time, a novel view on the role of the `universal current'
allowed Timokhin [113] (and later Timokhin and Arons [114]) to 
numerically simulate the process of particle creation near magnetic 
poles, for the first time including the possible nonstationarity of 
this process. The longitudinal current $j_{\parallel}^{({\rm us)}}$ 
was regarded as an external constant parameter. Pair creation was 
shown to be possible for a sufficiently large longitudinal current 
$j_{\parallel} > j_{\rm GJ}$  and, as already obtained earlier [103, 
104], impossible for lower currents. Moreover, particle creation was 
shown to occur in the inverse current region at the main magnetic 
field feet near the separatrix; in this case, the plasma density 
in the generation region should be not lower but higher than the 
Goldreich-Julian charge density. This becomes possible exactly 
because the particle creation is essentially nonstationary.

Here, however, we note that these calculations assumed one-dimensional 
plasma flow, and this did not allow believing that all essential 
points are taken into account. For example, the change in the 
toroidal magnetic field was ignored, which, as was shown already 
a long time ago [36], could significantly affect the plasma generation 
dynamics (see the Appendix). Nevertheless, these papers significantly 
contributed to the understanding of the very possibility of generation 
of the required longitudinal current, which is significantly different 
from the local Goldreich current. Here, in fact, we returned to the 
Ruderman-Sutherland model [115] because, despite the free particle 
ejection from the surface, the electric potential drop in the particle 
generation region turns out to be much larger than in the Arons model.

Summarizing, we can conclude that the `industrial revolution' enabled a 
leap forward in the understanding of the basic processes occurring in 
pulsar magnetospheres. It is very important that most of the results were 
confirmed by different research groups. Nevertheless, some key questions 
remained unanswered. One of the main issues was whether particle generation 
provides the necessary longitudinal current $j_{\parallel}^{({\rm us)}}$ 
determined by the `universal solution'. For example, for an orthogonal 
rotator (and for a period $P \sim$ 1 s), the longitudinal current should be
$(\Omega R/c)^{-1/2} \sim 100$ times as high as the local Goldreich current; 
this high parameter had not been used in numerical simulations.

\subsection{Modern times (after 2014)}
\label{chronicle7}

Formally, this is just the next stage in the numerical simulations related 
to the use of the particle-in-cell (PIC) method. In fact, a qualitative step 
forward has been made, because the kinetic treatment alone, unlike one-fluid 
MHD and moreover the force-free case, enabled self-consistently extending the 
models to particle generation, i.e., to a systematic description of both the 
region with the longitudinal electric field $E_{\parallel}$ and the particle 
injection into the calculation domain. This, in addition, enabled the description
of particle acceleration beyond the light cylinder (during the `industrial 
revolution' time, these effects were modeled by introducing an effective 
resistance [116, 117]). In other words, PIC simulations allowed {\it ab initio}
calculations, at least in principle [118].

\begin{figure}
\begin{center}
\includegraphics[width=200pt]{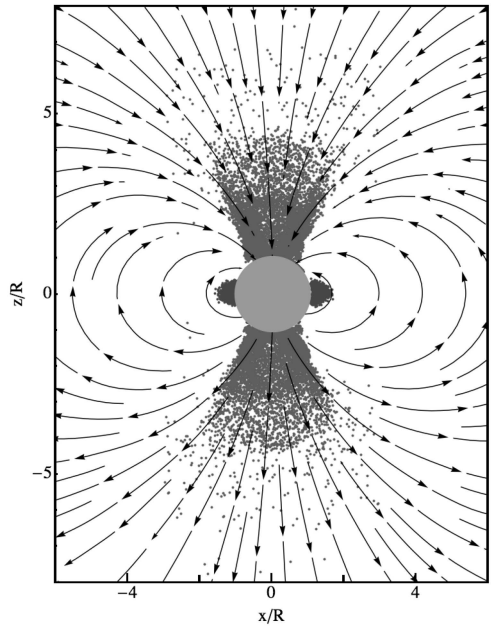}
\caption{First PIC calculation results: axisymmetric magnetosphere ignoring 
general relativistic effects [119].
}
\label{fig12}
\end{center}
\end{figure}

Of course, this epoch, in fact, has just begun, and therefore not all 
results mentioned below have been tested in independent calculations. 
Nevertheless, despite `teething problems' (natural in this case), the 
new possibilities offered by this method have already given several 
interesting results{\footnote{Again, about half of the active researchers 
in this field speak Russian.}}.

These studies started from two papers by Philippov and Spitkovsky [119] 
and by Chen and Beloborodov [120] (see also [121]). Their primary focus 
was on testing whether, in the absence of particle generation in the 
magnetosphere (i.e., only for free ejection of particles from the neutron 
star surface), no magnetized pulsar wind arises, but the `disk-dome' 
structure we already discussed in Section 1.4 is formed (Fig. 12). The 
force-free solution with pulsar wind arose only if the particle generation 
occurred inside the total magnetosphere volume.

\begin{figure}
\begin{center}
\includegraphics[width=220pt]{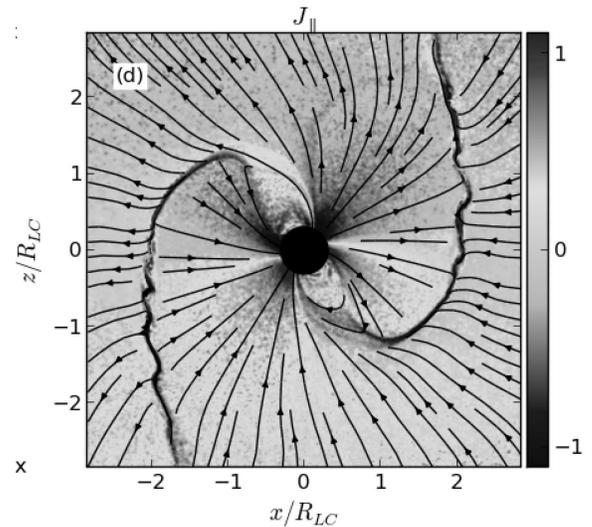}
\caption{First PIC calculation results: the structure of magnetic field lines
for an oblique rotator [122]. 
}
\label{fig13}
\end{center}
\end{figure}

Next, Philippov, Spitkovsky, and Cerutti [122], using a more realistic 
particle creation model, reproduced the structure obtained in the MHD 
simulations. In particular, as shown in Fig. 13, they confirmed the 
existence of the current sheet. Thus, the existence of the `universal 
solution' was also checked. In addition, effective particle acceleration
beyond the light cylinder to maximal energies $\gamma \approx \sigma_{\rm M}$ 
was shown to be possible [123, 124]. This process was made possible exactly 
because of the appearance of regions where the electric field exceeds the 
magnetic one (which is related to the decrease in the magnetic field inside 
the equatorial current sheet). Another reason for which particle acceleration 
turns out to be effective is the magnetic reconnection, because the electric 
field arising in this process leads to particle drift toward the current 
sheet.

As a result, according to the authors of [122], already at short distances
$r < 5\, R_{\rm L}$ from the light cylinder, up to 30\% of the total 
electromagnetic energy of the wind is transferred to the particles. However, 
it is currently difficult to say whether this solves the $\sigma$-problem; nonetheless, 
a certain step forward has undoubtedly been made. In any case, such effective
particle acceleration should definitely help to explain the high-energy 
radiation from pulsars detected by the Fermi gamma-ray observatory [125].

On the other hand, kinetic calculations posed more new questions than 
produced answers. Indeed, already the first results obtained in [119, 
120] for an axisymmetric magnetosphere unexpectedly showed the 
lack of particle generation near magnetic poles. Instead, as shown 
in Fig. 12, the free particle ejection from the neutron star surface 
(which, we recall, is postulated in the calculations!) leads not to 
the `universal solution' but to the appearance of the `disk-dome'
structure mentioned above. Later, such a solution was shown to also 
appear for not very large angles $\chi < 30^{\circ}$. This effect 
is directly related to the required longitudinal current
$j_{\parallel}^{({\rm us})}$ being less than $\chi < 30^{\circ}$; 
as mentioned in Section 1.4, no particle generation occurs in this 
case. Fortunately, this problem was successfully solved very soon 
afterwards [118]. It turned out that the general relativity effects 
(which, as we know, change the Goldreich-Julian charge density) 
lead to the reversal of the inequality 
($j_{\parallel}^{({\rm us})} > j_{\rm GJ}$), which enabled the particle 
generation process. As a result, as shown in Fig. 13, the `universal
solution' was generally reproduced in the framework of this approach. 

A second, more serious, problem proved to be connected to the inverse
current formation. By the present time, this current has been obtained 
or the fastest radio pulsars only, in which electromagnetic fields and 
article densities near the light cylinder are high enough for secondary 
lectron-positron plasma generation. In other words, presently, it is 
ot possible to close the current (and hence to provide the `universal 
olution') without additional particle creation beyond the light cylinder. 
In this case, as is also shown in [119, 120], a `disk-dome' like 
structure appears.

One way or another, studies in this field actively continue, and we can
hope that many points will be clarified in the nearest future.

\section{Several awkward questions}

\subsection{Old and forgotten...}

Success in studies of radio pulsar magnetospheres achieved in the last 
several years is quite impressive. Nevertheless, we should not forget 
that many explicit and implicit assumptions have been adopted in 
numerical simulations, which undoubtedly affect the generality of the 
results obtained. This especially relates to PIC simulations because, 
in fact, these calculations have just begun.

On the other hand, there are several questions that the author remembers 
from as early as the mid-1970s and which have not received reasonable 
answers so far. These, of course, include one of the principal questions 
of the radio pulsar theory: the question of the nature of coherent radio 
emission. Because of a lack of space, we do not consider it. But it is 
also impossible to ignore it among other unsolved issues.

Unfortunately, many of these questions are barely discussed now, although 
the corresponding theoretical and observational studies, in our opinion, 
would enable significant progress in the understanding of radio pulsar 
activity. Even more regrettable is the fact that radio pulsar observers
are not inclined to carry out test studies to check the predictions of the 
theory.

\subsubsection{Axis orientation.}
\label{3.1.1}

If we take virtually any catalogue with data on the inclination angle $\chi$
between the magnetic and rotation axes [24, 126-128], we discover that this 
angle lies in the range $0^{\circ}$--$90^{\circ}$. Of course, this by no 
means implies that there are no radio pulsars with the inclination angle 
$\chi$ in the range $90^{\circ}$--$180^{\circ}$. This is simply because 
most of the existing methods for determining this angle do not distinguish
between $\chi$ and $180^{\circ} - \chi$. Only additional information, for
example, from X-ray and gamma-ray observations, allows estimating the 
inclination angle in the entire possible range [129-133].

On the other hand, from the physical standpoint, acute and obtuse angles 
between the axes correspond to two principally different conditions because 
for small $\chi$, the longitudinal current outflowing from the polar cap 
area (with its sign determined by the sign of the Goldreich-Julian charge 
density $\rho_{\rm GJ} = - {\bf \Omega} \cdot {\bf B}/2 \pi c$) corresponds 
to negative charges, whereas for angles $\chi$ exceeding $90^{\circ}$ 
it corresponds to positive ones. Clearly in the models of secondary plasma
generation in which the particle ejection from the neutron star surface is 
important, these are two essentially different cases, at least because the 
work function of positive charges must be different from that of electrons. 
Indeed, the ejection of ions, and not positrons, which are absent in the 
neutron star crust, should be considered here. As we have seen, this relates 
to the Arons model first and foremost; in the Ruderman-Sutherland model, the 
particle creation is insensitive to the axis orientation.

Therefore, one of the most interesting questions actually facing the 
radio pulsar theory is understanding whether the necessary secondary 
electron-positron plasma generation can occur for arbitrary angles. 
If it is proved, as was shown in recent paper [133] using a small 
sampling of 26 pulsars, that the number of radio pulsars with 
inclination angles $\chi < 90^{\circ}$ is approximately equal 
to that with inclination angles $\chi > 90^{\circ}$ (with the 
properties of radio emission showing no sign of separation into 
two groups), then it might be possible to conclude that particle 
ejection is hardly significant for the secondary electron-positron 
plasma generation{\footnote{Of course, it is not ruled out that 
during the very formation of a radio pulsar, the inclination angle 
can take certain values. However, presently, there are no definitive 
claims regarding this issue.}}.

As stressed in Section 1.7, all numerical calculations of
pulsar magnetospheres carried out in recent years have
assumed the free particle ejection, i.e., a sufficiently small
work function $A_{\rm w}$. However, this also remains an open issue.
The point is not only in the accuracy achieved so far in
determining the work function [134, 135]; it turned out that
the chemical composition of the neutron star surface layer is
unknown: it may not consist of iron atoms, as most studies
have assumed.

The chemical composition of the surface layers of the
polar caps can significantly change due to their bombardment
by energetic particles accelerated by the longitudinal electric
field in the inner gap. In addition, according to [136], iron
atoms (which certainly are produced in the largest numbers
because they have the most stable nuclei) could `sink down' in
the gravitational field in the first several years after the
neutron star birth, when its surface is undoubtedly not solid.
Therefore, it is not ruled out that, in fact, the neutron star
surface layers consist of not iron but much lighter atoms like
hydrogen and helium. But because the melting temperature,
which can be roughly estimated by the formula [137]
\begin{equation}
T_{\rm m} \approx 3.4 \times 10^{7}
Z_{26}^{5/3} \, \rho_{6}
\label{Tsur}
\end{equation}
(where $\rho_{6}$  is the surface layer density in units $10^6$ g cm$^{-3}$ 
and $Z_{26} = Z/26$ depends on the atomic number $Z$, the neutron
star surface at a temperature $\sim 10^6$K, which is typical of
ordinary radio pulsars, should be liquid and, in any case,
should not prevent free particle ejection. Modern models of
thermal emission of radio pulsars are based just on this
picture [138, 139].

\subsubsection{Surface heating.}
\label{3.1.2}

The surface heating problem is also an old and very interesting question. 
As we noted above, in the 1960s, the smallness of the neutron star surface 
was one of the reasons why dedicated searches for them were not undertaken. 
Presently, the sensitivity of space X-ray telescopes enables successful 
observations of both accreting X-ray pulsars and proper thermal emission 
from the nearest isolated neutron stars. Thermal emission has been detected
not only from `active' radio pulsars but also from neutron stars that do 
not demonstrate radio emission for some reason [140-142].

We do not discuss the problems related to the interpretation of spectra 
here (a super strong magnetic field significantly changes the properties 
of the atmosphere and surface layers of neutron stars, which essentially 
distorts the thermal radiation spectrum [139]), but touch upon the total 
thermal power. Most of the radio pulsars from which thermal radiation is
observed were found to have sufficiently small spin periods. With decreasing 
the spin period $P$, the relative contribution of the X-ray luminosity $L_{\rm X}$  
with respect to the total energy losses $W_{\rm tot}$ generally decreases, 
and for spin periods $P \sim 0.3$ s, the ratio $L_{\rm X}/W_{\rm tot}$ is 
only a fraction of a percent. Incidentally, this dependence is the complete 
opposite of the one for gamma-ray emission (for fast radio pulsars, the 
gamma-ray luminosity is close to the total energy loss of neutron stars). 
However, this behavior has a quite simple explanation.

Indeed, the obvious heating source of the polar caps (and hence of the 
entire neutron star) is the flux of secondary particles accelerated in 
the inner gap directed toward the neutron star. Their total power is 
proportional to the product of the potential drop $\psi$ by the total 
flowing current $I$, whereas the full power $W_{\rm tot}$ due to current 
energy losses is equal to the product $I\psi_{\rm max}$. As we have seen, 
the ratio $\psi/\psi_{\rm max}$, independently of the particle generation 
model, decreases as the angular velocity $\Omega$ increases. Thus, the 
decrease in the ratio $L_{\rm X}/W_{\rm tot}$ with $\Omega$ can be regarded 
as a confirmation of the heating mechanism due to the inverse particle flux. 
In any case, thermal emission from neutron stars provides a clear upper 
bound for this flux.

Polar cap heating can significantly affect the particle
ejection rate from the neutron star surface. For example,
papers assuming the ejection of positively charged particles,
i.e., for angles $\chi > 90^{\circ}$ (see, e.g., [143, 144]) 
typically use the following formula for the thermal current [39]:
\begin{eqnarray}
j(T) = j_{\rm GJ} \, \exp\left(30 - \frac{A_{B}}{T}\right)
\, T_{\rm keV}^{1/2} \, P \, B_{12}^{-1} \, Z_{26} \, A_{56}^{-3/2} \, \rho_{4},
\label{CK80}
\end{eqnarray}
where $A_{B}$ is the binding energy of ions with the atomic number $A$ and 
charge $Z$; and $\rho_{4}$ is the surface density in units of $10^4$ g cm$^{-3}$; 
of course, this formula is valid only for $j(T) < j_{\rm GJ}$. As we see, 
the nonlinear exponential dependence of the thermal current $j$ on the 
temperature $T$ can significantly influence the particle generation rate.

As regards the various models of secondary plasma generation in the inner 
gap, we note that the original model of secondary plasma generation proposed 
by Ruderman and Sutherland, in which the longitudinal current is equal to the
Goldreich current ($j_{\parallel} \approx j_{\rm GJ}$) and the potential drop 
in the gap is $\psi \approx \psi_{\rm RS}$, Eqn (21), undoubtedly contradicts 
the observed surface heating, because in this model the inverse particle flux
is comparable to the particle flux outflowing from the pulsar magnetosphere. 
On the other hand, the Arons model [46], in which the inverse particle flux
\begin{eqnarray}
\frac{j_{\rm back}}{j_{\rm GJ}} & \approx  &
\left(\frac{\Omega R}{c}\right)^{1/2} 
\label{psiA}
\end{eqnarray}
and potential drop (24) amount to only a fraction of a percent of those in 
the Ruderman-Sutherland model, could seem to successively solve the surface 
heating problem. However, as stressed above, this model was rejected for 
other reasons.

The recent studies devoted to particle generation [113, 114], whose 
results, as already stressed, are very similar to the results of the 
Ruderman-Sutherland model, revived the surface heating problem, because 
for pulsars with sufficiently small periods the surface temperature 
should be much higher than actually observed. Presently, there is no
satisfactory answer to this question.

\subsubsection{Account for the potential drop in the `inner gap'.}
\label{3.1.3}

The next question is related to the need to account for the potential
$\psi$ drop in the inner gap above the neutron star surface in the
pulsar wind region. We recall that in our model (which we consider 
in Section 3 below), this potential plays a decisive role. However, 
in numerical simulations (with a possible exception of [113]), the 
additional potential $\psi$ has never been taken into account. Clearly, 
this assumption has been based on solid grounds, because for rapidly 
rotating pulsars, which, in fact, have been the only ones considered 
so far, the potential drop needed for the secondary plasma generation
has always been smaller than the maximum value $\psi_{\rm cmax}$ given
by formula (23). But for pulsars located near the `death line', this 
is clearly not the case, and the potential drop effect can be significant.

Indeed, as was already shown in [29], a nonzero potential $\psi$ in 
the region of open magnetic lines leads to additional plasma rotation 
around the magnetic axis (see Fig. 1). For an oblique rotator, this 
is observed as the subpulse drift phenomenon [145, 146]. A regular 
drift has already been found in 97 radio pulsars [147]; in 53 of them, 
subpulses shift in the negative direction relative to the pulse phase 
$\phi$ and in 44 in the positive direction. An approximate equality 
should indeed take place for an arbitrary orientation of the observer
relative to the magnetic axis. 

Everywhere below, we consider a model of an ideally conducting sphere, 
implying that the freezing-in condition
\begin{equation}
{\bf E}_{\rm in} + [{\pmb{\beta}}_{\rm R}\times{\bf B}_{\rm in}] = 0.
\label{corotat}
\end{equation}
holds inside the star. Here and below, by definition,
\begin{equation}
{\pmb{\beta}}_{\rm R} = \frac{[{\bf \Omega} \times {\bf r}]}{c}.
\label{betaR}
\end{equation}
On the other hand, assuming quasi-stationarity (with the coordinate $\varphi$ 
and time $t$ entering all expressions only in the combination $\varphi - \Omega t$), 
the Maxwell equation corresponding to the Faraday law is well known to admit 
the form $\nabla \times ({\bf E} + [{\pmb{\beta}}_{\rm R} \times {\bf B}]) =0$ [60], 
whence
\begin{eqnarray}
{\bf E} + [{\pmb{\beta}}_{\rm R}\times{\bf B}] = -\nabla \psi.
\label{E}
\end{eqnarray}
This implies that the potential $\psi = 0$ inside the sphere, and
therefore, above the inner gap in the outflowing plasma region, the 
potential $\psi$ is exactly equal to the potential drop in the gap 
itself. In this region, a full screening of the longitudinal electric 
field component is assumed, whence we have ${\bf{E}} \cdot {\bf B}  = 0$ 
there. Taking the scalar product with ${\bf B}$, we then immediately 
obtain $\nabla \psi \cdot {\bf B} = 0$, i.e., the potential $\psi$  
is constant along the magnetic field lines.

For illustration, we consider two examples visually showing how the 
pulsar wind structure can be significantly changed by taking the 
potential drop in the inner gap into account. We first note that a 
nonzero potential c alters the Goldreich-Julian charge density (if 
the latter is understood as the charge density needed to screen the 
longitudinal electric field). Then, in terms of the dimensionless 
potential [69]
\begin{eqnarray}
\beta_{0} = \frac{\psi}{\psi_{\rm max}},
\label{beta}
\end{eqnarray}
we have
\begin{eqnarray}
 \rho_{GJ} \approx - (1 - \beta_{0})\frac{{\bf \Omega} \cdot {\bf B}}{2 \pi c}.
\label{GJpsi}
\end{eqnarray}
near the neutron star surface. As we see, for pulsars near the
`death line', where $\beta_{0} \sim 1$, the Goldreich-Julian
charge density (and hence the Goldreich current) significantly 
decreases. The outgoing electromagnetic energy flux decreases 
accordingly. For example, in the axisymmetric case,
\begin{equation}
W_{\rm tot} \approx (1-\beta_{0})^2 W_{\rm tot}^{(0)},
\label{Wwtot}
\end{equation}
because both the electric and toroidal magnetic fields decrease
by the factor $(1 - \beta_{0})$. In view of the foregoing, it is 
not necessary to explain how important this point can be for
plasma generation near the neutron star surface.

As another example, we consider a generalization of asymptotic formula 
(46) for electromagnetic fields in a quasi-radial pulsar wind of an 
orthogonal rotator:
\begin{eqnarray}
B_{r} & \approx & B_{\rm L} \frac{R_{\rm L}^2}{r^2} \, 
\sin\theta \cos(\varphi -\Omega t + \Omega r/c - \varphi_{0}),  \\
B_{\theta} & \approx  & \frac{1}{r \sin\theta} \, \frac{\partial \psi}{\partial \varphi},
\\ 
B_{\varphi} & \approx  & - B_{\rm L} \, \frac{\Omega R_{\rm L}^2}{cr} 
\, \sin^2\theta  \, \cos(\varphi  -\Omega t + \Omega r/c - \varphi_{0})
\nonumber \\
&&- \frac{1}{r} \, \frac{\partial \psi}{\partial \theta}, \\
E_{r} & \approx & 0, \\
E_{\theta} & \approx  & - B_{\rm L} \, \frac{\Omega R_{\rm L}^2}{cr} 
\, \sin^2\theta  \, \cos(\varphi -\Omega t + \Omega r/c - \varphi_{0}) 
\nonumber \\
&&- \frac{1}{r} \, \frac{\partial \psi}{\partial \theta}, \\
E_{\varphi} & \approx &  
-  \frac{1}{r \sin\theta} \, \frac{\partial \psi}{\partial \varphi}.
\label{Eqn10newN}
\end{eqnarray}
Here,$\psi(\theta, \varphi - \Omega t + \Omega r/c)$ an arbitrary function of 
two arguments. Now, for example, determining the electromagnetic energy flux for
the potential{\footnote{This expression corresponds to the simplest dependence 
on the angles $\theta$ and $\varphi$, which correctly reproduces the sign of
$\psi$  above each magnetic pole.}}
\begin{equation}
\psi = \beta_{0}\psi_{\rm max} 
\sin\theta \cos\theta \cos(\varphi -\Omega t - \Omega r/c + \varphi_{0}), 
\end{equation}
for small $\beta_{0}$ we obtain 
\begin{equation}
W_{\rm tot} \approx \left(1-\sqrt{\frac{3}{10}}\beta_{0}\right) W_{\rm tot}^{(0)}.
\end{equation}
Therefore, also for an orthogonal rotator with a nonzero potential $\psi$, 
the energy losses decrease. However, in this case, the value $\psi$  itself 
cannot exceed $(\Omega R/c)^{1/2}\psi_{\rm max}$ ($\beta_{0} \ll 1$) and
therefore the role of the additional potential is insignificant.

\subsubsection{Orthogonal pulsars.}

To conclude this discussion, we make one more note about a possible 
avenue of research. Among more than 2600 presently known radio pulsars, 
there are about 30 sources that are classified as orthogonal rotators
due to the presence of an interpulse exactly in the middle of the main 
pulses. The presence of the interpulse in this case is easily explained 
by the emission from the opposite magnetic pole. Almost all morphological 
properties of pulsars with interpulses are indistinguishable from those 
of ordinary pulsars. The only distinctive feature of these pulsars is 
their sufficiently small spin periods \mbox{$P < 0.5$ s.} This is not 
surprising because for angles $\chi$ close to $90^{\circ}$, the plasma 
generation condition $\psi_{\rm RS}(\chi) < \psi_{\rm max}(\chi)$ can 
be satisfied only for sufficiently small periods $P$ (Fig. 14).

\begin{figure}
\begin{center}
\includegraphics[width=220pt]{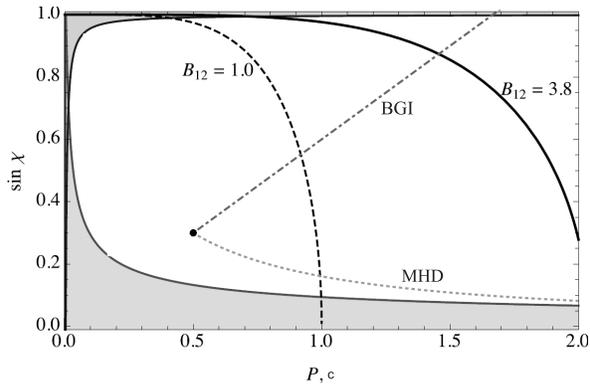}
\caption{`Death line' as a function of the angle $\chi$ and period $P$ 
for different values of the magnetic field $B_0$ [177]. Also shown are 
evolutionary trajectories in the `universal' MHD model and the BGI model. 
Grey color shows regions of possible interpulse observations for almost 
orthogonal (the upper region) and almost aligned (the bottom region) radio 
pulsars. 
}
\label{fig14}
\end{center}
\end{figure}

In our opinion, the analysis of interpulse pulsars could greatly help to 
understand pulsar magnetosphere processes. Indeed, the very existence of 
orthogonal radio pulsars, whose polarization and other radio emission 
properties do not differ from those of other pulsars, suggests `standard' 
particle generation. On the other hand, analysis of polarization properties 
of their radio emission shows that for these pulsars it is possible to 
reliably determine the angles $\chi > 90^{\circ}$ [148]. As a result, it 
turned out that the radio emission from some of them is connected to 
different secondary plasma generation regions, corresponding to positive 
and negative charge outflows.

\subsection{Longitudinal current}

\subsubsection{`Mestel's' current.}
\label{3.2.1}

We reiterate that the key point in the magnetosphere structure is the 
value of the longitudinal current circulating in the magnetosphere. 
Therefore, below, we make several notes concerning the procedure for 
determining this longitudinal current.

Already in the very first years of radio pulsar magnetosphere studies, 
it became clear that particle motion is mainly determined by the 
electric drift related to the electric field induced by neutron star rotation. 
Other drift velocities can be ignored. As a result, the density of the transverse 
electric current ${\bf j}_{\perp}$ can be written in the hydrodynamic form as 
${\bf j}_{\perp} = \rho_{\rm e} {\bf U}_{\rm dr}$ where 
${\bf U}_{\rm dr} = c [{\bf E}\times{\bf B}]/B^2$ is the electric drift
velocity. Such a simple dependence significantly simplified the problem, 
because the full current in the magnetosphere could be represented simply as 
${\bf j} = \rho_{\rm e} {\bf U}_{\rm dr} + a_{\parallel}{\bf B}$, where 
$a_{\parallel}$ is a scalar function. Using Eqn (62) (and disregarding potential
$\psi$), we finally arrive at
\begin{equation}
{\bf j} = \rho_{\rm e} [{\bf \Omega}\times{\bf r}] + i_{\parallel}{\bf B},
\label{jtot}
\end{equation}
where $i_{\parallel}$ is another scalar function.

Now, using another Maxwell equation written under the
quasi-stationary assumption in the form [60, 67]
\begin{equation}
\nabla \times ({\bf B} - [{\pmb{\beta}}_{\rm R}\times {\bf E}]) 
= \frac{4 \pi}{c}({\bf j} - c \rho_{\rm e}{\pmb{\beta}}_{\rm R}),  
\label{Maxwell4}
\end{equation}
we immediately conclude that $i_{\parallel}$ should be constant along the
magnetic field lines: ${\bf B} \cdot \nabla i_{\parallel} = 0$. As a result, 
for the zero potential $\psi$, we obtain{\footnote{The generalization of 
Eqn (76) to the case $\psi \neq 0$ was obtained as early as in [67].}}  
\begin{equation}
\nabla \times [{\bf B} 
+ {\pmb{\beta}}_{\rm R}\times [{\pmb{\beta}}_{\rm R} \times {\bf B}]] 
= i_{\parallel}{\bf B}.
\label{EqnMest}
\end{equation}
This is the so-called Mestel equation [60], which is valid, unlike pulsar 
equation (32), for any angle $\chi$.

In the Hellas and Rome times (1968-1983), the form of the current (74)
was very popular because it enabled the determination of the full current 
using only one additional scalar function $i_{\parallel}$. Moreover, in
the axisymmetric  case, this was a function of the magnetic flux $\Psi$ 
only. This equation was also helpful later, when the results of 3D 
numerical simulations were represented in the form of 2D plots (in such 
plots, the quantity $i_{\parallel}$ is usually shown in different colors).

Incidentally, exactly such a form enabled the determination of the 
dimensionless antisymmetric longitudinal current $i_{\rm a}^{\rm A}$,
that we discussed in Section 1.7, because the longitudinal current 
can be easily found in asymptotic solution (54), (55) at $r \gg R_{\rm L}$. 
Indeed, substituting field expressions (54), (55) for the orthogonal 
rotator in (75), we immediately obtain
\begin{equation}
i_{\parallel} = -3 \, \frac{\Omega}{c} \cos\theta.
\label{sasasa1}
\end{equation}
Now, comparing the full currents flowing across the upper hemisphere 
at $r \gg R_{\rm L}$ and through the north part of the polar cap on 
the neutron star surface, we finally obtain{\footnote{This formula 
would be precise for a circular polar cap.}  
\begin{equation}
i_{\rm a}^{\rm A} \approx f_{\ast}^{-1/2} \, 
\left(\frac{\Omega R}{c}\right)^{-1/2},
\label{sasasa2}
\end{equation}
where we recall that $f_{\ast} \approx 1,96$ is the dimensionless area 
of the polar cap ($s = f_{\ast} \pi R_{0}^2$) for the orthogonal rotator 
[69].

\subsubsection{`Gruzinov's' current.}
\label{3.2.2}

On the other hand, Eqn (74) had one essential shortcoming. For an oblique rotator, 
it was not `local' because the quantity $i_{\parallel}$ at a given point was not 
expressed through the fields and their derivatives at that point. In numerical 
simulations, this was important and did not allow fast calculations. This difficulty 
was overcome in 2005 by Gruzinov [94]. The scalar product of the Maxwell equation 
$\nabla \times {\bf B} = (1/c)\,\partial{\bf E}/\partial t + (4 \pi/c)\,{\bf j}$   
with ${\bf B}$ yields the scalar product ${\bf j}\cdot{\bf B}$ as a function of 
field derivatives. Using the screening condition ${\bf E}\cdot{\bf B} = 0$, i.e., 
replacing ${\bf B}\cdot \partial{\bf E}/\partial t$ with
$-{\bf E}\cdot\partial{\bf B}/\partial t$, 
after simple transformations we finally obtain
\begin{eqnarray}
&& {\bf j}_{\rm G}({\bf E},{\bf B}) = 
\label{jGruz} \\
&&  \frac{({\bf B}\cdot [\nabla \times {\bf B}] 
- {\bf E} \cdot [\nabla \times {\bf E}]){\bf B} 
+ (\nabla \cdot {\bf E})[{\bf E}\times{\bf B}]}{B^2},
\nonumber
\end{eqnarray}
where the charge density is expressed via the Maxwell equations 
$\nabla \cdot {\bf E} = 4 \pi \rho_{\rm e}$. This form turned 
out to be very convenient for numerical modeling and is currently 
in use in all simulations; of course, under certain conditions, 
the `classical' formulation (74) can be used (see, e.g., [149]).

Indeed, if we now use Eqn (62) (again with $\psi = 0$), the
quasi-stationary Maxwell equation (75) contains the mag-
netic field only:
\begin{eqnarray}
&& \nabla \times [{\bf B} + {\pmb{\beta}}_{\rm R}
\times [{\pmb{\beta}}_{\rm R} \times {\bf B}]] =
\label{EqnGruz} \\
&& \frac{4 \pi}{c} \, {\bf j}_{\rm G}(-[{\pmb{\beta}}_{\rm R} \times {\bf B}],{\bf B})
+ (\nabla \cdot [{\pmb{\beta}}_{\rm R} \times {\bf B}]) \, {\pmb{\beta}}_{\rm R}.
\nonumber
\end{eqnarray}
We recall that Eqn (80) remains valid for an oblique rotator. Jointly with the 
Maxwell equation $\nabla \cdot {\bf B} = 0$, they constitute a closed system of 
equations containing the magnetic field only. Equation (80), of course, is not 
as compact as Mestel's equation (76), but it does not contain any scalar functions.
Unfortunately, for a nonzero potential $\psi$, the locality is lost again, because 
the potential $\psi$ cannot be locally expressed through the field derivatives at 
a given point.

\subsection{So how do pulsars spin down?}

\subsubsection{The last time about magnetodipole losses.}
\label{3.3.3}

In our opinion, in its initial formulation, this question received a unique 
and decisive answer long ago: in the magnetosphere of radio pulsars, there 
are no magnetodipole losses. Nevertheless, magnetodipole losses are frequently 
invoked when discussing energy losses of radio pulsars. Indeed, because energy
losses (49) in the `universal solution' depend on the angle $\chi$  as 
$(1 + \sin^2\chi)$ and the addition to unity has the same dependence on $\chi$ 
as for magnetodipole losses, this would seem to indicate that the magnetodipole 
contribution can exist.

Numerical simulations carried out in the last decade have not clarified this 
issue. As we already stressed, Michel-Bogovalov analytic solution (47) certainly 
does not include a magnetodipole wave because, except for the current sheet
crossing moment, the electromagnetic fields are time-independent. However, as 
shown in Fig. 11, for an orthogonal rotator, there is a noticeable time-dependent 
field component. Accordingly, there is a significant time dependence in analytic
solution (54), (55), which reproduces the pulsar wind from an orthogonal rotator 
with good accuracy.

Apparently, the correct answer to this question can be formulated as follows. 
We can speak about magnetodipole radiation if the canonical conditions of the 
smallness of the emitting region size compared to the wavelength are satisfied,
i.e., if there is a wave zone. In the pulsar wind, this condition definitely 
does not hold, because both charges and currents {\it in situ} play a decisive 
role in the wind structure formation. Indeed, nobody would speak of the 
magnetodipole radiation of a wire with an alternating current, although the 
energy transfer outside the wire is exactly due to the electromagnetic energy 
flux. All this, however, occurs in the near zone, and hence both charges and 
currents are at distances comparable to the wavelength.

Thus, the pulsar wind should be viewed as an example of a relativistic 
magnetohydrodynamic wave with unusual properties, which we still know 
little about. For example, we have seen that the angular distribution 
of the energy flux changes from $\sin^{2}\theta$ for an axisymmetric 
rotator to $\sin^{4}\theta$ for an orthogonal one. This is already 
significantly different from magnetodipole losses, which are proportional 
to $1 + \cos^2\theta$ [150]. The absence of energy flux along the rotation 
axis (i.e., for $\theta = 0$) for the inclination angle $\chi = 60^{\circ}$ 
was already noted in [151], where the results of numerical simulations 
carried out in [105] were analyzed.

\subsubsection{Additional torque: a rotating magnetized ball.}
\label{3.3.1}

To finally clarify how the braking of a radio pulsar occurs, we return to
a problem apparently solved long ago, that of the braking of a uniformly 
magnetized ball rotating in a vacuum. We note first of all that in this 
problem, the rotating ball can be affected only by electromagnetic forces:
\begin{equation}
\label{F-lor}
{\rm d}{\bf F} = \rho_{\rm e} {\bf E}~ {\rm d}V
+ \frac{[{\bf j}\times {\bf B}]}{c} ~ {\rm d}V
+ \sigma_{\rm e} {\bf E}~ {\rm d}s
+ \frac{[{\bf J}_{\rm S} \times {\bf B}]}{c} ~ {\rm d}s,
\end{equation}
where the first two terms correspond to the volume contribution and 
the second to the surface one. However, if only corotation currents 
\mbox{${\bf j} = c \rho_{\rm e}{\pmb{\beta}}_{\rm R}$} are assumed to 
flow in the bulk, then it is easy to verify that the volume part of 
force (81) vanishes.

Now, using Eqns (60) and (62), it is easy to see that at $r = R + 0$, 
the vector $\nabla \psi$ is normal to the ball surface. As a result, 
the contribution of $\nabla \psi$ to the electromagnetic energy flux 
is zero, and after simple transformations we finally obtain
\begin{eqnarray}
W_{\rm tot} = \frac{c}{4 \pi} \oint [{\bf E} \times {\bf B}] {\rm d}s
=  - {\bf \Omega} \cdot {\bf K},
\label{WWtot}
\end{eqnarray}
where
\begin{equation}
{\bf K} = \frac{R}{4\pi} \oint \left[{\bf n}\times {\bf B}\right]
\left({\bf B}\cdot{\bf n}\right) {\rm d}s
\label{KKtot}
\end{equation}
is the braking torque due to the magnetic field only, as we see. Here, of course, 
the square brackets in (83) correspond to the surface current ${\bf J}_{\rm s}$, 
and the parentheses, to the magnetic field in the expression for the Amp\`ere force,
${\bf F} = [{\bf J}_{\rm s} \times {\bf B}]/c$. Thus, all energy losses are indeed 
determined by the surface currents ${\bf J}_{\rm s}$.

Surprisingly, even in this apparently absolutely clear question, there is one 
point that might be unexpected for the reader. It is clear that the total losses
\mbox{$W_{\rm tot} = - {\bf \Omega}\cdot {\bf K} = (2/3 \, c^3){\bf m}^2\Omega^4$} 
depend only on the pulsar magnetic moment ${\bf m}$. However, if we ask ourselves
which currents provide these losses, the answer is highly dependent on the fine 
details of the magnetic field structure near the neutron star surface.

To see this, we note that the braking torque ${\bf K}$ must be proportional 
to the third power of the angular velocity $\Omega$. More precisely, it must 
correspond to the third power of the expansion of the fields entering Eqn (83) 
for  ${\bf K}$ in the small parameter $\varepsilon = \Omega R/c$. It is clear 
that if we substitute the expression for the dipole magnetic field, the integral 
over the surface vanishes. In addition, it can be shown that the first-order 
terms $B^{(1)}$ in $\varepsilon$ are zero [152]. As a result, the general
expression for the torque becomes
\begin{equation}
{\bf K} 
= \frac{R}{4\pi}\oint \{[{\bf n}\times{\bf B}^{(3)}]
({\bf B}^{(0)}\cdot{\bf n}) 
+[{\bf n}\times{\bf B}^{(0)}]
({\bf B}^{(3)}\cdot{\bf n})\}
{\rm d}s,
\label{Ky13}
\end{equation}
where the indices (0) and (3) correspond to the power of expansion in the small 
parameter $\varepsilon$.

We now recall that in the pulsar magnetosphere theory, the Deutsch solution [153] 
obtained for a rotating magnetized ball in a vacuum is crucial; simple formulas 
for the fields on the star surface can be found in [154]. This solution was
constructed by assuming that the normal component of the magnetic field exactly 
coincides with the magnetic dipole field (the normal field component on the surface 
is the unique necessary boundary condition enabling a unique solution to be 
constructed). Clearly, in this case, by construction, $B_{n}^{(3)} = 0$; 
therefore, the first term only contributes to braking torque (84). But if 
we now use the solution for a rotating point-like orthogonal dipole as 
given in {\it Field Theory} by Landau and Lifshitz [150, {\S 72}] (see also [155]),
\begin{eqnarray}
{B}_{r}^{\perp} & = & \frac{|{\bf m}|}{r^3} \sin\theta \, {\rm Re} \,  
\left(2-2 i \frac{\Omega r}{c}\right)  \nonumber \\,
&& \times \exp\left(i \frac{\Omega r}{c} + i \varphi - i \Omega t \right),
\label{ll1a} \\
{B}_{\theta}^{\perp} & = & \frac{|{\bf m}|}{r^3} \cos\theta \, {\rm Re} \,
\left(-1+ i \frac{\Omega r}{c}+\frac{\Omega^2 r^2}{c^2}\right)  
\nonumber \\,
&& \times  
\exp\left( i \frac{\Omega r}{c} +i \varphi - i \Omega t \right),
\label{ll1b} \\
{B}_{\varphi}^{\perp} & = & \frac{|{\bf m}|}{r^3} \, {\rm Re} \,  
\left(-i -\frac{\Omega r}{c}+ i \frac{\Omega^2 r^2}{c^2}\right) 
\nonumber \\,
&& \times
\exp\left(i \frac{\Omega r}{c} + i\varphi -i \Omega t \right),
\label{ll1c} 
\end{eqnarray}    
we discover that only two thirds of the losses are determined by the first 
term in (84) as before, while one third is determined by the second term. 
Here, of course, neither the total losses nor the direction of the evolution 
of $\chi$, which are determined by the respective components of the braking
torque $K_{z'}$ and $K_{x'}$, depend on the choice of the solution
($x^{\prime}$, $y^{\prime}$, $z^{\prime}$ is the reference frame corotating 
with the star).

It is easy to identify the source of this discrepancy. As can be obtained 
directly from Eqns (85)-(87), the `Landau-Lifshitz solution' contains the 
magnetic field component independent of $r$ in the third order in $\varepsilon$:
\begin{equation}
{\bf B}^{(3)} = -\frac{2}{3} \, 
\frac{|{\bf m}|}{R^3}\left(\frac{\Omega R}{c}\right)^3 \, {\bf e}_{y^{\prime}}.
\label{B333}
\end{equation}
In other words, the `Landau-Lifshitz solution' differs from the 
Deutsch solution by adding the complementary magnetic dipole 
$\delta {\bf m}/|{\bf m}|  = (\varepsilon^3/3) {\bf e}_{y'}$ 
arising from the ball rotation. Clearly, such a small addition 
makes no contribution to the full electromagnetic losses; the 
corresponding losses are even smaller than electric quadrupole 
ones connected with the inevitable charge redistribution inside 
the rotating magnetized ball. However, as we see, the structure 
of braking currents is cardinally different here.

We note that such losses do not depend on whether the zeroth-approximation 
currents are concentrated on the star surface (the model of a uniformly 
magnetized ball) or at its center. This is because, as in Eqn (88), the 
braking torque ${\bf K}$ does not depend on the ball radius $R$ for a 
given magnetic dipole moment ${\bf m}$. Here, we should include the magnetic 
field perturbation over the entire neutron star surface, including in the 
closed magnetic field region.

\begin{figure}
\begin{center}
\includegraphics[width=200pt]{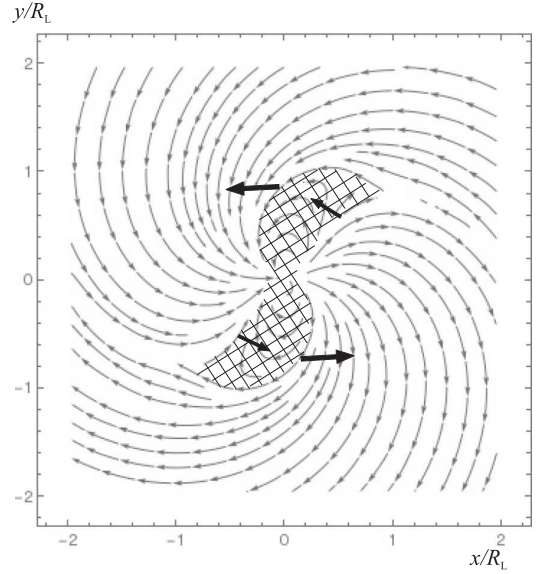}
\caption{Structure of magnetic field lines in the equatorial plane of an
orthogonal rotator $\chi = 90^{\circ}$ for nonzero longitudinal currents. 
The asymmetric form of the closed magnetosphere (hatched region) could
lead to an electromagnetic energy flow to the region of open field lines.
}
\label{fig15}
\end{center}
\end{figure}

We note one more point. Substituting Eqn (62) for the electric field (again 
for zero potential $\psi$) in the expression for the Poynting vector 
${\bf S}(r, \theta, \varphi) = (c/4 \pi) [{\bf E}\times {\bf B}]$, 
we immediately obtain
\begin{equation}
{\bf S} = 
\frac{\Omega r}{4 \pi} (-B_{\varphi} \sin\theta 
\, {\bf B} + B^2{\bf e}_{\varphi}).
\label{S_Poyn}
\end{equation}
It might seem that the second term in this equation cannot contribute to the 
energy losses because it is orthogonal to the normal vector. Consequently, 
the energy flux should be directed along the magnetic field lines only. 
However, as shown in Fig. 15, in the nonaxisymmetric case (due to the
possible difference in the magnetic field ${\bf B}$ at different boundaries 
of the closed magnetosphere), this term could be responsible for the total 
energy flux from the closed magnetosphere into the open field lines region, 
along which the energy is further transported beyond the light cylinder.

Thus, in the general case, in addition to the current losses discussed in 
Section 1, the neutron star spin-down can be related to perturbations of 
the normal magnetic field component $B_{n}$. This additional contribution 
to the braking torque $K_{\perp}$ can be due to a violation of the exact 
compensation between the magnetodipole radiation from the central star
and magnetosphere radiation that occurs for a zero longitudinal current.

\subsubsection{Additional torque: separatrix currents.}
\label{3.3.2}}

To discuss one more possible additional torque, we consider in more detail
the spin-down of an orthogonal rotator due to surface currents across polar 
caps, which corresponds to the first term in expansion (84). For this, it is 
convenient to rewrite the braking torque in the original form:{\footnote{We
must not forget that there are two poles in a radio pulsar!}}
\begin{equation}
{\bf K} = 
2 \, \int [{\bf r} \times \frac{[{\bf J}_{\rm s} \times {\bf B}]}{c}] \, {\rm d} s.
\label{tttorque}
\end{equation}
Hence, the full losses $W_{\rm tot} = -{\bf \Omega} \cdot {\bf K}$ become
\begin{equation}
W_{\rm tot} = 2 \, \frac{\Omega R}{c}\int J_{\theta} B_{n} {\rm d}s.
\label{tttot}
\end{equation}
From Eqn (90), it can be erroneously concluded [156] that the braking torque 
for the local Goldreich current $i_{\rm s}^{\rm A} = i_{\rm a}^{\rm A} = 1$ 
should not strongly depend on the inclination angle \mbox{$\chi$.} Indeed, 
for $\chi \approx 90^{\circ}$ the surface current $J_{\rm s}^{(90)}$ closing 
the volume currents flowing in the magnetosphere (and, hence, proportional to 
$\cos\theta_{\rm b}$) can 
be evaluated as $J_{\rm s}^{(90)} \sim \varepsilon^{1/2}J_{\rm s}^{(0)}$. But 
the characteristic distance $r_{\perp} = R$ from the axis to the polar cap, 
conversely, increases in the same proportion;
$r_{\perp}^{(90)} \sim \varepsilon^{-1/2}r_{\perp}^{(0)}$. 

However, the exact analysis [69] demonstrates that this argument, 
which is self-evident at first glance, ignores the real structure 
of the surface currents inside the polar cap. As shown in Fig. 16, 
the surface currents closing the volume currents should flow in such 
a way that the surface current  $J_{\theta}$ averaged over the surface 
vanishes (although just this component determines the neutron star 
energy losses). Therefore, when calculating the energy losses, the 
higher-order effects in the parameter $\varepsilon = \Omega R / c$
must be taken into account.

\begin{figure}
\begin{center}
\includegraphics[width=200pt]{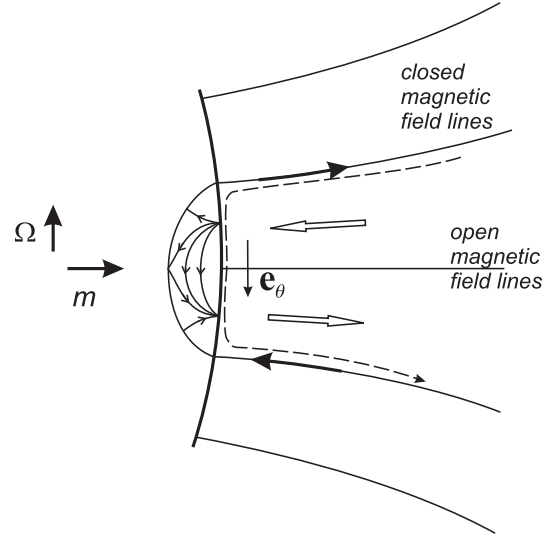}
\caption{Structure of the volume (light arrows), separatrix (dark arrows),
and surface (thin lines) currents near the polar cap of an orthogonal
rotator. The additional separatrix current is shown by the dashed line. 
}
\label{fig16}
\end{center}
\end{figure}

But if the surface current $<J_{\theta}>$ averaged over the polar 
cap is zero, then, along the separatrix dividing the region of open
and closed field lines, there should be a surface current comparable 
to the full current flowing in the open field line region, as is shown 
in Fig. 16. For example, for a circular polar cap and for the local 
Goldreich current (when the answer can be obtained analytically), 
the inverse current should be $3/4$ of the volume current [157].

One important note is in order here, however. The above conclusion 
on the energy losses was based on the assumptions that there are no 
volume longitudinal currents inside the closed magnetosphere and that
the surface currents closing the volume currents exist only inside 
the polar cap area and do not go beyond it [158]. If we relax these 
assumptions, the problem of the neutron star spin-down becomes 
ill-defined because it is impossible to calculate the additional 
current circulating inside the magnetosphere but not entering the
pulsar wind region. Indeed, as shown in Fig. 16, additional 
separatrix currents must result in a nonzero mean surface current 
$<J_{\theta}> \neq 0$ across the polar cap and hence in additional 
energy losses.

Thus, theoretical analysis has not resulted in sufficient clarity 
here. Some clarification became possible only after numerical 
modeling of the magnetosphere of an oblique rotator was carried 
out. As a result, it was shown that there are indeed no volume 
currents in a closed magnetosphere. It was also remarkable that 
the inverse currents along the separatrix were also obtained in 
numerical simulations [159]. But the inverse current was found 
to amount to only 20\% of the volume current. Of course, such a 
discrepancy could be explained by the fact that the radius of 
the star was only two to three times as small as the light cylinder 
in these calculations. However, the significant difference between
these parameters could also be related to additional separatrix 
currents ignored in the previous analysis. The `doomed uncertainty' 
persisted.

\subsubsection{Direct grasp of the truth.} 

We now return to the discussion of energy losses of a rotating neutron 
star. We recall that the direct current losses discussed above, Eqns 
(37) and (38), correspond to only the first term in expansion (84). 
They are due to the surface currents closing volume currents in the 
magnetosphere. Therefore, by the way, the action of this braking 
mechanism is concentrated in the polar cap area only. And the magnetic 
field corresponds to the zeroth-approximation magnetic field (the 
dipole magnetic field is close to a uniform field within the polar 
cap area). As we have seen, the direct current losses cannot explain 
the energy losses $W_{\rm tot}^{({\rm MHD})}$ in (49) for the 
'universal solution'.

On the other hand, as shown in Sections 2.3.2 and 2.3.3, there are two 
more reasons for the radio pulsar spin-down. First, the example of vacuum 
losses demonstrates that the second term in expansion (84) should not be 
ignored in general. In the expression ${\bf J}_{\rm s} \times {\bf B}$ 
for this term, the surface current ${\bf J}_{\rm s}$  corresponds to 
the zeroth-approximation current and the magnetic field is related to 
the rotation-induced third-order perturbation (in this case, the entire 
surface of the neutron star participates). Another possible reason could 
be the additional separatrix currents that are closed inside the polar 
cap area. Clearly, as before, they would correspond to the first term 
in expansion (84). 

As noted above, it was impossible not only to calculate analytically 
but even to estimate the corresponding contributions. Therefore, it 
is not surprising that virtually nothing was previously known about 
these losses. However, we can now answer this question by directly 
analyzing the numerical results.

But before showing the hand, we write the additional
braking torque in the general form as
\begin{eqnarray}
K_{\perp}^{\rm add}  =  - 
A \, \frac{B_{0}^{2}\Omega^{3}R^{6}}{c^{3}} \, i_{\rm a}
\label{17''}
\end{eqnarray}
and try to estimate the dimensionless constant $A$ from the result of 
numerical modeling of the `universal solution'. As shown in Section 
1.3, the dimensionless antisymmetric longitudinal current is 
$i_{\rm a}^{\rm A} \approx (\Omega R/c)^{-1/2}$; therefore, one can
evaluate the coefficient $A$ as
\begin{equation}
A \sim (\Omega R/c)^{1/2}.
\label{A}
\end{equation}
For such a small $A \ll 1$, we can disregard the additional
torque $K_{\perp}^{\rm mag}$ in Eqn (92) for the local 
Goldreich current $i_{\rm a}^{\rm A} \sim 1$, which 
was in fact done in the BGI model (see Section 3.1.1).

\begin{figure}
\begin{center}
\includegraphics[width=\columnwidth]{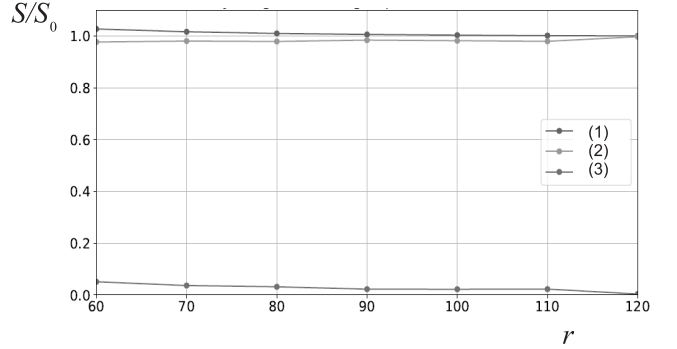}
\caption{Electromagnetic energy flux $S$ from the closed (bottom 
curve) and open (middle curve) magnetic field line region for the 
angle $\chi = 60^{\circ}$. The upper curve corresponds to the total 
power. The neutron star radius \mbox{(50 units)} is 10 times smaller than 
the light cylinder radius ($\varepsilon = 0.1$). $S_{0}$ is the energy 
flux corresponding to the energy losses $W_{\rm tot}^{({\rm MHD})}$ 
in Eqn (49).     
}
\label{fig17}
\end{center}
\end{figure}

And here is the correct answer inferred from the processing of 
the numerical calculation data [160]. Figure 17 shows the 
electromagnetic energy flux as a function of the radius $r$ for
the angle $\chi = 60^{\circ}$ from the closed (bottom curve)
and open (middle curve) line regions. The sizes of the star and
the light cylinder are 50 and 500 units ($\varepsilon = 0.1$).

As we see, almost all the energy flux is concentrated in the
open magnetic line region. This means that in the numerical
simulations reproducing the `universal solution', the energy
losses from an oblique rotator are likely to be due to
additional separatrix currents. This solution does not reveal
any energy flux from the closed magnetosphere, which would
otherwise be present. We do not even need to mention that
these results leave no room for magnetodipole losses (in that
case, the energy flux would also be distributed quite
homogeneously and would not be concentrated in the open
field line region only).

Now, using Eqn (91), it is possible to obtain the expression
for the surface current $<J_{\theta}>$ averaged over the polar 
cap area that provides the MHD losses $W_{\rm tot}^{({\rm MHD})}$
given by formula (49). In particular, for an orthogonal rotator, 
we obtain
\begin{equation}
<J_{\theta}> = \frac{c}{4 \pi f_{\ast}} \, B_{0} \, 
\left(\frac{\Omega R}{c}\right)^2.
\label{end1}
\end{equation}
Accordingly, the averaged magnetic field $<B_{\varphi}>$ in the open
field line region at a distance $r$ is
\begin{equation}
<B_{\varphi}> = 
\frac{1}{f_{\ast}} \, B_{0} \, 
\left(\frac{\Omega R}{c}\right)^2 \frac{R}{r}.
\label{end2}
\end{equation}
This can be easily obtained by calculating the electromagnetic 
energy flux through the corresponding area: 
\mbox{$s(r) = f_{\ast} \pi (\Omega/c)^{1/2} r^{3/2}$.} 
As we see, the field $B_{\varphi}$ behaves as if in a
magnetodipole wave. The difference is that this radial
dependence takes place inside the light cylinder and not 
only in the wave zone. Finally, we note that the full 
separatrix current should be $(\Omega R/c)^{1/2}$ times 
smaller than the full current circulating in the magnetosphere.

\subsubsection{Anomalous torque.}
\label{3.3.4}

To conclude this section, we note another curious point. 
Besides the torque considered above, which leads to spin-down 
and the inclination angle $\chi$ evolution of a rotating 
magnetized ball, there is the so-called anomalous torque 
directed along the $y^{\prime}$ axis perpendicular to the 
plane formed by the magnetic moment ${\bf m}$ and the spin 
axis ${\bf \Omega}$. This torque is called anomalous because 
of its amplitude,
\begin{equation}
K_{y^{\prime}} = \xi \, 
\frac{{\bf m}^2}{R^3}\left(\frac{\Omega R}{c}\right)^2 \sin\chi \cos\chi,
\label{Ky1}
\end{equation}
where $\xi$ is a numerical factor of the order of unity, which is
$(\Omega R/c)^{-1}$ times the braking torque components $K_{x'}$ 
and $K_z$. Various authors give different values of $\xi$: $\xi = 1$ 
[161, 162], $\xi = 3/5$ [163], $\xi = 2/5$ [164], $\xi = 1/5$ [156, 
165], $\xi = 0$ [166, 167] (in decreasing order).

Clearly, the situation where no full concordance is achieved on a 
problem that is elementary at first glance was surprising. We recall 
that neutron star precession due to the anomalous torque, being 
superimposed on the secular pulsar spin-down, should change the 
observed braking index $n_{\rm br} = \ddot\Omega \Omega/{\dot \Omega}^2$ 
[163, 168, 169]. Thus, this question is of both theoretical and 
purely practical interest. 

The analysis performed in [152] showed that unlike the braking 
torque considered above, the determination of the anomalous 
torque requires taking both the electric force ${\bf E} \, \sigma_{n}$ 
(where $\sigma_{n}$ is the surface charge density) and the angular
momentum of the electromagnetic field inside the star into account. 
As a result, the anomalous torque depends on the internal structure 
of the magnetic field. 

Indeed, in almost all the papers mentioned above, the anomalous 
torque ${\bf K}$ was calculated as the flux 
\mbox{$K_{i} = - \oint \varepsilon_{ijk}r_{j}T_{kl}{\rm d}s_{l}$} of the 
electromagnetic stress tensor $T_{kl}$ using the formula [65]
\begin{eqnarray}
{\bf K}^{M} =\frac{1}{4\pi}\oint \Bigl(
\left[{\bf r}\times{\bf B}\right]
\left({\bf B}\cdot{\rm d}{\bf s}\right)+\left[{\bf r}\times{\bf E}\right]
\left({\bf E}\cdot{\rm d}{\bf s}\right)
\nonumber \\
-\frac{1}{2}~({\bf E}^2+{\bf B}^2)\left[{\bf n}\times{\rm d}{\bf s}\right]
\Bigr).
\label{KKK}
\end{eqnarray}
Integrating over the sphere yields
\begin{equation}
{\bf K}^{M}_{y^{\prime}} =
\frac{R}{4\pi}\oint\Bigl( \left[{\bf n}\times{\bf B}\right]_{y^{\prime}}
\left({\bf B}\cdot{\bf n}\right)+\left[{\bf n}\times{\bf E}\right]_{y^{\prime}}
\left({\bf E}\cdot{\bf n}\right)\Bigr) {\rm d}s.
\label{KMel}
\end{equation}

However, the correct derivation of the electromagnetic torque for a ball 
requires caution. Indeed, using general relation (81) and expressing the 
surface charges and currents through the corresponding jumps $\{{\bf E}\}$ 
and $\{{\bf B}\}$ in the electric and magnetic fields, we obtain the total 
torque in the form
\begin{eqnarray}
\label{Ky}
&& {\bf K} = \int{\bf r}\times{\rm d}{\bf F} 
\\
&& = \frac{R}{4\pi}\oint\Bigl( \left[{\bf n}\times\left\{{\bf B}\right\}\right]
\left({\bf B}\cdot{\bf n}\right)+\left[{\bf n}\times{\bf E}\right]
\left(\left\{{\bf E}\right\}\cdot{\bf n}\right)\Bigr)
{\rm d}s.
\nonumber
\end{eqnarray}
This formula differs from (98) in having the electric and magnetic field 
jumps on the surface of the ball, which exactly determine the surface 
charges and currents. As shown in [152], exactly this expression should 
be used when determining the torque for a rotating magnetized ball.

The point is that the angular momentum vector flux of the electromagnetic 
field is related to the angular momentum of forces acting on matter via 
the angular momentum conservation law for the electromagnetic field, which 
has the form
\begin{equation}
\label{AMconserv}
\frac{{\rm d}\,{\bf L}_{\rm field}}{{\rm d}t} 
 + \int [{\bf r}\times {\bf F}] {\rm d}V = {\bf K}^{M},
\end{equation}
where
\begin{equation}
{\bf L}_{\rm field} = 
\int \frac{[{\bf r}\times [{\bf E}\times{\bf B}]]}{4\pi c}{\rm d}V 
\label{Lfield}
\end{equation}
is the angular momentum of the electromagnetic field in a 
volume $V$, ${\bf K}^{M}$ is the angular momentum flux of 
the field through the surface bounding this volume, and 
${\bf F} = \rho_{\rm e}{\bf E} + [{\bf j} \times {\bf B}]/c$  
is the Lorentz force density. The last term in Eqn (100) 
plays the role of the source or sink and is therefore
responsible for the transfer of the electromagnetic field
angular momentum to matter,
\begin{equation}
\frac{{\rm d\,{\bf L}_{\rm mat}}}{{\rm d}t} 
= \int [{\bf r}\times {\bf F}] {\rm d}V.
\end{equation}
It is this term, and not ${\bf K}^{M}$, as most of the papers 
assumed, that has the meaning of the torque applied to a rotating 
body. Therefore, to calculate the anomalous torque (in the second 
order in the parameters $\varepsilon = \Omega R/c$), we should 
indeed take the angular momentum of the electromagnetic field 
into account. Some of the stresses related to the electromagnetic
field should affect the angular momentum of the field itself,
and some others should affect the interaction with the rotating 
body. According to Eqn (100), this means that for the angular 
momentum of forces applied to the ball, we should use the formula
\begin{equation}
{\bf K} \equiv \frac{{\rm d} {\bf L}_{\rm mat}}{{\rm d}t} 
= {\bf K}^{M} - \frac{{\rm d}\,{\bf L}_{\rm field}}{{\rm d}t}.
\end{equation}
The correct expression for ${\bf K}^{M}$  (independent of the 
internal structure of the field!) was first derived by Melatos 
[163]:
\begin{equation}
K_{y^{\prime}}^{M} = 
\frac{3}{5} \, \frac{{\bf m}^2}{R^3}
\left(\frac{\Omega R}{c}\right)^2 \sin\chi \cos\chi.
\end{equation}
The quantity ${\rm d}\,{\bf L}_{\rm field}/{\rm d}t$, turned out, 
as was already mentioned, to be dependent on the internal structure 
of the star.

As a result, the anomalous torque applied to a homogeneously magnetized 
ball was found to be [152]
\begin{equation}
\label{result1}
{K}_{y^{\prime}} = 
\frac{1}{3}\,\frac{{\bf m}^2}{R^3}
\left(\frac{\Omega R}{c}\right)^2 \sin\chi \cos\chi.
\end{equation}
On the other hand, for a rotating hollow sphere, when charges
and currents are concentrated inside the spherical shell $r = R$,
we obtain 
\begin{equation}
\label{result2}
{K}_{y^{\prime}} = 
\frac{31}{45}\,\frac{{\bf m}^2}{R^3}
\left(\frac{\Omega R}{c}\right)^2 \sin\chi \cos\chi.
\end{equation}
Finally, if the homogeneous magnetic field is present 
only inside the inner radius $R_{\rm in}$, and for 
$R_{\rm in} < r < R$ (also outside the ball) the 
zeroth-order magnetic field is identical to that of
a point-like dipole, we have
\begin{equation}
{K}_{y^{\prime}} =  \left(\frac{8}{15} -\frac{1}{5}\frac{R}{R_{\rm in}}\right)
\frac{{\bf m}^2}{R^3}\left(\frac{\Omega R}{c}\right)^2 \sin\chi \cos\chi.
\label{result3}
\end{equation}
For $R_{\rm in} = R$, we return to the previous value $\xi = 1/3$.

As regards the spin-down of the ball (i.e., the third-order effect 
in the parameter $\varepsilon$), the angular momentum of the
electromagnetic field was found to be insignificant in this
case. Therefore, formula (83) for the torque applied to a
rotating ball is valid and independent of the internal structure
of the field.

\section{BGI theory thirty years later}

We take this opportunity to present a detailed analysis of the
theory of magnetospheres and radio emission of pulsars
formulated in the 1980s in a series of papers [67, 170-172]
and later collected in monograph [69]; several significant
improvements were added later in [151, 173-177].

\subsection{Theory of the pulsar magnetosphere}

\subsubsection{Basics.}
\label{4.1.1}

The main results of the BGI theory on the
magnetosphere structure can be formulated as follows.
\begin{itemize}
\item 
The longitudinal current $j_{\parallel}$, outflowing from magnetic
poles (more precisely, its symmetric part $j_{\rm s}$), does not exceed
the local Goldreich current
$j_{\rm GJ} \approx \Omega B_{0} \cos\chi/2 \pi$.
\item
The amplitude of the local current is determined by the
drop in the electric potential $\psi$; the dimensionless symmetric
current amplitude $i_{\rm s}^{\rm A}$ is
\begin{equation}
i_{\rm s}^{\rm A} (\Omega, B_{0}, \chi) \approx 
\frac{1}{2}\left(\frac{\psi}{\psi_{\rm max}}\right)^{1/2},  
\label{Ohm}
\end{equation}
where, again, $\psi_{\rm max}$ is the maximum potential difference (23).
\item
The potential $\psi$ is given by the Ruderman-Sutherland
model $\psi_{\rm RS}$, Eqn (21).
\item
The dependence of the inclination angle $\chi$ on both the maximum 
potential drop $\psi_{\rm max} = \psi_{\rm max}(0) \cos\chi$ and the 
potential $\psi_{\rm RS}$ is taken into account; the analysis of the 
evolution of radio pulsars takes the dependence on the angle $\chi$  
and on the location of the `death line' into account.
 \item
Due to the relatively small value of the longitudinal current, there 
is a light surface beyond the light cylinder (see Fig. 7) at which 
the outflowing plasma is accelerated up to energies 
$\sigma_{\rm M} m_{\rm e}c^2$.
\end{itemize}

As a result, the key parameter in the BGI theory is the
dimensionless quantity $Q \approx i_{\rm s}$, which can 
be represented in the form [170]
\begin{equation}
Q = 2 \, P^{11/10} {\dot P}_{-15}^{-2/5},
\label{Q1}
\end{equation}
or more precisely, in terms of the angle $\chi$ 
\begin{equation}
Q = P^{15/14}B^{-4/7}_{12} \cos^{2d-2}\chi,
\label{Q}
 \end{equation}
where $d \approx 0,75$. The condition $Q = 1$ for Eqn (110) exactly corresponds 
to the relation $\psi_{\rm max}(\chi) = \psi_{\rm RS}(\chi)$ defining the
'death line'. Therefore, in pulsars for which definition (109) gives $Q > 1$, 
it is possible to expect an irregular generation of the secondary electron-positron 
plasma and hence a decreased value of $\dot P$, which results in the observed 
values $Q > 1$. Thus, for these pulsars, we should set $Q = 1$. The parameter 
$Q$ is convenient because it determines some key characteristics, such as the 
total energy losses:
\begin{equation}
W_{\rm tot}^{({\rm BGI})} \approx Q \frac{B_{0}^2 \Omega^4 R^6}{c^3} \cos^2\chi,
\label{WQ}
 \end{equation}
and the energy flux carried by particles inside the light cylinder,
\begin{equation}
W_{\rm tot}^{\rm out} \sim Q^2 \, W_{\rm tot}.
\label{WpartQ}
\end{equation}
The last equation coincides with the total particle flux toward
the neutron star surface in the inner gap:
\begin{equation}
W_{\rm part}^{\rm in} \sim Q^2 W_{\rm tot}.
\label{WinQ}
\end{equation}
We keep in mind, finally, that the angle $\chi$ tends to 90$^{\circ}$ in this
model: $\Omega \sin \chi = {\rm const}$.

\subsubsection{Magnetosphere structure.}
\label{4.1.2}

One of the features that distinguish the BGI theory from numerical results
obtained in recent years lies in the assumption that the longitudinal current 
flowing in the open line region is determined by the magnetosphere structure 
inside the line cylinder and not beyond it. It is the matching of solutions in
the closed and open parts of the magnetosphere that determined the `Ohm law' 
$i_{\rm s} \propto \psi_{\rm RS}^{1/2}$ given by formula (108) underlying the 
theory [67]. It was also shown that the minimal energy corresponds to a solution 
in which the null point lies on the light cylinder and not at shorter distances
from the neutron star [178].

Here, however, we assumed that the longitudinal current $i_{\parallel}$ 
(except in the orthogonal rotator case, of course) remains almost constant 
everywhere within the open line region and does not change sign near the 
separatrix, as the `universal solution' now requires. This, in particular, 
has led to a different topology of the magnetic surfaces at the crossing of
the equator with the light cylinder (an X, but not a Y point). However (and 
we wish to especially stress this here), a much more significant difference 
between the BGI theory and the `universal solution' considered at present 
was the existence of the light surface changing the entire structure of the 
flow beyond the light cylinder.

\begin{figure}
\begin{center}
\includegraphics[width=\columnwidth]{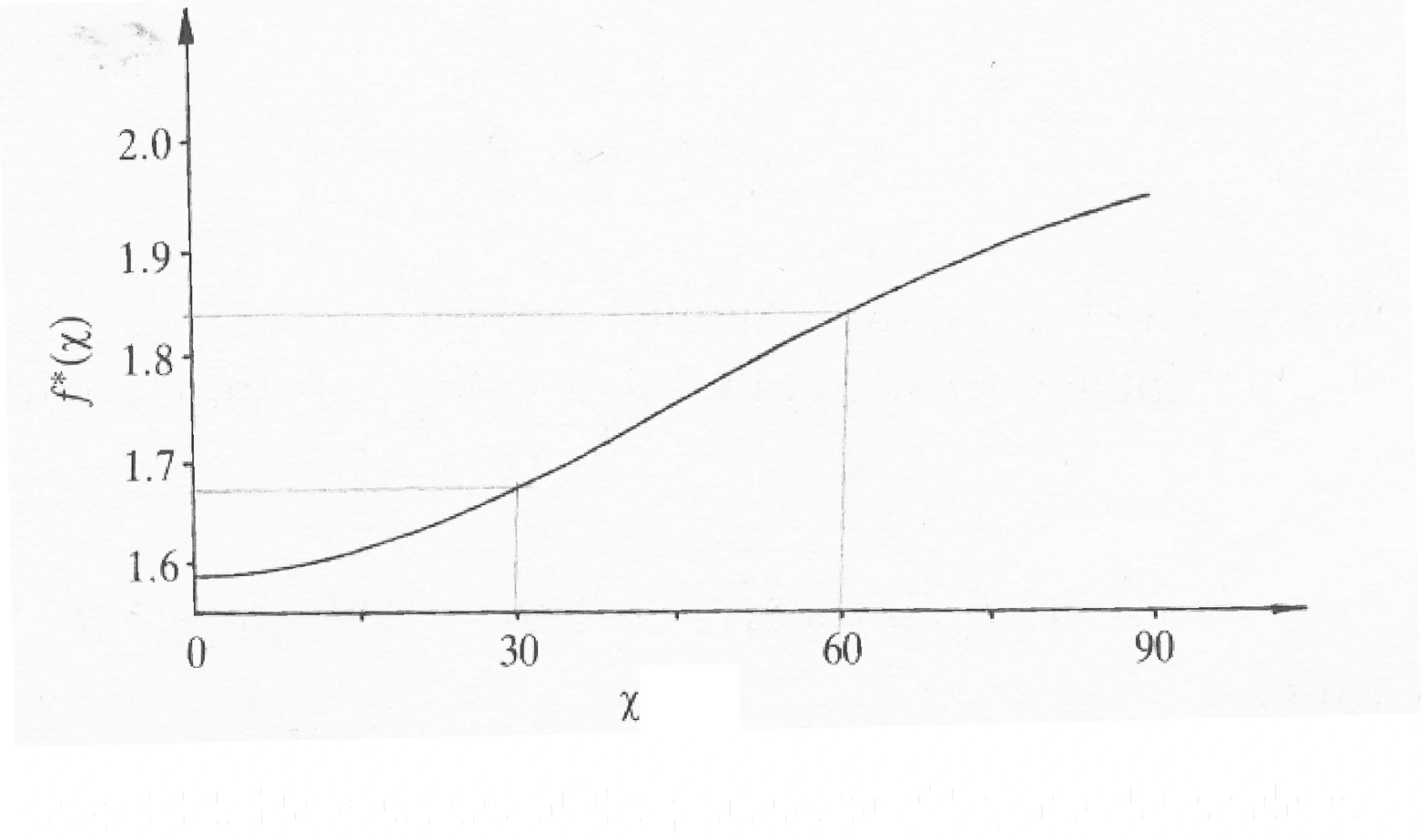}
\caption{Dimensionless polar cap area $f_{\ast}(\chi)$ as a function of 
the angle $\chi$ [91]. It coincides up to 10\% with the numerical 
modeling in [112].
}
\label{fig18}
\end{center}
\end{figure}

Nevertheless, the analytic solutions obtained more than thirty years ago 
for an oblique rotator in the BGI theory proved to be surprisingly close 
in many respects to the characteristics obtained for the `universal solution'. 
For example, the form of the closed magnetosphere shown in Figure 4 almost 
coincides with the result presented in Fig. 10. Moreover, the dependence of 
the dimensionless polar cap area $f_{\ast}(\chi)$ on the angle $\chi$ (Fig. 18) 
found in monograph [91] coincides up to 10\% with the dependence 
$f_{\ast}(\chi) \approx (1 + 0.2 \, \sin^2 \chi) f_{\ast}(0)$ derived quite 
recently in [112]. This means that the longitudinal currents in the 
magnetosphere do not strongly disturb the closed field line region.

As regards the prediction of an effective acceleration of outflowing plasma 
near the light surface, this conclusion, as stressed above, has not been 
confirmed so far by numerical calculations. However, as noted in Section 
1.7, a need for particle creation beyond the light cylinder is being widely
discussed at present. Moreover, it is often claimed that beyond the light 
cylinder, there are regions where the electric field exceeds the magnetic 
field. However, here, one usually means a current sheet and not the light 
surface that is shown in Fig. 7. The existence of the light surface 
$|{\bf E}| = |{\bf B}|$ near the light cylinder was also mentioned in [179]. 
However, the `Aristotle' approach discussed in this paper in fact limits 
the values of the longitudinal current from above, which actually
leads to diminishing the toroidal magnetic field.

\begin{figure}
\begin{center}
\includegraphics[width=220pt]{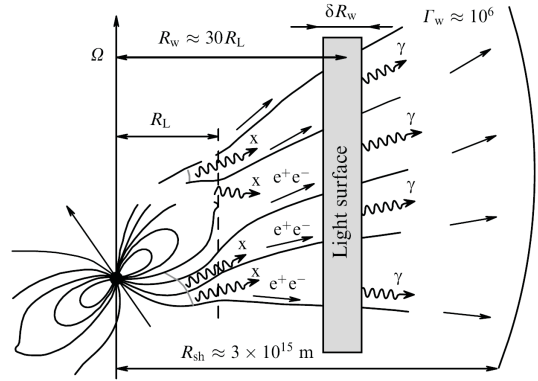}
\caption{Light surface near the light cylinder proposed in [180] to explain
the observed variability of the Crab Nebula pulsar at TeV energies. 
}
\label{fig19}
\end{center}
\end{figure}

Nevertheless, we would like to draw attention to the paper by Aharonian, 
Bogovalov, and Khangulyan [180], in which, in our opinion, direct 
observational clues of the effective particle acceleration on light 
surfaces were found. By the way, this paper is entitled `'Abrupt acceleration 
of a `cold' ultrarelativistic wind from the Crab pulsar''. As shown in
Fig. 19, this model is geometrically surprisingly similar to our model 
(see Fig. 7). In particular, it is shown in [180] that the observed 
variability at TeV energies [181] can be explained if, at distances of 
the order of $30 \, R_{\rm L}$ from the neutron star, acceleration of 
particles to energies corresponding to the Lorentz factor $\gamma \approx 10^{6}$
occurs. As noted above, the value $\gamma \sim 10^{6}$ exactly corresponds 
to estimate (19) of the magnetization parameter $\sigma_{\rm M}$ of the 
pulsar magnetosphere in the Crab Nebula. Last, the scale $30 \, R_{\rm L}$ 
is definitely smaller than the fast magnetosonic surface distance 
$r_{\rm F} \approx 100 \,R_{\rm L}$, and therefore the particle acceleration
process based on the force-free solution considered in [62, 182] can be 
applied here.

\subsubsection{Braking index.}
\label{4.1.3}

\begin{table*}[ht]
\caption{Braking index $n_{\rm br} = {\ddot \Omega} \Omega/{\dot \Omega}^2$ for fast pulsars [186].}
\vspace{0.2cm}
\centering
\begin{tabular}{|c|c|c|c|c|c|c|c|}
    \hline
{\rm PSR} & J1734+3333 & B0833$-$45 & J1833$-$1034 & B0540$-$69 & B0531+21 & B1509$-$59 & J1640$-$4631    \\
          &            &   Vela     &              &            &   Crab   &            &            \\ 
\hline
$n_{\rm br}$  & 0.9(2)   &  1.4(2)  &   1.857(1) &  2.14(1) &  2.51(1) & 2.839(1) &   3.15(3)  \\
\hline
\end{tabular}
\label{table1}
\end{table*}

As is well known, direct information on the spin-down mechanism should be 
obtained from the value of the so-called `braking index' [183, 184]
\begin{equation}
n_{\rm br} = \frac{{\ddot \Omega} \Omega}{{\dot \Omega}^2}.
\label{nbr}
\end{equation}
However, as is also well known, this parameter cannot be directly used 
to analyze the radio pulsar evolution, unfortunately: the regular pulsar 
spin-down is superimposed with additional fluctuations on time scales much 
shorter than the dynamical age of pulsars $\tau_{\rm D} \sim P/{\dot P}$. 
This results in ${\ddot \Omega}$ being not indicative of the secular pulsar
spin-down. As already noted in Section 2.3.5, such perturbations can most 
likely be related to the neutron star precession due to its nonspherical
shape (see, e.g., [163, 169]). For this reason, measurements of most of 
the pulsars give \mbox{$n_{\rm br} = \pm (10^{4} $--$10^{5})$}[169], which 
does not relate to the secular pulsar spin-down mechanism. As a result, only
braking indices of a handful of young pulsars can be used to analyze the 
energy loss mechanism; for other radio pulsars, longer observations are 
needed to measure the average braking index.

Table 1 shows that observations of most of the young pulsars yield 
$n_{\rm br} \sim 3$, which, in the zeroth approximation, does not 
contradict any of the pulsar spin-down theories:{\footnote{These 
formulas are written with the magnetic field evolution ignored, 
which can significantly change the value of $n_{\rm br}$.}} 
\begin{eqnarray}
n_{\rm br}^{({\rm V)}} & = & 3 + 2 \, {\rm ctg}^2\chi, \\
n_{\rm br}^{({\rm BGI})} & \approx & 1,93 + 1,5 \, {\rm tg}^2\chi, \\
n_{\rm br}^{({\rm MHD})} & \approx &
3 + 2\frac{\sin^2\chi \, \cos^2\chi}{(1 + \sin^2\chi)^2}.
\label{Eqn11} 
\end{eqnarray}

On the other hand, to infer the evolution law from observations, it is 
necessary, as we have seen, to measure the braking index $n_{\rm br}$ 
with a precision of two significant digits by segregating this quantity 
from the background of much larger fluctuations. Presently this cannot 
be done, unfortunately. Nevertheless, we note that for most young pulsars, 
$n_{\rm br} < 3$. As we see again, there is good correspondence with 
the BGI theory predictions. As regards the `universal model', for which 
$3 < n_{\rm br}^{({\rm MHD})}  < 3.25$, it was shown in [185] that the
account for the additional precession enables bringing this theory into 
accordance with observations as well.

\subsubsection{Evolution of the inclination angle.}
\label{4.1.4}

Of course, the spin-down theory could be tested by measurements of 
the angle $\chi$ evolution, i.e., by the sign of the time derivative 
$\dot\chi$. Unfortunately, so far there are no reliable estimates 
of this quantity (see, however, [187]). This is mainly due to the 
lack of a quantitative theory of radio emission from pulsars, which
does not allow connecting the secular changes in the mean pulse 
profiles with the angle between the magnetic and rotation axes.

Nevertheless, it can not be said that there are no resultë devoted 
to the analysis of the evolution of the inclination angle $\chi$.
In particular, the Crab pulsar observations carried out over several 
decades reported in recent paper [187] might favor the counter-alignment 
(the angle $\chi$ tending to $90^{\circ}$). However, this
paper made several model assumptions (for example, that the
gamma-ray pulse profile changed with time in the same way
as in the radio range), which precludes us from considering
this evidence as final.

Many attempts have also been made to infer the angle $\chi$
evolution from analysis of the pulsar distributions over 
$\chi$ itself and the mean pulse profile widths [189-194]. 
Here, alignment (the angle $\chi$ tending to zero) was 
frequently obtained. However, as has been noted many times 
(see, e.g., [151]), the dependence of the `death line' on 
the angle $\chi$ has been ignored. As shown in Fig. 14, 
for sufficiently large periods $P$, plasma generation is 
possible only for small $\chi$. Hence, irrespective of the
evolution of individual pulsars, the mean value of the
inclination angle \mbox{$<\chi>$}  should decrease with the 
spin period.

Clearly, fully taking the `death line' dependence on the angle 
$\chi$  into account can be done using only the kinetic approach, 
in which two-parameter (multi-parameter) distributions are 
analyzed. Unfortunately, up to now most studies have ignored 
the $\chi$ angle evolution [195-199]. Only recently have papers 
appeared where it was systematically taken into account [200-203]. 
However, these studies have not made definitive conclusions 
about the direction of the inclination angle evolution either.

Our recent paper [177] also failed to answer this question.
In that paper, we attempted to formulate a test to determine
the $\chi$ angle evolution direction. The idea, as was already
mentioned above, was to use the analysis of the fraction of 
interpulse pulsars, i.e., pulsars that show, besides the main 
pulse, another emission pulse located roughly in the middle 
between consecutive main pulses (i.e., at the phase 
$\phi \approx 180^{\circ}$). 

As shown in Fig. 14, the interpulse can be observed in pulsars either 
with $\chi \approx 90^{\circ}$ (when the interpulse is due to emission 
from another magnetic pole) or with $\chi \approx 0^{\circ}$ (when the 
main pulse and the interpulse are due to crossing by the line of sight 
of `hollow cone' radio emission). Clearly, the number of pulsars with 
$\chi \sim 0^{\circ}$ and $\chi \sim 90^{\circ}$ depends significantly 
on the $\chi$ angle evolution direction. However, it turned out that 
the observed number of interpulse pulsars can be explained by both 
the BGI model (counter-alignment) and the MHD model (alignment). This
is due to a large uncertainty in the initial pulsar distribution over 
the angle $\chi$ and the spin period $P$. In any case, the observed 
angle $\chi$ distribution was shown to be consistent with the BGI model.

\subsubsection{Surface heating.}
\label{4.1.5}

To conclude this section, we return to the problem of neutron star 
heating by the inverse particle flux $W_{\rm part}^{\rm in}$ in the 
plasma generation region. According to (113), in the BGI model, the 
energy of these particles should be $Q^{2}$ times smaller than the 
total energy losses $W_{\rm tot}$. Therefore, we can expect that 
the X-ray luminosity $L_{\rm X}$ would also depend in a similar way 
on the total energy losses of the pulsar.

Figure 20 shows the ratio $L_{\rm X}/W_{\rm tot}$ as a function of 
parameter $Q$ in Eqn (109). The thermal X-ray luminosity $L_{\rm X}$ 
of the polar caps was taken from [140-142]; the errors are related 
to both uncertainty in the pulsar distances (pulsars PSR J0205+6449 and 
J2021+3651) and the flaring activity (PSR J1119-6127 and J1846-0258) 
unrelated to the polar cap heating. The line corresponds to the 
$Q^{2}$ law. As we see, within the measurement errors, the X-ray 
luminosity from the polar cap thermal heating does not contradict 
theoretical predictions.

\begin{figure}
\begin{center}
\includegraphics[width=220 pt]{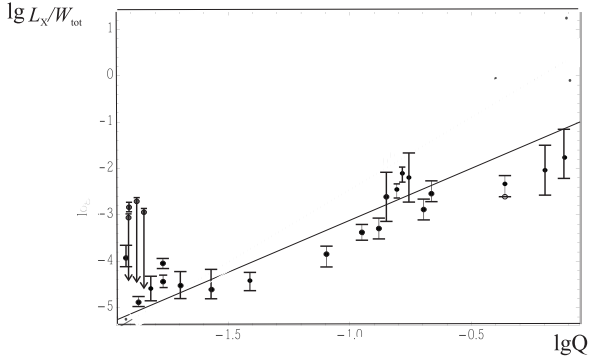}
\caption{Ratio $L_{\rm X}/W_{\rm tot}$ as a function of the parameter $Q$. 
Thermal luminosities of the polar caps $L_{\rm X}$ are taken from [140-142]; 
the line corresponds to the $Q^{2}$ dependence.  
}
\label{fig20}
\end{center}
\end{figure}

\subsection{Theory of radio emission}

\subsubsection{Basic.}
\label{4.2.0}

The principal statements of the BGI theory about pulsar radio emission can be 
formulated as follows [69, 172]. 
\begin{itemize}
\item 
The main instability determining the observed coherent radio emission is the 
curvature plasma instability of `mode 4' (Fig. 21).
\item
A saturation of the instability is due to three-wave decay of `mode 4' 
into extraordinary ('mode 1', or X-mode) and ordinary ('mode 2', or 
O-mode) waves capable of leaving the neutron star magnetosphere.
\item
The ordinary wave propagates with deviation from the magnetic axis and 
forms the outer (conal) radio beam, whereas the extraordinary wave 
propagates along a straight line and forms the inner (core) beam component.
\item
The total radio luminosity makes up the fraction a $\alpha = 10^{-4}$--$10^{-5}$ 
of the particle energy flux ejected from the neutron star magnetosphere.
\item
The radio emission has a power-law spectrum with the spectral index ranging 
from $-1$ to $-3$; at the frequency
$\nu_{\rm max} \approx 3 \, P^{-1/2} \, \Gamma_{100}^{7/4}$ GHz, there is a 
spectral break with the change in the spectral index by approximately 1, 
and a low-frequency `cut-off' occurs at the frequency
\begin{equation}
\nu_{\rm min} \approx 120 \, P^{-1/2} \Gamma_{100}^{-3/4}B_{12}^{1/2} \, {\rm MHz}. 
\label{numin}
\end{equation}
Here, $\Gamma \approx <\gamma>$ is the characteristic Lorentz factor of the
plasma and $\Gamma_{100} = \Gamma/100$.
\end{itemize}

Below, we formulate the main properties of these modes in more detail, 
discuss theoretical predictions, and compare them with observations. 

\subsubsection{Mode classification.}
\label{4.2.1}

\begin{figure}
\begin{center}
\includegraphics[width=220pt]{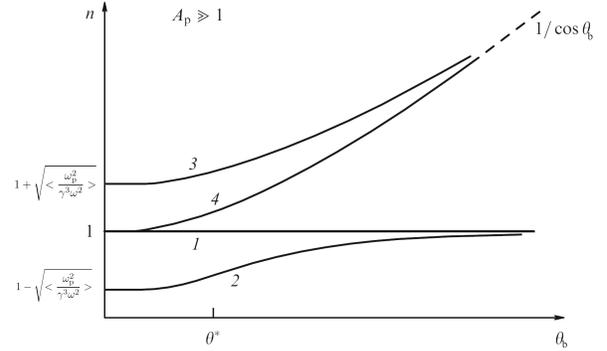}
\caption{Normal modes with the refractive indices $n \approx 1$ propagating
from the neutron star surface as a function of the angle $\theta_{\rm b}$ 
between the wave vector ${\bf k}$ and the external magnetic field ${\bf B}$. 
Waves 1 (extraordinary X-mode) and 2 (ordinary O-mode) can escape from the 
pulsar magnetosphere.
}
\label{fig21}
\end{center}
\end{figure}

The radio waves propagating in pulsar magnetospheres can be easily classified 
by analyzing the dielectric permittivity tensor in the limit of the infinite
magnetic field when particles can move only along magnetic field lines:
\begin{equation}
\varepsilon_{ij}(\omega, {\bf k}) =
\begin{bmatrix}
    1  &  0  & 0  \\
    0  &  1  & 0 \\
    0  &  0  & 1 - <\dfrac{\omega_{\rm pe}^2}{\gamma^3(\omega - {\bf kv})^2}>
\end{bmatrix}.
\end{equation}
Here, $\omega_{\rm pe} = (4 \pi e^2 n_{\rm e}/m_{\rm e})^{1/2}$ 
is the nonrelativistic electron plasma frequency and the angular 
brackets denote averaging over a distribution function. It turns 
out that this simple model allows discovering quite unusual 
properties. Indeed, we consider the limit case
\begin{equation}
A_{\rm p} = \frac{\omega_{\rm pe}^2}{\omega^2} \, \Gamma \gg 1,
\label{Ap}
\end{equation}
realized at sufficiently small distances from the neutron star,
$r < r_{\rm A}$, where [69]
\begin{equation}
r_{\rm A} \approx
10^2 R \cdot \lambda_4^{1/3}
\Gamma_{100}^{1/3}B_{12}^{1/3}\nu_\mathrm{GHz}^{-2/3}P^{-1/3}.
\label{eq:rO}
\end{equation}
In this parameter region, the solution of the corresponding dispersion 
equation yields the following four normal modes at small angles 
$\theta_{\rm b}$ between the wave vector ${\bf k}$ and the external 
magnetic field ${\bf B}$ (see Fig. 21):
\begin{eqnarray}
n_1 & = & 1,
\label{nj1} \\
n_2 & \approx & 1 + \frac{\theta_{\rm b}^2}{4}
- \left(<\frac{\omega_{\rm pe}^2}{\gamma^{3}\omega^2}>
+ \frac{\theta_{\rm b}^4}{16}\right)^{1/2},
\label{nj2}
\\
n_3 & \approx & 1 + \frac{\theta_{\rm b}^2}{4}
+ \left(<\frac{\omega_{\rm pe}^2}{\gamma^{3}\omega^2}>
+ \frac{\theta_{\rm b}^4}{16}\right)^{1/2},\\
n_4 & \approx & 1 + \frac{\theta_{\rm b}^2}{2}.
\label{nj4}
\end{eqnarray}

We note that in [204], where such waves were first investigated, 
'mode 2' was omitted. There, `mode 4' was called an `ordinary' 
wave,  which in fact corresponds to the ordinary wave at small 
angles $\theta_{\rm b}$. Subsequently, in many papers studying 
the theory of radio emission, only three types of radio waves 
were considered [205-207].

However, as we see, when the condition $A_{\rm p} \gg 1$ is satisfied
(see (120)) for large angles $\theta_{\rm b} > \theta^{\ast}$, where
\begin{equation}
\theta^{\ast} =  \left(<\frac{\omega_{\rm pe}^2}{\gamma^{3}\omega^2}>\right)^{1/4},
\label{THETA}
\end{equation} 
the dispersion curve of `mode 4' is close to that of a 
relativistic Alfv\'en wave with the refractive index 
\mbox{$n \approx 1/\cos\theta_{\rm b}$,}  and this means 
that at large angles $\theta_{\rm b}$ `mode 4' propagates 
along the magnetic field lines and cannot leave the 
neutron star magnetosphere  (as correctly stated in many
papers). Therefore, it is natural to refer to `mode 4' 
and not `mode 2' as ordinary. For small angles 
$\theta_{\rm b}$,  `mode 2' is a longitudinal plasma wave 
propagating toward the neutron star in the plasma rest 
frame; in the laboratory frame, of course, it moves in 
the opposite direction. For large angles $\theta_{\rm b}$,
it transforms into a transverse wave with $n_2 \approx 1$. 
It is exactly the difference between $n_{2}$ and unity at 
small angles $\theta_{\rm b} < \theta^{\ast}$ that leads 
to the ordinary wave deviating from the magnetic axis. 
Only very recently was the status of `mode 2' as being
ordinary finally recognized [15, 208].

We also note that the theory of polarized emission propagation 
that we developed in the last few years [175, 209] allows us to 
determine which of the modes, ordinary or extraordinary, forms 
the mean radio pulse profile. As was shown in [175, 209],  the 
same signs of the circular polarization (Stokes parameter V) 
and the derivative ${\rm d}p.a./{\rm d}\phi$ (the rate of change 
of the position angle of the linear polarization $p.a.$ with the 
pulse phase $\phi$) correspond to the X-mode, and different signs 
of these quantities to the O-mode. Lack of space prevents us from 
discussing this in greater detail.

\subsubsection{Radio luminosity as a function of spin period.}
\label{4.2.2}

In the BGI theory, as noted in Section 3.2.1, the radio luminosity $L_{\rm rad}$
amounts to a fraction $\alpha \sim 10^{-4}$--$10^{-5}$ of the total particle
energy flux. On the other hand, by Eqn (112), the energy carried by the particles 
for sufficiently fast pulsars (i.e., those with $Q < 1$) is $Q^{2}$ times the full 
energy losses  $W_{\rm tot} = -J_{\rm r}\Omega \dot\Omega$. Therefore, the ratio 
$L_{\rm rad}/W_{\rm tot}$ for pulsars with $Q < 1$ must scale as $\alpha \, Q^2$. 
As shown in Fig. 22, such a dependence is indeed observed (see also [210]). Here, 
both the slope of the curve (the power-law exponent $2.1 \pm 0.1$) and the 
transformation coefficient $\alpha^{\rm obs} = (2.0 \pm 0.2) \times 10^{-5}$ 
are consistent with the theory with good accuracy.

\begin{figure}
\begin{center}
\includegraphics[width=200pt]{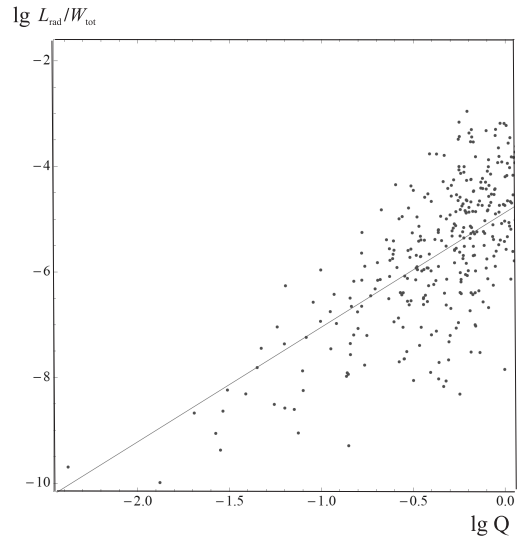}
\caption{Ratio $L_{\rm rad}/W_{\rm tot}$ as a function of $Q$ in Eqn (110) 
for radio pulsars with $Q < 1$. The data are from the ATNF catalogue [212]; 
the line corresponds to the dependence 
\mbox{$(2.0 \pm 0.2) \times 10^{-5} \, Q^{2.1 \pm 0.1}$.}  
}
\label{fig22}
\end{center}
\end{figure}

Incidentally, using the explicit expression (109) for $Q$, we can find
\begin{equation}
L_{\rm rad} \propto P^{-0.8}{\dot P}_{-15}^{0.2}.
\label{LRad}
\end{equation}
The dependence $L_{\rm rad} \propto P^{p}{\dot P}^{q}$ has been analyzed many times.
E.g., the authors of [211], based on data for 242 pulsars, derived 
$p = -0.86 \pm 0.20$ and \mbox{$q = 0.38 \pm 0.08$.} In [213], $p = -0.8 \pm 0.2$ 
was obtained. Clearly, these values are consistent with (127). As regards
the values $p = -1.39 \pm 0.09$ and $q = 0.48 \pm 0.04$ obtained in [214], 
the difference can be related to the analysis of all pulsars and not a 
subset with $Q < 1$.

\subsubsection{Pulse width a frequency dependence.}
\label{4.2.3}

As mentioned above, the main difference between two orthogonal modes
capable of leaving the neutron star magnetosphere is that the ordinary 
wave, unlike the extraordinary one, deviates from the magnetic axis. 
However, according to (121), this can occur only at sufficiently short 
distances $r < r_{\rm A}$ from the neutron star, at which the condition 
$A_{\rm p} > 1$ is satisfied in accordance with (120). As we see, the 
region $r < r_{\rm A}$ overlaps with the radio emission generation 
region, and therefore the ordinary wave refraction should be taken 
into account in the analysis of the mean radio pulse profiles. We 
note that in spite of a vast literature devoted to ordinary wave 
refraction in the neutron star magnetosphere [204, 215-218], in the 
`hollow cone' model the propagation effects have hardly been taken 
into account to date.

\begin{figure}
\begin{center}
\includegraphics[width=210pt]{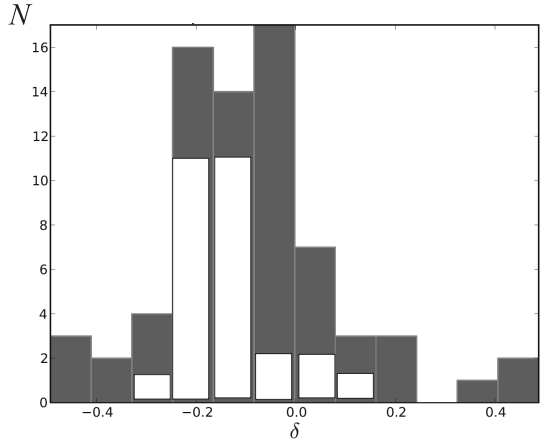}
\caption{Distribution of pulsars over the exponent $\delta$ [219] 
determining the dependence of the mean profile width on the frequency 
$W_{\rm r} \propto \nu^{\delta}$. The isolated part corresponds to 
pulsars in which the mean pulse is formed by an ordinary wave according 
to the method described in [175]. 
}
\label{fig23}
\end{center}
\end{figure}

Of course, the beam width $W_{\rm r}$ depends not only on the propagation 
effects but also on the radiation generation level $r_{\rm rad}$. The BGI 
model (which consistently takes the propagation effects into account) 
suggested that the radiation at a given frequency $r_{\rm rad}$ can be 
generated in a wide region up to the distance [172]
\begin{equation}
r_{\rm rad} \sim 3.5 \, R  \, \nu_{\rm GHz}^{-1} P^{-0.5} \Gamma_{100}^{1.75}, 
\label{rnu}
\end{equation}
depending, obviously, on the wavelength. As a result, the following expressions 
for the beam width $W_{\rm r}$ of two normal modes were obtained (here, we omit 
the weak dependence on the multiplicity  $\lambda$, the magnetic field $B_{0}$, 
and the characteristic Lorentz factor $\Gamma$ of the outflowing plasma):
\begin{eqnarray}
W_{\rm r}^{\rm O} & \approx & 7.8^{\circ} P^{-0.43} \nu_{\rm GHz}^{-0.14},
\label{Wr1} \\
W_{\rm r}^{\rm O}  & \approx & 10.8^{\circ} P^{-0.5} \nu_{\rm GHz}^{-0.29},
\label{Wr2} \\
W_{\rm r}^{\rm X}  & \approx & 3.6^{\circ} P^{-0.5} \nu_{\rm GHz}^{-0.5}. 
\label{Wr3}
\end{eqnarray}
The two expressions for the beam width of the ordinary wave correspond 
to two cases where the main contribution is made by the most and least 
distant regions from the neutron star. The accuracy of the BGI theory, 
in which, we recall, the radio emission generation is due to the three-wave 
decay of `mode 4', did not allow us to answer this question. 

Thus, the BGI theory predicts a power-law dependence of the mean radio pulse 
profile, $W_{\rm r}(\nu) \propto \nu^{\delta}$, with the power-law exponents
$\delta = -0.14$ and $\delta = -0.29$ for the ordinary and $\delta = -0.5$ 
for the extraordinary waves. Figure 23 shows the pulsar distribution over 
$\delta$ obtained quite recently by the new LOFAR radio telescope [219]. The 
isolated part corresponds to pulsars in which, according to the method described 
in our paper [175], the mean pulse is formed by an ordinary wave. It is 
seen that the pulsar distribution indeed has a sharp maximum at 
$\delta \approx -0.2$. On the other hand, for pulsar PSR B0943+10, 
in which the radio emission, according to our method, is formed by 
an extraordinary wave, the value $\delta \approx -0.5$ was obtained in [220].
Obviously, the BGI theory predictions are consistent with observations 
here as well.

\subsubsection{Low-frequency cut-off.}
\label{4.2.4}

As we have seen, the accuracy of the determination of spectral indices 
in the BGI model (which are expected to fall within the range from $-1$ 
to $-3$) is insufficient to carry out a detailed comparison between the
theory and observations. However, no precise measurements have been 
obtained from observations; these indices are found to be indeed 
distributed in a wide range (see, e.g., [220]). 

By contrast, the  low-frequency cut-off at $\nu_{\rm min}$ given by 
formula (118) is well defined and can be used to compare theory with 
observations. We note that in the BGI theory, the low-frequency cut-off 
is due to the very existence of the ordinary wave with the refractive 
index $n_{2}$ given by formula (123): at frequencies $\nu < \nu_{\rm min}$ 
[see (118)], this wave cannot escape the neutron star magnetosphere anymore.

Figure 24 presents a comparison between the theory and observations taken 
from two recent reviews [220, 221]. The line corresponds to the BGI prediction 
in Eqn (118). Apparently, here again, there is sufficiently good agreement.

\begin{figure}
\begin{center}
\includegraphics[width=\columnwidth]{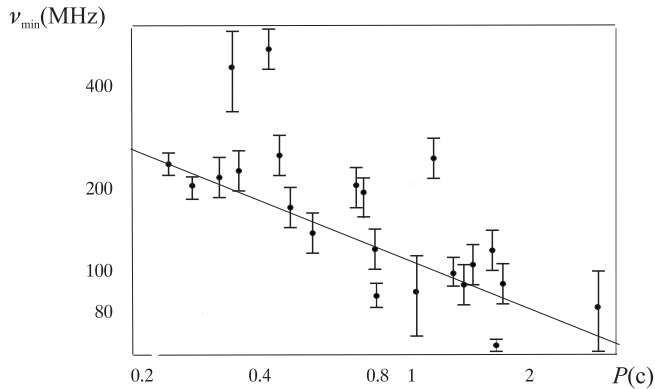}
\caption{Low-frequency cut-off $\nu_{\rm min}$ as a function 
of the period $P$. The data are from reviews [220, 221]. The 
line shows the prediction in Eqn (118).
}
\label{fig24}
\end{center}
\end{figure}

\subsubsection{Statistics of O- and X-modes.}

Finally, recent paper [222] enabled theoretical predictions about the 
O- and X-mode statistics to be tested using a sufficiently large number 
of data. This paper presents the most complete catalogue to date of
polarization properties of 600 radio pulsars at a frequency of 1.4 GHz. 
In 170 cases the polarization angle of the linear polarization, $p.a.$, 
and circular polarization (Stokes parameter V) were reliably determined, 
which allowed finding which of the polarization modes forms the mean radio 
pulse profile. It was found that in 100 pulsars the mean profile is formed 
by the X-mode (the same signs of V and ${\rm d} p.a./{\rm d}\phi$), and 
in 70 pulsars, by the O-mode (different signs of these quantities). Here, 
as was predicted in [175], most of the X-mode pulsars (86 out of 100) 
exhibited single-humped mean profiles, and most of the O-mode pulsars 
(48 out of 70) had double-humped profiles. Both the X- and O-mode pulsars 
had widths of double-humped profiles a factor of one and a half larger 
than single-humped ones. This should be the case, because in both cases 
double-humped and single-humped profiles correspond to the respective 
central and peripheral crossing of the emission beam.

\begin{figure}
\begin{center}
\includegraphics[width=220pt]{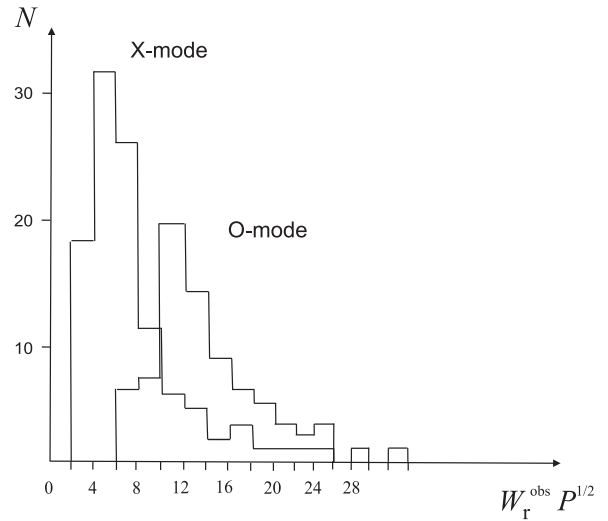}
\caption{Distribution of 170 O- and X-pulsars over $W_{\rm r} P^{1/2}$ 
(in degrees) taken from [222]. The distribution maxima exactly correspond
to predictions (129)-(131).  
}
\label{fig25}
\end{center}
\end{figure}

Moreover, the present statistics are sufficient to reliably argue 
that predictions (129)-(131) about the mean profile width $W_{\rm r}$ 
for these two modes are also in remarkable agreement with observations. 
Figure 25 shows distributions of O- and X-pulsars over the quantity 
$W_{\rm r} P^{1/2}$. Clearly, these histograms are indeed significantly 
shifted relative to each other, with the distribution maxima being 
consistent with predictions (129)-(131). As regards the sources 
with large values of $W_{\rm r} P^{1/2}$, they can easily be explained
by widening due to the nonorthogonality of rotators (the observed pulse
width is $W_{\rm r}^{\rm obs} = W_{\rm r}/\sin\chi$). As shown in [177], 
for large $W_{\rm r}^{\rm obs}$, the differential distribution should have 
the form $N(W_{\rm r}^{\rm obs}) \propto (W_{\rm r}^{\rm obs})^{-3}$.

\section{Conclusion}

The 50 years since the discovery of radio pulsars that have passed smoothly, 
leave a sense of distinct incompletion. Undoubtedly, much work should be done 
to understand even the general picture of the physical processes occurring
in neutron star magnetospheres. Unfortunately, pulsars do not provide us with 
clear experiments, and quantitative predictions are frequently difficult to 
check with observations, as is the case, for example, with the evolution of the
inclination angle ${\dot \chi}$ or with the mean braking index $n_{\rm br}$.

We note that Physics-Uspekhi has always quickly responded to all important 
events related to radio pulsar physics. As early as 1969, the first results 
of the national observations of pulsars both in the radio (Pushchino) [223-225] 
and in the optical (Crimean Astrophysical Observatory) [226, 227] ranges were 
published. In the same years, papers by the patriarchs of Russian (Soviet) 
astrophysicists Shklovskii [228, 229], Ginzburg [230-232], and Zel'dovich 
[233] were published without delay; for the first time, they performed a
serious analysis of physical processes in neutron star magnetospheres. The 
possibility of pulsar observations in different electromagnetic ranges were 
discussed in [234-236].

In later years, Physics-Uspekhi published papers devoted to both observations 
[237-240] and the theory of radio pulsar emission [241-243]. Reviews and 
conference reports were published about the internal structure of neutron stars 
[138, 139, 244-249], their evolution [152, 250], and general relativity effects 
[254-258]. In addition, it should be remembered that B B Kadomtsev, the 
editor-in-chief of Physics-Uspekhi for many years, wrote the book `On the 
Pulsar' [259].

Anyway, at present, the theory of pulsar magnetospheres and pulsar wind is 
a rapidly developing field in which dozens of researchers are working. 
Substantial progress has been achieved, including some quantitative predictions 
(for example, formulas for the total energy loss), which, we hope, will be
directly tested in the nearest future.

As regards our theory, it has not been much in demand to date, although, 
as shown above, many statements made more than thirty years ago are at 
least consistent with observations. However, as mentioned above, direct 
tests that could shed light on energy release and coherent radio emission 
mechanisms are not being carried out at present.

In conclusion, the author thanks D Barsukov, A Beloborodov, Ya N Istomin, 
O Kargaltsev, G Pavlov, A Potekhin, and especially A Philippov for the 
fruitful discussions of topics in this review. The author would also like 
to acknowledge A Jessner and J P\'etri, who carefully read the text of my
popular lecture at the Gamma-2016 conference (Heidelberg, Germany) devoted 
to similar topics, which enabled me to make several important improvements, 
and also L Arzamasskiy, A Galishnikova, E Novoselov, and M Rashkovetskiy
for help in the calculations. The author separately acknowledges the referee 
for numerous remarks that greatly improved the style of the paper.

This study was supported by the Russian Foundation for Basic Research grant 
17-02-00788.

\section{Appendix}
\label{appendix}

Following [36] (see also [50]), we recall how accounting for the
toroidal magnetic field time dependence inside the inner gap can influence 
the plasma generation dynamics. In contrast to the one-dimensional case 
considered in [113, 114] (which, by the way, does not enable one to take 
this into account), we here return to the classical idea of `sparks' 
proposed as early as 1975 by Ruderman and Sutherland [29].

We consider a strongly elongated quasi-cylindrical spark with 
a cross diameter $r_{0}$ in which a current $I$ due to secondary particle 
generation grows according to the law $I = I_{0} \exp(t/\tau)$. Here, 
$\tau$ is the characteristic time of the current growth that can be 
estimated as $\tau \approx {H}/{c}$, where $H$ is the vertical size 
of the acceleration region. Now, by substituting the time-dependent 
toroidal magnetic field $B_{\varphi}(t) = 2I(t)/(cr_{\perp})$ in the 
Maxwell equation $c \, \nabla \times {\bf E} = - \partial {\bf B}/\partial t$,
we obtain for the induced electric field
\begin{equation}
E_{z} \approx \frac{I(t)}{c^2\tau}.
\label{Et}
\end{equation}

Naturally, the direction of this field is opposite to the field 
inside the gap $E_{\rm RS} \approx 4 \pi \rho_{\rm GJ}H$. Therefore, 
after reaching some value $I_{\rm max}$ at which the electric 
field in the acceleration region is screened, further particle 
creation is impossible. Thus, we come to the conclusion that 
the longitudinal current in an individual spark cannot exceed 
$I_{\rm max} \approx c^2 \tau E_{\rm RS}$:
\begin{equation}
 I_{\rm max} \approx \frac{c\tau}{H} c \rho_{\rm GJ} H^2.
\label{IEt}
\end{equation}

It is very important that, as shown in [36], this electric 
field exists up to distances $r_{\perp} \sim H$ from the spark axis. 
Therefore, this effect was called `dynamical screening of the acceleration
region'. Because the total number of sparks inside the polar cap can 
be estimated as \mbox{$N \approx R_{0}^2/H^2$} in this case, we obtain the
total electric current
\begin{equation}
 I_{\rm tot} \approx \frac{c\tau}{H} I_{\rm GJ}.
\label{Itottot}
\end{equation}
With the estimate $\tau \approx {H}/{c}$, this means that 
the total longitudinal current cannot significantly exceed the local 
Goldreich current.

\vspace{1cm}

{\bf References}

\vspace{0.5cm}

1. Baade W, Zwicky F {\it Proc. Natl. Acad. Sci.} USA {\bf 20} 254 (1934)

2. Landau L {\it Phys. Z. Sowjetunion} {\bf 1} 285 (1932); Translated into
Russian: {\it Sobranie Trudov} (Collected Works) Vol. 1 (Moscow:
Nauka, 1969) p. 86

3. Yakovlev D G et al. {\it Phys. Usp.} {\bf 56} 289 (2013); {\it Usp. Fiz. Nauk} {\bf 183}
307 (2013)

4. Pacini F {\it Nature} {\bf 216} 567 (1967)

5. Kardashev N S {\it Sov. Astron.} {\bf 8} 643 (1965); {\it Astron. Zh.} {\bf 41} 807 (1964)

6. Hewish A et al. {\it Nature} {\bf 217} 709 (1968)

7. Zel'dovich Ya B {\it Sov. Phys. Dokl.} {\bf 9} 195 (1964); {\it Dokl. Akad. Nauk
SSSR} {\bf 155} 67 (1964)

8. Shklovsky I S  {\it Astrophys.~J.}  {\bf 148} L1 (1967)

9. Salpeter E E  {\it Astrophys.~J.}  {\bf 140} 796 (1964)

10. Giacconi R et al.  {\it Astrophys.~J.}  {\bf 167} L67 (1971)

11. Gold T {\it Nature} {\bf 218} 731 (1968)

12. Ostriker J P, Gunn J E  {\it Astrophys.~J.}  {\bf 157} 1395 (1969)

13. Gunn J E, Ostriker J P  {\it Astrophys.~J.}  {\bf 160} 979 (1970)

14. Gunn J E, Ostriker J P {\it Phys. Rev. Lett.} {\bf 22} 728 (1969)

15. Lyne A G, Graham-Smith F {\it Pulsar Astronomy} 3rd ed. (Cambridge:
Cambridge Univ. Press, 2006)

16. Max C, Perkins F {\it Phys. Rev. Lett.} {\bf 27} 1342 (1971)

17. Ass\'eo E, Kennel C F, Pellat R {\it Astron. Astrophys.} {\bf 65} 401 (1978)

18. Sturrock P A {\it Astrophys.~J.} {\bf 164} 529 (1971)

19. Zheleznyakov V V {\it Radiation in Astrophysical Plasmas} (Dordrecht:
Kluwer, 1996); {\it Elektromagnitnye Volny v Kosmicheskoi Plazme}
({\it Electromagnetic Waves in Cosmic Plasma}) (Moscow: Nauka,
1977)

20. Berestetskii V B, Lifshitz E M, Pitaevskii L P {\it Quantum 
Electrodynamics} (Oxford: Butterworth-Heinemann, 1999); Translated
from Russian: {\it Kvantovaya Elektrodinamika} (Moscow: Nauka,
1989)

21. Brice N M, Ioannidis G A {\it Icarus} {\bf 13} 173 (1970)

22. Radhakrishnan V, Cooke D J {\it Astrophys. Lett.} {\bf 3} 225 (1969)

23. Oster L, Sieber W {\it Astrophys.~J.} {\bf 210} 220 (1976)

24. Rankin J M {\it Astrophys.~J.} {\bf 274} 333 (1983)

25. Goldreich P, Julian W H {\it Astrophys.~J.} {\bf 157} 869 (1969)

26. Michel F C {\it Astrophys.~J.} {\bf 158} 727 (1969)

27. Beskin V S {\it Phys. Usp.} {\bf 53} 1199 (2010); {\it Usp. Fiz. Nauk} {\bf 180} 1241
(2010)

28. Popov M V, Rudnitskii A G, Soglasnov V A {\it Astron. Rep.} {\bf 61} 178
(2017); {\it Astron. Zh.} {\bf 94} 194 (2017)

29. Ruderman M A, Sutherland P G {\it Astrophys.~J.} {\bf 196} 51 (1975)

30. Al'ber Ya I, Krotova Z N,  Eidman V Ya {\it Astrophysics} {\bf 11} 189 (1975);
{\it Astrofiz.} {\bf 11} 283 (1975)

31. Kadomtsev B B, Kudryavtsev V S {\it JETP Lett.} {\bf 13} 9 (1971); {\it Pis'ma
Zh. Eksp. Teor. Fiz.} {\bf 13} 15 (1971)

32. Ginzburg V L, Usov V V {\it JETP Lett.} {\bf 15} 196 (1972); {\it Pis'ma Zh. Eksp.
Teor. Fiz.} {\bf 15} 280 (1972)

33. Chen H-H, Ruderman M A, Sutherland P G {\it Astrophys.~J.} {\bf 191} 473
(1974)

34. Hillebrandt W, M\"uller E {\it Astrophys.~J.} {\bf 207} 589 (1976)

35. Flowers E G et al. {\it Astrophys.~J.} {\bf 215} 291 (1977)

36. Beskin V S {\it Sov. Astron.} {\bf 26} 443 (1982); {\it Astron. Zh.} {\bf 59} 726 (1982)

37. Ghosh A, Chakrabarty S J {\it Astrophys. Astron.} {\bf 32} 377 (2011)

38. Fowler R H, Nordheim L {\it Proc. R. Soc. London A} {\bf 119} 173 (1928)

39. Cheng A F, Ruderman M A {\it Astrophys.~J.} {\bf 235} 576 (1980)

40. M\"uller  E{\it  Astron. Astrophys}. {\bf 130} 415 (1984)

41. Jones P B {\it Mon. Not. R. Astron. Soc.} {\bf 216} 503 (1985)

42. Neuhauser D, Langanke K, Koonin S E {\it Phys. Rev. A} {\bf 33} 2084 (1986)

43. Neuhauser D, Koonin S E, Langanke K {\it Phys. Rev. A} {\bf 36} 4163 (1987)

44. Fawley W M, Arons J, Scharlemann E T {\it Astrophys.~J.} {\bf 217} 227 (1977)

45. Arons J, Scharlemann E T {\it Astrophys.~J.} {\bf 231} 854 (1979)

46. Arons J {\it Astrophys.~J.} {\bf 248} 1099 (1981)

47. Daugherty J K, Harding A K {\it Astrophys.~J.} {\bf 273} 761 (1983)

48. Gurevich A V, Istomin Ya N {\it Sov. Phys. JETP} {\bf 62} 1 (1985); {\it Zh. Eksp.
Teor. Fiz.} {\bf 89} 3 (1985)

49. Hibschman J A, Arons J {\it Astrophys.~J.} {\bf 554} 624 (2001)

50. Istomin Ya N, Sob'yanin D N {\it JETP} {\bf 109} 393 (2009); {\it Zh. Eksp. Teor.
Fiz.} {\bf 136} 458 (2009)

51. Medin Z, Lai D {\it Mon. Not. R. Astron. Soc.} {\bf 406} 1379 (2010)

52. Barsukov D P, Kantor E M, Tsygan A I {\it Astron. Rep.} {\bf 51} 469 (2007);
{\it Astron. Zh.} {\bf 84} 523 (2007)

53. Gralla S E, Lupsasca A, Philippov A {\it Astrophys.~J.} {\bf 851} 137 (2017)

54. de Jager O C et al. {\it Astrophys.~J.} {\bf 457} 253 (1996)

55. Bucciantini N, Arons J, Amato E {\it Mon. Not. R. Astron. Soc.} {\bf 410} 381
(2011)

56. de Jager O C {\it Astrophys.~J.} {\bf 658} 1177 (2007)

57. Lyutikov M {\it Mon. Not. R. Astron. Soc.} {\bf 353} 1095 (2004)

58. Hooper D, Blasi P, Dario Serpico P J {\it Cosmology Astropart. Phys.}
{\bf 01} 025 (2009)

59. Malyshev D, Cholis I, Gelfand J {\it Phys. Rev. D} {\bf 80} 063005 (2009)

60. Mestel L {\it Astrophys. Space Sci.} {\bf 24} 289 (1973)

61. Michel F C {\it Astrophys.~J.} {\bf 180} 207 (1973)

62. Okamoto I {\it Mon. Not. R. Astron. Soc.} {\bf 167} 457 (1974)

63. Scharlemann E T, Wagoner R V {\it Astrophys.~J.} {\bf 182} 951 (1973)

64. Mestel L, Wang Y-M {\it Mon. Not. R. Astron. Soc.} {\bf 188} 799 (1979)

65. Landau L D, Lifshitz E M {\it Electrodynamics of Continuous Media}
(Oxford: Pergamon Press, 1984); Translated from Russian: {\it Elektrodinamika 
Sploshnykh Sred} (Moscow: Nauka, 1978)

66. Michel F C {\it Astrophys.~J.} {\bf 180} L133 (1973)

67. Beskin V S, Gurevich A V, Istomin Ya N {\it Sov. Phys. JETP} {\bf 58} 235
(1983); {\it Zh. Eksp. Teor. Fiz.} {\bf 85} 401 (1983)

68. Mestel L, Panagi P, Shibata S {\it Mon. Not. R. Astron. Soc.} {\bf 309} 388
(1999)

69. Beskin V S, Gurevich A V, Istomin Ya N {\it Physics of the Pulsar
Magnetosphere} (Cambridge: Cambridge Univ. Press, 1993)

70. Cerutti B, Beloborodov A M {\it Space Sci. Rev.} {\bf 207} 111 (2017)

71. Okamoto I {\it Mon. Not. R. Astron. Soc.} {\bf 173} 357 (1975)

72. Heinemann M, Olbert S J {\it Geophys. Res.} {\bf 83} 2457 (1978)

73. Ardavan H {\it Mon. Not. R. Astron. Soc.} {\bf 189} 397 (1979)

74. Heyvaerts J, Norman C {\it Astrophys.~J.} {\bf 347} 1055 (1989)

75. Bogovalov S V {\it Sov. Astron. Lett.} {\bf 16} 362 (1990); {\it Pis'ma Astron. Zh.}
{\bf 16} 844 (1990)

76. Pelletier G, Pudritz R E {\it Astrophys.~J.} {\bf 394} 117 (1992)

77. Tomimatsu A {\it Publ. Astron. Soc. Jpn.} {\bf 46} 123 (1994)

78. Beskin V S {\it Sov. Astron. Lett.} {\bf 16} 286 (1990); {\it Pis'ma Astron. Zh.} {\bf 16}
665 (1990)

79. Muslimov A G, Tsygan A I {\it Sov. Astron.} {\bf 34} 133 (1990); {\it Astron. Zh.}
{\bf 67} 263 (1990) 

80. Muslimov A G, Tsygan A I {\it Mon. Not. R. Astron. Soc.} {\bf 255} 61 (1992)

81. Dermer C D {\it Astrophys.~J.} {\bf 360} 197 (1990)

82. Sturner S J {\it Astrophys.~J.} {\bf 446} 292 (1995)

83. Jessner A, Lesch H, Kunzl T {\it Astrophys.~J.} {\bf 547} 959 (2001)

84. Coroniti F V {\it Astrophys.~J.} {\bf 349} 538 (1990)

85. Michel F C {\it Astrophys.~J.} {\bf 431} 397 (1994)

86. Kennel C F, Coroniti F V {\it Astrophys.~J.} {\bf 283} 694 (1984)

87. Kennel C F, Coroniti F V {\it Astrophys.~J.} {\bf 283} 710 (1984)

88. Krause-Polstorff J, Michel F C {\it Mon. Not. R. Astron. Soc.} {\bf 213} 43P
(1985)

89. Rylov Yu A {\it Astrophys. Space Sci.} {\bf 51} 59 (1977)

90. Jackson E A {\it Astrophys.~J.} {\bf 206} 831 (1976)

91. Contopoulos I, Kazanas D, Fendt C {\it Astrophys.~J.} {\bf 511} 351 (1999)

92. Ogura J, Kojima Y {\it Prog. Theor. Phys.} {\bf 109} 619 (2003)

93. Goodwin S P et al. {\it Mon. Not. R. Astron. Soc.} {\bf 349} 213 (2004)

94. Gruzinov A {\it Phys. Rev. Lett.} {\bf 94} 021101 (2005)

95. Contopoulos I {\it Astron. Astrophys.} {\bf 442} 579 (2005)

96. Komissarov S S {\it Mon. Not. R. Astron. Soc.} {\bf 367} 19 (2006)

97. McKinney J C {\it Mon. Not. R. Astron. Soc.} {\bf 368} L30 (2006)

98. Timokhin A N {\it Mon. Not. R. Astron. Soc.} {\bf 368} 1055 (2006)

99. Lovelace R V E, Turner L, Romanova M M {\it Astrophys.~J.} {\bf 652} 1494
(2006)

100. Bogovalov S V {\it Astron. Astrophys.} {\bf 349} 1017 (1999)

101. Ingraham R L {\it Astrophys.~J.} {\it 186} 625 (1973)

102. Lyubarsky Y, Kirk J G {\it Astrophys.~J.} {\bf 547} 437 (2001)

103. Shibata S {\it Mon. Not. R. Astron. Soc.} {\bf 287} 262 (1997)

104. Beloborodov A M {\it Astrophys.~J.} {\bf 683} L41 (2008)

105. Spitkovsky A {\it Astrophys.~J.} {\bf 648} L51 (2006)

106. Kalapotharakos C, Contopoulos I {\it Astron. Astrophys.} {\bf 496} 495
(2009)

107. Kalapotharakos C, Contopoulos I, Kazanas D {\it Mon. Not. R.
Astron. Soc.} {\bf 420} 2793 (2012)

108. P\'etri J {\it Mon. Not. R. Astron. Soc.} {\bf 424} 605 (2012)

109. Li J G, Spitkovsky A, Tchekhovskoy A {\it Astrophys.~J.} {\bf 746} L24 (2012)

110. Tchekhovskoy A, Spitkovsky A, Li J G {\it Mon. Not. R. Astron. Soc.}
{\bf 435} L1 (2013)

111. Philippov A, Tchekhovskoy A, Li J G {\it Mon. Not. R. Astron. Soc.} {\bf 441}
1879 (2014)

112. Tchekhovskoy A, Philippov A, Spitkovsky A {\it Mon. Not. R. Astron.
Soc.} {\bf 457} 3384 (2016)

113. Timokhin A N {\it Mon. Not. R. Astron. Soc.} {\bf 408} 2092 (2010)

114. Timokhin A N, Arons J {\it Mon. Not. R. Astron. Soc.} {\bf 429} 20 (2013)

115. Timokhin A N, Harding A K {\it Astrophys.~J.} {\bf 810} 144 (2015)

116. Li J, Spitkovsky A, Tchekhovskoy A {\it Astrophys.~J.} {\bf 746} 60 (2012)

117. Kalapotharakos C et al. {\it Astrophys.~J.} {\bf 749} 2 (2012)

118. Philippov A A et al. {\it Astrophys.~J.} {\bf 815} L19 (2015)

119. Philippov A A, Spitkovsky A {\it Astrophys.~J.} {\bf 785} L33 (2014)

120. Chen A Y, Beloborodov A M {\it Astrophys.~J.} {\bf 795} L22 (2014)

121. Yuki S, Shibata S {\it Publ. Astron. Soc. Jpn.} {\bf 64} 43 (2012)

122. Philippov A A, Spitkovsky A, Cerutti B {\it Astrophys.~J.} {\bf 801} L19
(2015)

123. Belyaev M A {\it Mon. Not. R. Astron. Soc.} {\bf 449} 2759 (2015)

124. Cerutti B et al. {\it Mon. Not. R. Astron. Soc.} {\bf 448} 606 (2015)

125. Kalapotharakos C, Harding A K, Kazanas D {\it Astrophys.~J.} {\bf 793} 97
(2014)

126. Lyne A G, Manchester R N, Taylor J H {\it Mon. Not. R. Astron. Soc.}
{\bf 213} 613 (1985)

127. Maciesiak K, Gil J, Ribeiro V A R M {\it Mon. Not. R. Astron. Soc.} {\bf 414}
1314 (2011)

128. Malov I F {\it Radiopul'sary} ({\it Radio Pulsars}) (Moscow: Nauka, 2004)

129. Narayan R, Vivekanand M {\it Astron. Astrophys.} {\bf 113} L3 (1982)

130. von Hoensbroech A, Xilouris K M {\it Astron. Astrophys.} {\bf 324} 981
(1997)

131. Everett J E, Weisberg J M {\it Astrophys.~J.} {\bf 553} 341 (2001)

132. Mitra D, Rankin J M {\it Astrophys.~J.} {\bf 727} 92 (2011)

133. Rookyard S C, Weltevrede P, Johnston S {\it Mon. Not. R. Astron. Soc.}
{\bf 446} 3367 (2015)

134. Usov V V, Melrose D B {\it Astrophys.~J.} {\bf 464} 306 (1996)

135. Medin Z, Lai D {\it Mon. Not. R. Astron. Soc.} {\bf 382} 1833 (2007)

136. Lai D, Salpeter E E {\it Astrophys.~J.} {\bf 491} 270 (1997)

137. Shapiro S L, Teukolsky S A {\it Black Holes, White Dwarfs, and Neutron
Stars: The Physics of Compact Objects} (New York: Wiley, 1983);
Translated into Russian: {\it Chernye Dyry, Belye Karliki i Neitronnye
Zvezdy} (Moscow: Mir, 1985)

138. Yakovlev D G, Levenfish K P, Shibanov Yu A {\it Phys. Usp.} {\bf 42} 737
(1999); {\it Usp. Fiz. Nauk} {\bf 169} 825 (1999)

139. Potekhin A Yu {\it Phys. Usp.} {\bf 57} 735 (2014); {\it Usp. Fiz. Nauk} {\bf 184} 793
(2014)

140. Kargaltsev O et al. {\it Astrophys.~J. Suppl.} {\bf 201} 37 (2012)

141. Szary A, PhD Thesis (Zielena G ora: Univ. of Zielena G ora, 2015);
arXiv:1304.4203

142. Vigan\`o D et al. {\it Mon. Not. R. Astron. Soc.} {\bf 434} 123 (2013)

143. Gil J, Melikidze G I, Geppert U Astron. Astrophys. {\bf 407} 315 (2003)

144. Gil J A, SendykM {\it Astrophys.~J.} {\bf 541} 351 (2000)

145. Rankin J {\it Astrophys.~J.} {\bf 301} 901 (1986)

146. Rankin J, Rosen R {\it Mon. Not. R. Astron. Soc.} {\bf 439} 3860 (2014)

147. Weltevrede P, Stappers B W, Edwards R T {\it Astron. Astrophys.} {\bf 469}
607 (2007)

148. Keith M J et al. {\it Mon. Not. R. Astron. Soc.} {\bf 402} 745 (2010)

149. Gruzinov A {\it Astrophys.~J. Lett.} {\bf 647} L119 (2006)

150. Landau L D, Lifshitz E M The {\it Classical Theory of Fields} (Oxford:
Butterworth-Heinemann, 2000); Translated from Russian: {\it Teoriya
Polya} (Moscow: Nauka, 1973)

151. Beskin V S, Istomin Ya N, Philippov A A {\it Phys. Usp.} {\bf 56} 164 (2013);
{\it Usp. Fiz. Nauk} {\bf 183} 179 (2013)

152. Beskin V S, Zheltoukhov A A {\it Phys. Usp.} {\bf 57} 799 (2014); {\it Usp. Fiz.
Nauk} {\bf 184} 865 (2014)

153. Deutsch A J {\it Ann. d'Astrophys.} {\bf 18} 1 (1955)

154. Michel F C, Li H {\it Phys. Rep.} {\bf 318} 227 (1999)

155. Beskin V S {\it Osesimmetrichnye Statsionarnye Techeniya v Astrofizike}
({\it Axisymmetric Stationary Flows in Astrophysics}) (Moscow:
Fizmatlit, 2005)

156. Good M L, Ng K K {\it Astrophys.~J.} {\bf 299} 706 (1985)

157. Beskin V S {\it MHD Flows in Compact Astrophysical Objects} (Heidelberg: Springer, 2009)

158. Beskin V S, Nokhrina E E {\it Astron. Lett.} {\bf 30} 685 (2004); {\it Pis'ma
Astron. Zh.} {\bf 30} 754 (2004)

159. Bai X-N, Spitkovsky A {\it Astrophys.~J.} {\bf 715} 1282 (2010)

160. Beskin V S et al. {\it J. Phys. Conf. Ser.} {\bf 932} 012012 (2017)

161. Davis L, GoldsteinM {\it Astrophys.~J.} {\bf 159} L81 (1970)

162. Goldreich P {\it Astrophys.~J.} {\bf 160} L11 (1970)

163. Melatos A {\it Mon. Not. R. Astron. Soc.} {\bf 313} 217 (2000)

164. Ostriker J P, Gunn J E {\it Astrophys.~J.} {\bf 157} 1395 (1969)

165. Mestel L, Moss D {\it Mon. Not. R. Astron. Soc.} {\bf 361} 595 (2005)

166. Michel F C {\it Theory of Neutron Star Magnetospheres} (Chicago: Univ.
of Chicago Press, 1991)

167. Istomin Ya N, in {\it Progress in Neutron Star Research} (Ed. A P Wass)
(New York: Nova Science Publ., 2005) p. 27

168. Barsukov D P, Tsygan A I {\it Mon. Not. R. Astron. Soc.} {\bf 409} 1077
(2010)

169. Biryukov A, Beskin G, Karpov S {\it Mon. Not. R. Astron. Soc.} {\bf 420} 103
(2012)

170. Beskin V S, Gurevich A V, Istomin Ya N {\it Astrophys. Space Sci.} {\bf 102}
301 (1984)

171. Istomin Ya N {\it JETP} {\bf 67} 1380 (1988); {\it Zh. Eksp. Teor. Fiz.} {\bf 94} 148
(1988)

172. Beskin V S, Gurevich A V, Istomin Ya N {\it Astrophys. Space Sci.} {\bf 146}
205 (1988)

173. Beskin V S, Nokhrina E E {\it Astrophys. Space Sci.} {\bf 308} 569 (2007)

174. Istomin Ya N, Philippov A A, Beskin V S {\it Mon. Not. R. Astron. Soc.}
{\bf 422} 232 (2012)

175. Beskin V S, Philippov A A {\it Mon. Not. R. Astron. Soc.} {\bf 425} 814 (2012)

176. Prokofev V V, Arzamasskiy L I, Beskin V S {\it Mon. Not. R. Astron.
Soc.} {\bf 454} 2146 (2015)

177. Arzamasskiy L I, Beskin V S, Pirov K K {\it Mon. Not. R. Astron. Soc.}
{\bf 466} 2325 (2017)

178. Beskin V S, Malyshkin L M {\it Mon. Not. R. Astron. Soc.} {\bf 298} 847
(1998)

179. Gruzinov A, arXiv:1303.4094

180. Aharonian F A, Bogovalov S V, Khangulyan D {\it Nature} {\bf 482} 507
(2012)

181. Aleksic J et al. {\it Astrophys.~J.} {\bf 742} 43 (2011)

182. Beskin V S, Rafikov R R {\it Mon. Not. R. Astron. Soc.} {\bf 313} 433 (2000)

183. Smith F G {\it Pulsars} (Cambridge: Cambridge Univ. Press, 1977);
Translated into Russian: {\it Pul'sary} (Moscow: Mir 1979)

184. Manchester R N, Taylor J H {\it Pulsars} (San Francisco:
W.H. Freeman, 1977); Translated into Russian: {\it Pul'sary} (Moscow: Mir, 1980)

185. Arzamasskiy L, Philippov A, Tchekhovskoy A {\it Mon. Not. R. Astron.
Soc.} {\bf 453} 3540 (2015)

186. Archibald A et al. {\it Astrophys.~J.} {\bf 819} L16 (2016)

187. Istomin Ya N, Shabanova T V {\it Astron. Rep.} {\bf 51} 119 (2007); {\it Astron.
Zh.} {\bf 84} 139 (2007)

188. Lyne A et al. {\it Science} {\bf 342} 598 (2013)

189. Rankin J M {\it Astrophys. J.} {\bf 352} 247 (1990)

190. Tauris T M, Manchester R N {\it Mon. Not. R. Astron. Soc.} {\bf 298} 625
(1998)

191. Faucher-Gigu\`ere C-A, Kaspi V M {\it Astrophys.~J.} {\bf 643} 332 (2006)

192. Weltevrede P, Johnston S {\it Mon. Not. R. Astron. Soc.} {\bf 387} 1755 (2008)

193. Young M D T et al. {\it Mon. Not. R. Astron. Soc.} {\bf 402} 1317 (2010)

194. Gullon M et al. {\it Mon. Not. R. Astron. Soc.} {\bf 443} 1891 (2014)

195. Vivekanand M, Narayan R J. Astrophys. Astron. {\bf 2} 315 (1981)

196. Stollman G M Astron. Astrophys. {\bf 178} 143 (1987)

197. Lorimer D R et al. {\it Mon. Not. R. Astron. Soc.} {\bf 263} 403 (1993)

198. Popov S B et al. {\it Mon. Not. R. Astron. Soc.} {\bf 401} 2675 (2010)

199. Igoshev A P, Popov S B {\it Mon. Not. R. Astron. Soc.} {\bf 444} 1066 (2014)

200. Barsukov D P, Polyakova P I, Tsygan A I {\it Astron. Rep.} {\bf 53} 86 (2009);
{\it Astron. Zh.} {\bf 86} 95 (2009)

201. Goglichidze O A, Barsukov D P, Tsygan A I {\it Mon. Not. R. Astron.
Soc.} {\bf 451} 2564 (2015)

202. Tong H, Kou F F {\it Astrophys.~J.} {\bf 837} 117 (2017)

203. Eksi K Y et al. {\it Astrophys.~J.} {\bf 823} 34 (2017)

204. Barnard J J, Arons J {\it Astrophys.~J.} {\bf 302} 137 (1986)

205. Lyutikov M J. {\it Plasma Phys.} {\bf 62} 65 (1999)

206. Usov V V, in {\it On the Present and Future of Pulsar Astronomy},
26th Meeting of the IAU, Joint Discussion 2, 16 + 17 August, 2006,
Prague, Czech Republic, JD02, id.\#3

207. Lyubarsky Yu {\it AIP Conf. Proc.} {\bf 983} 29 (2008)

208. Noutsos A et al. {\it Astron. Astrophys.} {\bf 576} A62 (2015)

209. Hakobyan H L, Beskin V S, Philippov A A {\it Mon. Not. R. Astron.
Soc.} {\bf 469} 2704 (2017)

210. Szary A et al. {\it Astrophys.~J.} {\bf 784} 59 (2014)

211. Proszynski M, Przybycien D, in {\it Birth and Evolution of Neutron
Stars.} Issues Raised by Millisecond Pulsars. Proc. of the NRAO
Workshop, Green Bank, West Virginia, June 6 - 8, 1984 (Eds
S P Reynolds, D R Stinebring) (Green Bank: National Radio
Astronomy Observatory, 1984) p. 151

212. Manchester R N et al. {\it Astron. J.} {\bf 129} 1993 (2005)

213. Malov I F, Malov O I {\it Astron. Rep.} {\bf 50 483} (2006); {\it Astron. Zh.} {\bf 83} 542
(2006)

214. Bates S D et al. {\it Mon. Not. R. Astron. Soc.} {\bf 439} 2893 (2014)

215. Lyubarskii Y E, Petrova S A {\it Astron. Astrophys.} {\bf 333} 181 (1998)

216. Petrova S A, Lyubarskii Y E {\it Astron. Astrophys.} {\bf 355} 1168 (2000)

217. Wang C, Lai D, Han J {\it Mon. Not. R. Astron. Soc.} {\bf 403} 569 (2010)

218. Wang C, Han J L, Lai D {\it Mon. Not. R. Astron. Soc.} {\bf 417} 1183 (2011)

219. Pilia M et al. {\it Astron. Astrophys.} {\bf 586} A92 (2016)

220. Bilous A V et al. {\it Astron. Astrophys.} {\bf 591} A134 (2016)

221. Murphy T et al. {\it Publ. Astron. Soc. Australia} {\bf 34} e020 (2017)

222. Johnston S, Kerr M {\it Mon. Not. R. Astron. Soc.} {\bf 474} 4629 (2018)

223. Alekseev Yu I, Vitkevich V V, Shitov Yu P {\it Sov. Phys. Usp.} {\bf 12} 805
(1970); {\it Usp. Fiz. Nauk} {\bf 99} 522 (1969)

224. Alekseev Yu I et al. {\it Sov. Phys. Usp.} {\bf 12} 806 (1970); {\it Usp. Fiz. Nauk} {\bf 99}
523 (1969)

225. Vitkevich V V et al. {\it Sov. Phys. Usp.} {\bf 12} 806 (1970); {\it Usp. Fiz. Nauk} {\bf 99}
523 (1969)

226. Shakhovskoi N M, Efimov Yu S, Pronik V I {\it Sov. Phys. Usp.} {\bf 12} 804
(1970); {\it Usp. Fiz. Nauk} {\bf 99} 520 (1969)

227. Pronik V I, Pronik I I, Chuvaev K K {\it Sov. Phys. Usp.} {\bf 12} 805 (1970);
{\it Usp. Fiz. Nauk} {\bf 99} 521 (1969)

228. Shklovskii I S {\it Sov. Phys. Usp.} {\bf 11} 435 (1968); {\it Usp. Fiz. Nauk} {\bf 95} 249
(1968)

229. Shklovskii I S {\it Sov. Phys. Usp.} {\bf 12} 808 (1970); {\it Usp. Fiz. Nauk} {\bf 99} 526
(1969)

230. Ginzburg V L {\it Sov. Phys. Usp.} {\bf 12} 800 (1970); {\it Usp. Fiz. Nauk} {\bf 99} 514
(1969)

231. Ginzburg V L {\it Sov. Phys. Usp.} {\bf 14} 83 (1971); {\it Usp. Fiz. Nauk} {\bf 103} 393
(1971)

232. Ginzburg V L {\it Sov. Phys. Usp.} {\bf 14} 229 (1971); {\it Usp. Fiz. Nauk} {\bf 103} 770
(1971)

233. Zel'dovich Ya B {\it Sov. Phys. Usp.} {\bf 16} 559 (1974); {\it Usp. Fiz. Nauk} {\bf 110}
441 (1973)

234. Kardashev N S {\it Sov. Phys. Usp.} {\bf 12} 808 (1970); {\it Usp. Fiz. Nauk} {\bf 99} 526
(1969)

235. Slysh V I {\it Sov. Phys. Usp.} {\bf 12} 808 (1970); {\it Usp. Fiz. Nauk} {\bf 99} 526 (1969)

236. Stepanyan A A et al. {\it Sov. Phys. Usp.} {\bf 12} 806 (1970); Usp. Fiz. Nauk
{\bf 99} 523 (1969)

237. Vitkevich V V, Malov I F, Shitov Yu P {\it Sov. Phys. Usp.} {\bf 12} 809
(1970); {\it Usp. Fiz. Nauk} {\bf 99} 527 (1969)

238. Shitov Yu P {\it Sov. Phys. Usp.} {\bf 16} 288 (1973); {\it Usp. Fiz. Nauk} {\bf 109} 775
(1973)

239. Kuz'min A D {\it Sov. Phys. Usp.} {\bf 31} 881 (1988); {\it Usp. Fiz. Nauk} {\bf 156} 181
(1988)

240. Shabanova T V Phys. Usp. {\bf 37} 618 (1994); {\it Usp. Fiz. Nauk} {\bf 164} 662
(1994)

241. Erukhimov L M {\it Sov. Phys. Usp.} {\bf 12} 806 (1970); {\it Usp. Fiz. Nauk} {\bf 99}
523 (1969)

242. Zheleznyakov V V {\it Sov. Phys. Usp.} {\bf 12} 807 (1970); {\it Usp. Fiz. Nauk} {\bf 99}
524 (1969)

243. Zheleznyakov V V {\it Sov. Phys. Usp.} {\bf 16} 289 (1973); {\it Usp. Fiz. Nauk} {\bf 109}
777 (1973)

244. Kirzhnits D A {\it Sov. Phys. Usp.} {\bf 14} 512 (1972); {\it Usp. Fiz. Nauk} {\bf 104} 489
(1971)

245. Beskin V S {\it Sov. Phys. Usp.} {\bf 30} 733 (1987); {\it Usp. Fiz. Nauk} {\bf 152} 683
(1987)

246. Sedrakyan D M, Shakhabasyan K M {\it Sov. Phys. Usp.} {\bf 34} 555 (1991);
Usp. Fiz. Nauk {\bf 16} 3 (1991)

247. Kirzhnits D A Phys. Usp. {\bf 38} 791 (1995); {\it Usp. Fiz. Nauk} {\bf 165} 829
(1995)

248. Kirzhnits D A, Yudin S N {\it Phys. Usp.} {\bf 38} 1283 (1995); {\it Usp. Fiz. Nauk}
{\bf 165} 1335 (1995)

249. Yakovlev D G {\it Phys. Usp.} {\bf 44} 823 (2001); {\it Usp. Fiz. Nauk} {\bf 171} 866
(2001)

250. Bisnovatyi-Kogan G S {\it Phys. Usp.} {\bf 49} 53 (2006); {\it Usp. Fiz. Nauk} {\bf 176}
59 (2006)

251. Beskin V S, Gurevich A V, Istomin Ya N {\it Sov. Phys. Usp.} {\bf 26} 1006
(1983); {\it Usp. Fiz. Nauk} {\bf 141} 539 (1983)

252. Beskin V S, Gurevich A V, Istomin Ya N {\it Sov. Phys. Usp.} {\bf 29} 946
(1986); {\it Usp. Fiz. Nauk} {\bf 150} 257 (1986)

253. Beskin V S {\it Phys. Usp.} {\bf 42} 1071 (1999); {\it Usp. Fiz. Nauk} {\bf 169} 1169
(1999)

254. Weisberg J M, Taylor J H, Fowler L A {\it Sci. Am.} 245 (10) 66 (1981);
{\it Usp. Fiz. Nauk} {\bf 137} 707 (1982)

255. Grishchuk L P {\it Sov. Phys. Usp.} {\bf 31} 940 (1988); {\it Usp. Fiz. Nauk} {\bf 156} 297
(1988)

256. Hulse R A {\it Rev. Mod. Phys.} {\bf 66} 699 (1994); {\it Usp. Fiz. Nauk} {\bf 164} 743
(1994)

257. Taylor J H (Jr.) {\it Rev. Mod. Phys.} {\bf 66} 711 (1994); {\it Usp. Fiz. Nauk} {\bf 164}
757 (1994)

258. Will C M {\it Usp. Fiz. Nauk} {\bf 164} 765 (1994); {\it Phys. Usp.} {\bf 37} 697 (1994)

259. Kadomsev B B {\it On the Pulsar} (Singapore: World Scientific, 2009);
Translated from Russian: Kadomtsev B B {\it Na Pul'sare} (Izhevsk:
Regulyarnaya i Khaoticheskaya Dinamika, 2001); Translated into
Italian: Kadomtsev B B {\it Sulla Pulsar} (Ferrara: Akousmata, 2013)

\end{document}